\documentclass[twocolumn]{aastex62}

\graphicspath{{./}{figures/}}
\usepackage{graphicx}
\usepackage{amsmath}
\usepackage{amssymb}
\usepackage{lineno}
\DeclareGraphicsExtensions{.,.png}
\accepted{to ApJ on March 24, 2021}

\shorttitle{A search for off-nuclear AGN with ZTF}
\shortauthors{Ward et al.}

\begin{document}

\begin{abstract}
A supermassive black hole (SMBH) ejected from the potential well of its host galaxy via gravitational wave recoil carries important information about the mass ratio and spin alignment of the pre-merger SMBH binary.  Such a recoiling SMBH may be detectable as an active galactic nucleus (AGN) broad line region offset by up to 10\,kpc from a disturbed host galaxy. We describe a novel methodology using forward modeling with \texttt{The Tractor} to search for such offset AGN in a sample of 5493 optically variable AGN detected with the Zwicky Transient Facility (ZTF). We present the discovery of 9 AGN which may be spatially offset from their host galaxies and are candidates for recoiling SMBHs. Five of these offset AGN exhibit double-peaked broad Balmer lines which may arise from unobscured accretion disk emission and four show radio emission indicative of a relativistic jet. The fraction of double-peaked emitters in our spatially offset AGN sample is significantly larger than the 16\% double-peaked emitter fraction observed for ZTF AGN overall. In our sample of variable AGN we also identified 52 merging galaxies, including a new spectroscopically confirmed dual AGN. Finally, we detected the dramatic rebrightening of SDSS1133, a previously discovered variable object and recoiling SMBH candidate, in ZTF. The flare was accompanied by the re-emergence of strong P-Cygni line features indicating that it may be an outbursting luminous blue variable star. 

\vspace{1cm}
\end{abstract}

\title{AGN on the move: A search for off-nuclear AGN from recoiling SMBHs and ongoing galaxy mergers with the Zwicky Transient Facility}

\correspondingauthor{Charlotte Ward}
\email{charlotteward@astro.umd.edu}

\author[0000-0002-4557-6682]{Charlotte Ward}
\affil{Department of Astronomy, University of Maryland, College Park, MD  20742, USA} 
\author[0000-0002-0786-7307]{Suvi Gezari}
\affil{Space Telescope Science Institute, 3700 San Martin Dr., Baltimore, MD 21218, USA}
\affil{Department of Astronomy, University of Maryland, College Park, MD  20742, USA} 
\author[0000-0001-9676-730X]{Sara Frederick}
\affiliation{Department of Astronomy, University of Maryland, College Park, MD  20742, USA}
\author[0000-0002-5698-8703]{Erica Hammerstein}
\affil{Department of Astronomy, University of Maryland, College Park, MD  20742, USA}
\author[0000-0002-3389-0586]{Peter Nugent}
\affiliation{Lawrence Berkeley National Laboratory, 1 Cyclotron Road, Berkeley, CA 94720, USA}
\affiliation{Department of Astronomy, University of California, Berkeley, Berkeley, CA 94720, USA}
\author[0000-0002-3859-8074]{Sjoert van Velzen}
\affiliation{Department of Astronomy, University of Maryland, College Park, MD 20742, USA}
\affiliation{Center for Cosmology and Particle Physics, New York University, NY 10003}
\author{Andrew Drake}
\affiliation{Division of Physics, Mathematics, and Astronomy, California Institute of Technology, Pasadena, CA 91125, USA}
\author{Abigail García-Pérez}
\affiliation{Instituto Nacional de Astrofísica, Óptica y Electrónica, Tonantzintla, Puebla 72840, México}
\author{Immaculate Oyoo}
\affiliation{Prince George’s Community College, Largo, MD 20774, USA}
\author[0000-0001-8018-5348]{Eric C. Bellm}
\affiliation{DIRAC Institute, Department of Astronomy, University of Washington, 3910 15th Avenue NE, Seattle, WA 98195, USA}
\author[0000-0001-5060-8733]{Dmitry A. Duev}
\affiliation{Division of Physics, Mathematics, and Astronomy, California Institute of Technology, Pasadena, CA 91125, USA}
\author[0000-0002-3168-0139]{Matthew J. Graham}
\affiliation{Division of Physics, Mathematics, and Astronomy, California Institute of Technology, Pasadena, CA 91125, USA}
\author[0000-0002-5619-4938]{Mansi M. Kasliwal}
\affil{Division of Physics, Mathematics, and Astronomy, California Institute of Technology, Pasadena, CA 91125, USA}
\author{Stephen Kaye}
\affil{Caltech Optical Observatories, California Institute of Technology, Pasadena, CA  91125, USA}
\author[0000-0003-2242-0244]{Ashish A. Mahabal}
\affiliation{Division of Physics, Mathematics and Astronomy, California
Institute of Technology, Pasadena, CA 91125, USA}
\affiliation{Center for Data Driven Discovery, California Institute of
Technology, Pasadena, CA 91125, USA}
\author[0000-0002-8532-9395]{Frank J. Masci}
\affiliation{IPAC, California Institute of Technology, 1200 E. California
             Blvd, Pasadena, CA 91125, USA}
\author[0000-0001-7648-4142]{Ben Rusholme}
\affiliation{IPAC, California Institute of Technology, 1200 E. California
             Blvd, Pasadena, CA 91125, USA}
\author[0000-0001-6753-1488]{Maayane T. Soumagnac}
\affiliation{Lawrence Berkeley National Laboratory, 1 Cyclotron Road, Berkeley, CA 94720, USA}
\affiliation{Department of Particle Physics and Astrophysics, Weizmann Institute of Science, Rehovot 76100, Israel}
\author[0000-0003-1710-9339]{Lin Yan}
\affil{Caltech Optical Observatories, California Institute of Technology, Pasadena, CA 91125, USA}

\section{Introduction} \label{sec:intro}
Supermassive black holes (SMBHS) reside in the center of most galaxies \citep{Kormendy1995, Ferrarese2005}. 
Galaxy growth via hierarchical mergers therefore results in the formation of SMBH binaries. The time taken for these SMBH binaries to merge depends on the nature of their host galaxies. While binaries in gas-poor galaxies may stall at 1\,pc separations \citep[e.g.][]{Milosavljevic2001}, SMBH binaries in gas-rich environments may merge on timescales of $10^6-10^7$ years  \citep{Escala2005}. 

A consequence of SMBH mergers in gas-rich environments may be the gravitational wave recoil of coalesced SMBHs after merger. In this process, the asymmetric emission of gravitational waves during SMBH coalescence imparts momentum to the coalesced black hole, ejecting it from the central potential well to wander about the  galaxy halo for $10^6$ to $10^9$ years \citep{Volonteri2005DynamicalUniverse,Campanelli2007,Loeb2007,Volonteri2008,Blecha2008}.  The `recoiling black hole' is expected to carry broad line gas with it and continue to undergo regulated accretion, allowing it to be observable as an active galactic nucleus (AGN) spatially offset from the center of its host galaxy \citep{Blecha2008}. Simulations by \citet{Volonteri2008}, for example, show that an AGN with a 500 km/s kick velocity could still be accreting and observable as an off-center quasar at 30 kpc from its host center.

Other observable signatures of a recoil event include evidence of recent galaxy merging activity. Since the recoiling AGN may continue to accrete for a $10^6$ year timescale after recoil \citep{Blecha2008}, tidal structures may still be visible while the recoiling SMBH is active and detectable. The ejected AGN could also leave behind a trail of feedback evidence in the form of enhanced H$\alpha$ emission leading to the galaxy centre \citep{Loeb2007}. 

Simulations show that the recoil velocity and maximum host-AGN spatial offset of the recoiling SMBH depends on the mass ratio and spin alignment of the black hole binary prior to merger \citep{Campanelli2007, Blecha2016}. Binaries with perfectly aligned spins can produce a maximum recoil velocity of only 200 km s$^{-1}$. A binary with misaligned spins can produce kicks up to 5000 km s$^{-1}$. A population of recoiling black holes would therefore provide strong constraints on the distribution of masses and spins in SMBH binaries. This would, in turn, inform simulations of SMBH spin alignment based on torques in the circumbinary gas disk \citep{Bogdanovic2007,Lodato2013} and stellar interactions during inspiral \citep{Berczik2006}. 

A confirmed sample of recoiling SMBHs could be used to test the predictions of numerical relativity simulations on the fraction of massive black holes ejected via recoil from their host galaxies at different redshifts and the effects of this on observed black hole occupation fractions and the $M_{\mathrm{BH}}-\sigma_{*}$ relation  \citep{Volonteri2007,Volonteri2010GravitationalTime,Blecha2011}. 
Such a sample would also allow us to study the effect of displaced AGN feedback on the evolution of merger remnants, such as the expected increase in star formation rates and lengthening of the starburst phase \citep{Blecha2011}.

Despite the many motivations to search for recoiling SMBHs, only a few good candidates have been found to date. One such object is the radio-loud QSO 3C 186 which has an 11\,kpc spatial offset from its host galaxy and a $-2140 \pm 390$ km s$^{-1}$ velocity offset between the broad and narrow emission lines \citep{Chiaberge2017}. The tidal features of the host galaxy indicate recent merger activity. 
Integral field spectroscopy was performed to study the complex kinematics and determine if the velocity offset could result from a peculiar outflow \citep{Chiaberge2018}. The results were consistent with the recoiling SMBH scenario but final confirmation will require both James Webb Space Telescope IFU spectroscopy to map the H$\beta$ region with 0.1" resolution and deep imaging from HST to rule out the presence of a second low mass galaxy \citep{Chiaberge2018}.

SMBH recoil may also be the origin of the variable object SDSS J113323.97+550415.8 (SDSS1133) \citep{Koss2014}. This object is 800\,pc from the center of a low redshift dwarf galaxy and has displayed AGN-like stochastic variability over $>63$ years. However, AGN-like variability can be mimicked by long-lived stellar transients \citep{Burke2020} and giant stellar outbursts or supernovae such as SN2009ip \citep{2009CBET.1928....1M} and UGC 2773 OT2009-1 \citep{2009CBET.1931....1B}. Spectra of SDSS1133 show the presence of blue-shifted Balmer absorption lines and [Fe\,{\sc II}] $\lambda$7155 and [Ca\,{\sc II}] $\lambda \lambda$ 7291, 7324 forbidden emission lines which are highly unusual for an AGN. It is therefore possible that SDSS1133 is a luminous blue variable (LBV) star continuing to demonstrate non-terminal outbursts \citep{Koss2014}. 

The lack of many recoiling SMBH candidates has motivated a number of systematic searches for offset AGN using different techniques. A Gaia analysis of a sample of low redshift, unobscured broad line AGN from SDSS showed that at least 99\% were within 1\,kpc of the host, 90\% within 500pc and 40\% within 100\,pc \citep{Shen2019}. This study used a technique called varstrometry \citep{Hwang2020} to measure AGN-host spatial offsets via the astrometric jitter of the photocenter induced by the AGN flux variability, allowing them to rule out the existence of a substantial offset AGN population on 10\,pc to 1 \,kpc scales at redshifts $0.3<z<0.8$. 

From this study, it appears that unobscured, accreting, recoiling SMBHs with $>10$\,pc separations must be very rare at low redshifts, if they exist. This may be because SMBH spin alignment is always very efficient, inducing only small velocity recoils with small maximum separations. Recoiling black holes may also be more common in the early universe due to higher merger rates and lower galaxy masses. Cold gas inflow during merger may increase the required escape velocity for many galaxies and gas drag could play a role in keeping recoiling black holes close to the galaxy center \citep{Blecha2016}. The level of accretion may be too small for recoiling SMBHs to be detectable and offset AGN may be frequently obscured by the gas environment induced by the merger \citep{Shen2019}. 

While these results appear discouraging, another systematic search by \citet{Lena2014} undertook careful isophotal modeling of archival HST images of 14 nearby core elliptical galaxies and found that 10 of the 14 had small $\sim1-10$ pc displacements between the AGN and host galaxy center, 6 of which were considered confident detections because the galaxy profiles were not asymmetric. 4 of the 6 galaxies showed alignment between the AGN-photocenter displacements and the radio jet axis. This correlation is predicted for gravitational recoil of SMBHs but may instead indicate that the spatial offset was induced by radio jet acceleration of the SMBH. This radio axis correlation would not be produced by interactions with massive perturbers or orbital motion prior to SMBH binary coalescence \citep{Lena2014}.

Other searches have used a multi-wavelength approach to successfully find offset AGN candidates. \citet{Skipper2018} searched for radio-optical spatial offsets in a sample of 345 SDSS galaxies with nearby compact radio sources detected in the Cosmic-Lens All Sky Survey (CLASS) catalogue, finding 3 sources with offsets greater than 0.6 arcseconds. \citet{Condon2017} found one offset AGN candidate amongst 492 radio point sources from the NRAO VLA Sky Survey \citep[VLASS;][]{Condon1998} when crossmatched to extended sources in the Two Micron All-Sky Survey (2MASS) Extended Source Catalog \citep{Jarrett20002MASSAlgorithms}.

A recent study by \citet{Reines2019} found that the majority of a sample of 13 radio AGN in dwarf galaxies were off-nuclear, likely because the lower escape velocities in dwarf galaxies make it easier for black holes to wander from the central potential. \citet{Kim2017} found a recoiling SMBH candidate in a systematic search for spatially offset X-ray AGN in a sample of 2542 sources with optical/near-infrared counterparts in archival HST images from the \textit{Chandra} Source Catalogs - Sloan Digital Sky Survey Cross-match Catalog \citep{Evans2010TheCatalog,Rots2011DeterminingPositions}.

Even though these search strategies have yielded some recoiling SMBH candidates, there are many challenges to confirming the nature of these objects. Candidates with broad line gas at $>1000$\,km s$^{-1}$ velocities relative to narrow emission lines but no observable AGN-host spatial offset can often be explained by outflowing winds \citep{Allen2015,Robinson2010}, scattered broad line emission from an SMBH binary \citep{Robinson2010}, asymmetric double-peaked emission from an elliptical accretion disk \citep{Steinhardt2012}, or two superposed AGN \citep{Shields2009,Shields2009a}.

For recoiling SMBH candidates with observable host-AGN spatial offsets, it must be ruled out that they are not QSOs with an additional undermassive host \citep{Chiaberge2017}. In this case, the true host galaxy may be on the lower end of the luminosity - SMBH mass scaling relation \citep{McLure2002} and very compact such that the extended galaxy emission around the QSO is not detectable, resulting in a false association with the brighter, offset companion galaxy in a merging system. This can occur when the AGN's host galaxy was tidally stripped as it merged with the larger galaxy \citep{Bellovary2010} and has been proposed as the nature of the \citet{Jonker2010} recoiling SMBH candidate and the origin of a number of off-nuclear ultraluminous X-ray sources such as HLX-1 \citep{Farrell2009}. Compact dwarf galaxies hosting SMBHs may be very common at low redshifts. Four ultra-compact dwarf galaxies with masses $M>10^7M_{\odot}$ in the Virgo and Fornax clusters have been shown to host SMBHs through analysis of their velocity dispersion and mass profiles \citep{Seth2014AGalaxy,Ahn2017DetectionGalaxies,Afanasiev2018AUCD3,Ahn2018TheM59-UCD3}. These systems were likely produced through tidal stripping of a larger galaxy hosting an SMBH. \citet{Voggel2019TheUniverse} estimate that such stripped nuclei may host 8\% to 32\% of local SMBHs.

AGN in merging galaxies also have typical relative velocities of $10-400$\,km s$^{-1}$ \citep{Liu2017, Comerford2009, Comerford2014OffsetGrowth} and these velocities are comparable to the predicted velocities of recoiling SMBHs from spin-aligned binaries. \citet{Comerford2014OffsetGrowth} estimate that 4\% to 8\% of Type 2 AGNs are in galaxy mergers, so these systems may be quite common. Triple SMBH systems, in which a merger with a third galaxy occurs before the initial SMBH binary forms, can also be difficult to distinguish from recoiling SMBHs \citep{Civano2010,Kalfountzou2017}.

Offset AGN in merging galaxies are nonetheless important to find because they provide a way to study AGN fueling by galaxy merger triggered gas inflows \citep{Surace1998Galaxies, Canalizo2001QuasiStellarMergers,Treister2012MajorNuclei}. The increased incidence of galaxy mergers amongst X-ray selected AGN and increasing X-ray luminosity with decreasing AGN separation in dual AGN suggests that black hole accretion peaks during the merging process \citep{Koss2012}. \citet{Comerford2014OffsetGrowth} also found that the fraction of AGN in galaxy mergers increases from 0.7\% to 6\% over the AGN bolumetric luminosity range of $43<\log(L_{bol})\text{[erg/s]}<46$ which suggests that galaxy mergers trigger high luminosity AGNs. High-resolution hydrodynamical simulations by \citet{VanWassenhove2012ObservabilityGalaxies} predict that AGN triggering and the likelihood of dual AGN activity is strongest at $<10 kpc$ separations and that most merger-triggered AGN activity is non-simultaneous such that 90\% of SMBHs in mergers appear as single or offset AGN instead dual AGN at $L_{\text{bol}}>10^{44}$ erg s$^{-1}$ and separations $>1-10$ kpc. These predictions are supported by observations of the relative occurrence of offset vs dual AGN \citep[e.g.][]{Comerford2009}. Discoveries of AGN in tidally stripped dwarf galaxies in mergers may also yield IMBH candidates which can be used to constrain models of BH seed formation in the early universe \citep[e.g.][]{Volonteri2009JourneyUniverse,Reines2016ObservationalSeeds}.

Discoveries of AGN in mergers have occurred both serendipitously and in targeted searches. Binary AGN were found in X-ray imaging spectroscopy of ultraluminous infrared galaxy NGC 6240 \citep{Komossa2003} and Mrk 739, a galaxy with two optically distinguishable bulges \citep{Koss2011}. A search for AGN companions to a sample of ultra-hard X-ray-selected AGNs from the all-sky Swift Burst Alert Telescope (BAT) survey with Chandra, XRT and XMM imaging combined with emission line diagnostics with SDSS and Gemini spectroscopy revealed 16 dual AGN \citep{Koss2012}. 

Many searches for AGN in mergers on $<10$\,kpc scales have looked for double peaked narrow [O\,{\sc iii}] $\lambda$5007 emission lines in large spectroscopic datasets such as SDSS \citep[e.g. ][]{Xu2009,Wang2009,Liu2010,Smith2010}. \citet{Comerford2014OffsetGrowth} found 351 offset AGN candidates amongst a sample of 18,314 Type 2 AGNs by measuring velocity offsets between the forbidden and Balmer emission lines relative to the stellar absorption lines.  \citet{Fu2011} found 16 dual AGN candidates with high-resolution near-infrared images of 50 double-peaked [O\,{\sc iii}] $\lambda$5007 AGNs and one of these was confirmed as a\,kpc scale binary AGN with high resolution radio images \citep{Fu2011a}. 

Searches for dual AGN via X-ray, radio, and optical imaging suffer from different selection effects, and there are ongoing efforts to understand the affects of AGN obscuration by gas and dust during merger \citep{Glikman2015,Kocevski2015,Koss2011}. \citet{Koss2018} found that obscured luminous black holes, with X-ray emission but not visible broad $H\beta$ lines, were significantly more likely be in a later stage nuclear merger than a comparable sample of inactive galaxies.

The lack of a large sample of dual AGN and even a small sample of confirmed recoiling SMBHs motivates the development of new search strategies to find AGN offset from their host galaxies and from companion galaxies. A large transient survey which identifies offset AGN candidates via their variability provides one such approach.

In this paper we present a new method for a systematic search for offset AGN - both recoiling SMBHs and AGN in galaxy mergers - using the Zwicky Transient Facility \citep[ZTF;][]{Bellm2019,Graham2019,Dekany2020TheSystem}. In Section 2, we present our techniques for filtering ZTF transients to make a sample of 5493 optically variable AGN. We present a new version of \texttt{The Tractor} forward modeling software for confirmation of AGN-host spatial offsets and describe the candidate selection strategy used to obtain 9 offset AGN candidates and 52 AGN in mergers. In Section 3, we describe the multi-wavelength and spectroscopic properties of these new samples and present the rebrightening of the previously discovered recoiling SMBH candidate SDSS1133.

\section{Sample selection}
\subsection{The Zwicky Transient Facility}
ZTF is a wide-field optical transient survey of the northern sky which observes in g, r and i bands with an average 3 day cadence. It uses a 47 square degree field of view camera mounted on the Samuel Oschin 48-inch Schmidt telescope at Palomar Observatory. ZTF Phase 1 ran from November 2017 to October 2020. 
ZTF identifies variable sources by difference imaging, by which a change in flux is detected by subtracting a high-quality reference image from the data obtained at each epoch. ZTF provides a unique way to search for offset AGN via their optical variability by giving three main advantages for a systematic AGN search.

Firstly, ZTF provides an opportunity to find new AGN by applying light curve modeling techniques to new transients and identifying AGN-like stochastic variability. Light curve modeling of variable sources to find previously undiscovered AGN has been demonstrated for SDSS Stripe82 difference imaging and for transients in the Palomar Transient Factory \citep{Baldassare2018, Baldassare2020}. 

Secondly, image subtractions containing the AGN can be used to locate the position of AGN-like variability relative to the host galaxy. Because spectroscopic surveys tend to have large plate sizes (for example, the SDSS spectroscopic plate size is 1.49"), it can be difficult to confirm that the location of the AGN broad lines is associated with an offset point source. By detecting AGN-like variability from an offset point source, we can confirm the spatially offset nature of AGN candidates. 

Finally, the sky coverage of ZTF allows us to search a very large area for offset AGN, which is important given the apparent rarity of $>10$\,pc spatial offset recoiling SMBHs at low redshifts.

\subsection{Selection of variable AGN in ZTF}
We obtain our sample of variable AGN using data from the ZTF alert stream \citep{Patterson2019}. ZTF alerts contain details of the photometry and astrometry of single epoch transient detections \citep{Masci2019}. The ZTF pipeline produces approximately 100,000 alerts every night, so we implement a filter with the alert broker and analysis framework \texttt{AMPEL} \citep[Alert Management, Photometry and Evaluation of Lightcurves;][]{Nordin2019} to detect variable AGN amongst other transient phenomena. 

Our method to filter out poor subtractions, moving sources and variable stars and find only extragalactic transients is similar to the approach used in the Tidal Disruption Event (TDE) filter of \citet{vanVelzen2020}. We apply a liberal cut of $<0.8$ on the star-galaxy score \citep{Tachibana2018} to find transients associated with galaxies, and a cut of $<0.3$ on the real-bogus score \citep{Mahabal2019, Duev2019} to remove bogus transients. We remove objects in busy stellar fields by crossmatching to the Gaia and PanSTARRS catalogs with \texttt{catsHTM} \citep{Soumagnac2018} to ensure that there are no more than 30 Gaia objects and 100 PanSTARRS objects within a 15" radius. We also require at least 3 significant detections $>0.01$ days apart and a minimum flux increase of 2.5 magnitudes. 

This filtering strategy primarily finds two kinds of common extragalactic transient: supernovae and variable AGN. To select AGN and remove supernovae within the \texttt{AMPEL} filter, we require that our transients either match an object in a series of AGN catalogs or have variability which is more characteristic of an AGN than a supernova. We use \texttt{catsHTM} and \texttt{Extcats}\footnote{https://github.com/MatteoGiomi/extcats} look for a 2" crossmatch with objects in The Million Quasar Catalog \citep{Flesch2015}, a machine learning based catalog of photometric AGN candidates \citep{Brescia2015}, and a catalog of 720,961 variable sources from the Palomar Transient Factory (PTF) and intermediate Palomar Transient Factory (iPTF) found between 2009 and 2016 which were not classified as a star \citep{Miller2017} and had $>5$ detections over $>24$ hours.

For sources which do not have an AGN catalog crossmatch, we model their full ZTF light curve history within the \texttt{AMPEL} filter. We use the \texttt{SNCOSMO} supernova modeling tool \citep{kyle_barbary_2016_168220} to fit the `salt2' SN Ia model to the g and r band light curves and extract the reduced $\chi^2_{\text{SN}}$ goodness of fit for the best fit SN Ia model. 

For comparison to the SN Ia goodness of fit, we implement the \citet{Butler2011} quasar modeling routine. This routine calculates the structure function for input light curves and compares this to the ensemble quasar structure function for Sloan Stripe 82 g- and r-band AGN light curves. The goodness of fit of the ensemble structure function model $\chi^2_{Q}$ gives a measure of how likely the ZTF light curve shows AGN-like variability. 

The \citet{Butler2011} routine also calculates the reduced $\chi^2$ for the null hypothesis that the source shows non-AGN like variability (such as from a variable star). We denote this as $\chi^2_{Q0}$. For the purposes of separating AGN from supernovae, we found that AGN generally have much lower $\chi^2_{Q0}$ values than $\chi^2_{\text{SN}}$, so this variability statistic was also effective at separating SN from AGN.

Our filter accepted any source where either the g- or r-band light curve had a $\chi^2_{Q}$ or $\chi^2_{Q0}$ value less than the SN Ia goodness of fit $\chi^2_{\text{SN}}$. Based on tests with a sample of 111 spectroscopically classified supernovae with $>20$ ZTF epochs, we determined that this method removes approximately 95\% of SN Ia and  60\% of Type II supernovae. With a sample of 166 spectroscopically confirmed AGN with $20-100$ ZTF epochs, we were able to classify 80\% correctly. 

Objects which pass either the AGN crossmatch criteria or light curve fitting criteria pass the \texttt{AMPEL} filter and are pushed to the GROWTH Marshal science portal for arrangement of spectroscopic followup \citep{Kasliwal_2019}. The AGN candidates are then confirmed either with existing SDSS spectroscopy, follow-up spectroscopic observations with the DeVeny spectrograph on the Lowell Discovery Telescope, or by their WISE color or variability history. To classify AGN based on their WISE W1-W2 color we use the criteria:
\begin{equation} 
W1-W2>0.662 \text{ exp } \{0.232(W2-13.97)^2\}
\end{equation}
from \cite{Assef2013}, and to classify AGN based on significant variability in their WISE light curve we require the $\chi^2$ relative to a flat light curve to satisfy $\chi^2/\text{dof}>10$.

We apply this procedure to ZTF alerts from a 2.5 year period between 2018-01-01 to 2020-07-06 and obtain a sample of 5493 AGN. This constitutes our final sample of strongly variable ZTF AGN with spectroscopic or WISE color/variability confirmation. 

\subsection{Selection of AGN spatially offset from their host galaxy}
\subsubsection{Image modeling with \texttt{The Tractor}}
In order to model the positions of the variable AGN relative to their host galaxy, we apply \texttt{The Tractor} \citep{2016ascl.soft04008L} to forward model the host galaxy profile and transient point source emission across ZTF images. \texttt{The Tractor} forward models in pixel space by parametrizing the astrometry solution, sky noise and point spread function of each image and modeling this simultaneously with the shape, flux and position of each source across images in multiple bands and surveys. 

We developed a version of \texttt{The Tractor} to fit a host galaxy profile and an overlapping point source with a position in or around the host. The point source position and the host galaxy shape, flux and position are assumed constant over all epochs, while the point source flux may vary across single epochs. The version of \texttt{The Tractor} that we apply determines the best fitting one of two galaxy profile models: a pure de Vaucouleurs profile described by $I(r) = I_0 \text{ exp}(-7.67 [(r/r_e)^{1/4}])$ and a pure exponential profile described by $I(r) = I_0 \exp(-1.68r/r_e)$. 
\texttt{The Tractor} is a highly advantageous tool for this analysis because it allows for forward modeling across images of different bands and different instruments. ZTF images taken when the transient is bright can be simultaneously modeled with higher resolution, deeper images from a different telescope to improve modeling of the host galaxy. Subtraction of the model from the coadded data can also help to reveal irregularities in galaxy structure in the residuals. 

We model 3422 out of the sample of 5493 AGN with \texttt{The Tractor}. Due to limitations with archival storage of ZTF images on our filesystem and the availability of overlapping Legacy Survey \citep{Dey2019} and PanSTARRS \citep{Chambers2016TheSurveys} images for astrometric source matching, we do not model the remaining 2197. As the AGN are isotropically distributed in the sky, choosing to model a subset of the full sample with \texttt{The Tractor} does not introduce any biases with respect to the distribution of host-AGN offsets. 

We remove 126 of the 3422 AGN because they are duplicates of existing AGN in the sample. The ZTF alerts for particular transients sometimes demonstrate such a large scatter in position that they are considered to be 2 or 3 different transients with separate ZTF names by the alert pipeline. As such, the alerts associated with a single transient can be distributed across 2 or 3 different transient objects. By applying an 8" cone search to all AGN which pass our \texttt{AMPEL} AGN filter, we find transients which are associated with one another and select the transient ID with the most alert packets containing real detections. 

We model the sample of 3296 unique AGN by selecting the 30 ZTF g- and r-band images taken closest to peak magnitude. This allows us to reach a median depth in AB magnitude of 22.4 in the r-band and 22.2 in the g band. We chose to model only 30 ZTF images due to computational and time constraints, but future work could model the whole sample with \texttt{The Tractor} to greater depths by modeling larger numbers of ZTF images and including higher resolution, deeper DECam imaging. 

Of our 3296 AGN, 186 do not have sufficient S/N in the 30 ZTF images to model a host galaxy and point source with \texttt{The Tractor} above the limiting magnitude. This leaves 3110 objects with measured AGN-host offsets. 

\subsubsection{Determination of statistically significant spatial offsets}
Since the ZTF camera has 1" pixels but much of the simulated recoiling black hole population is only observable at sub-arcsecond AGN-host spatial offsets, it is important to understand the positional accuracy that can be obtained from ZTF image subtractions. 

In order to determine which AGN-host spatial offsets are statistically significant, we studied the distribution of offsets from the sample of 3110 objects with host galaxy and point source positions determined by simultaneously modeling 30 g- and r-band ZTF images with \texttt{The Tractor} (see Section 2.3.1). If the uncertainties in observed RA and Dec of the host galaxies are normally distributed with standard deviation $\sigma_{\text{ref}}$, and the uncertainties in observed RA and Dec of the variable point sources are normally distributed with standard deviation $\sigma_{\text{sci}}$, the radial distribution of spatial offsets between them will follow a Rayleigh distribution with $\sigma^2_{\text{R}}=\sigma_{\text{ref}}^2+\sigma_{\text{sci}}^2$.

In a study of radio AGN from the Cosmic-Lens All Sky Survey (CLASS) catalogue, \citet{Skipper2018} found that the population of offsets between the radio AGN and optical galaxies in SDSS followed a mixture distribution consisting of a Rayleigh component and an exponential tail component, where the latter component may represent real AGN-host spatial offsets. Similarly, we find that the shape of our offset distribution from \texttt{Tractor} modeling of the ZTF AGN sample is described by the expected Rayleigh distribution at offsets $\lessapprox$ 1" and at offsets $\gtrapprox 1$" the distribution is better described by a decaying exponential.

\begin{figure*}
\gridline{\fig{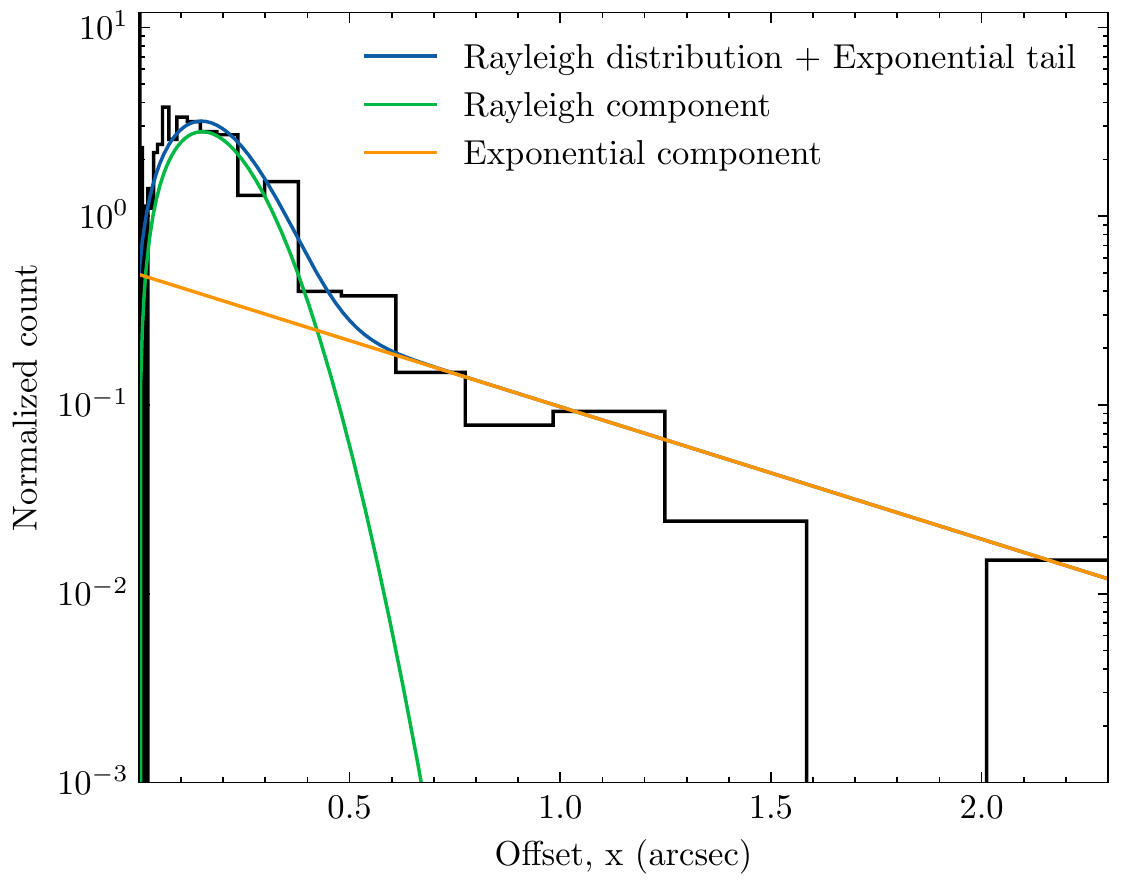}{0.45\textwidth}{(a) AGN with a peak difference magnitude between 15 and 18.}
          \fig{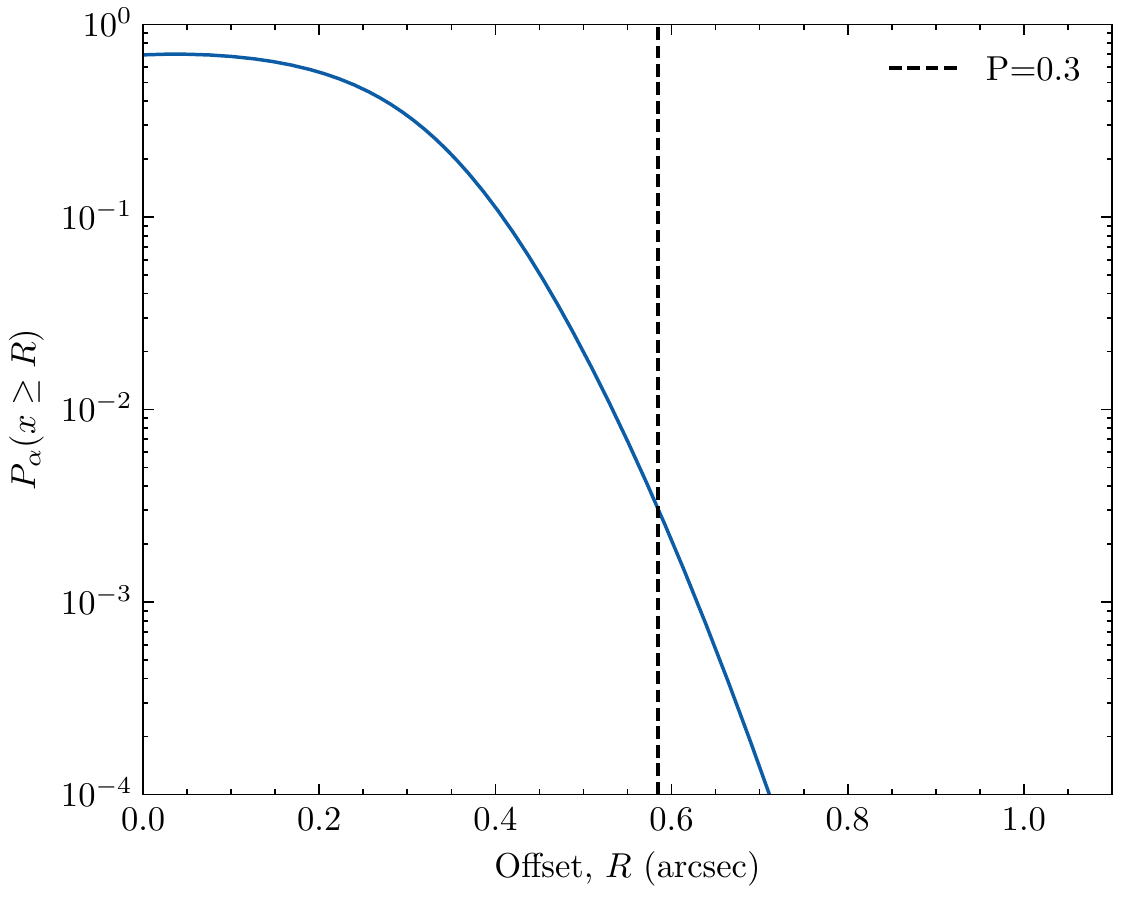}{0.45\textwidth}{(b) 3$\sigma$ offset = 0.574".}}
\gridline{\fig{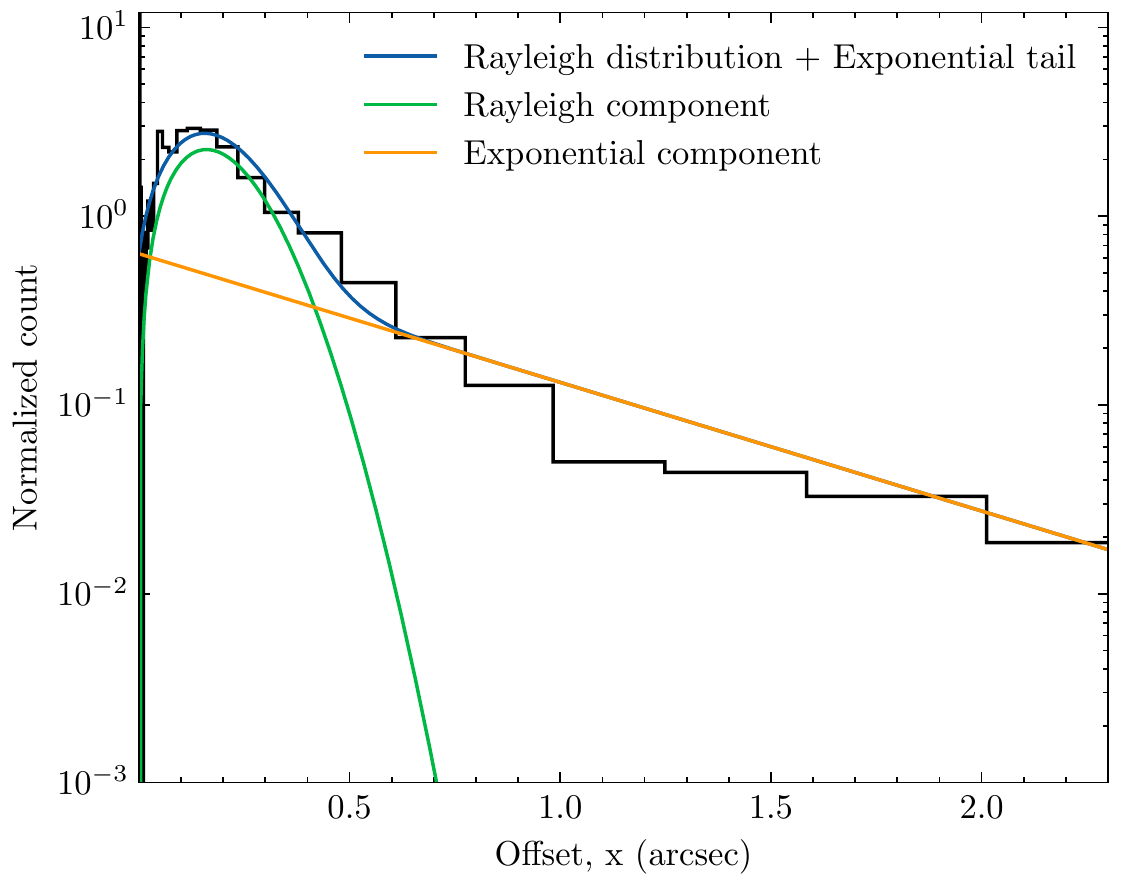}{0.45\textwidth}{(c) AGN with a peak difference magnitude between 18 and 19.5.}
          \fig{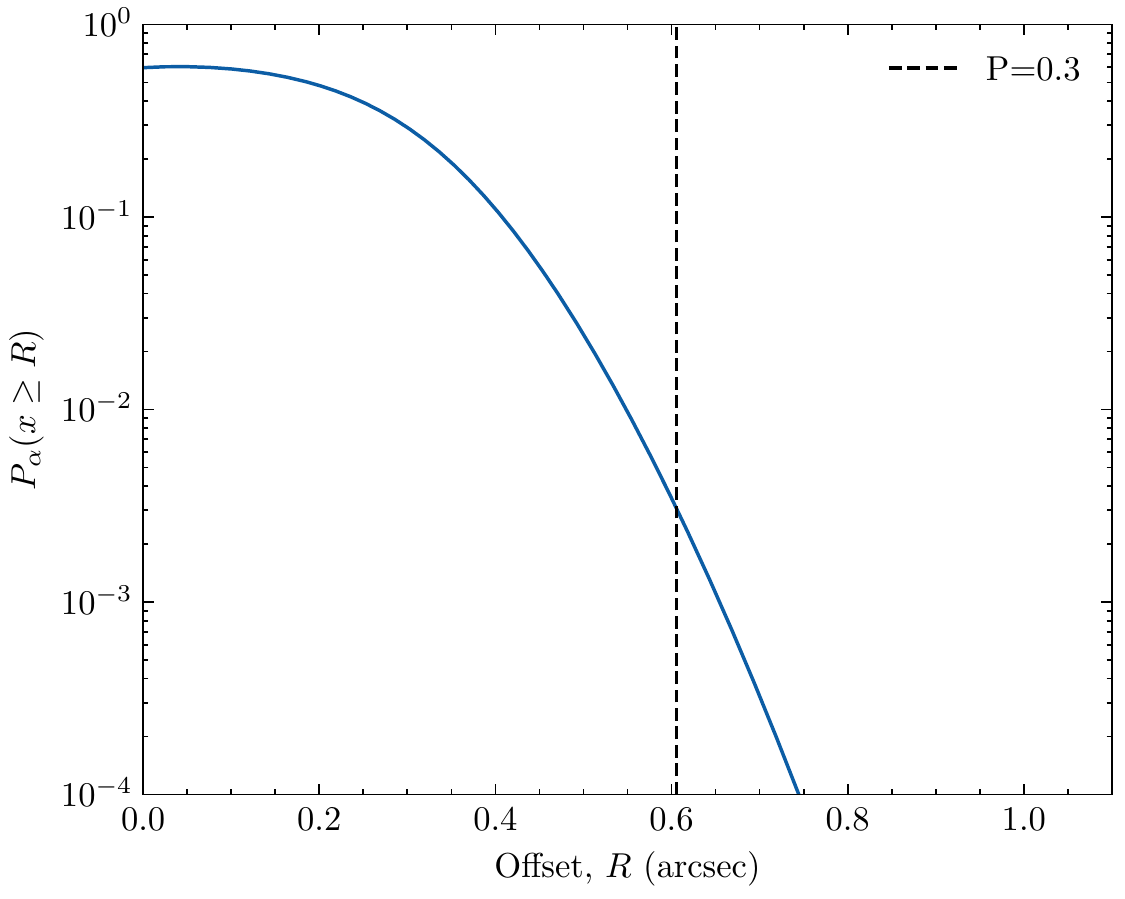}{0.45\textwidth}{(d) 3$\sigma$ offset = 0.605".}}
\gridline{\fig{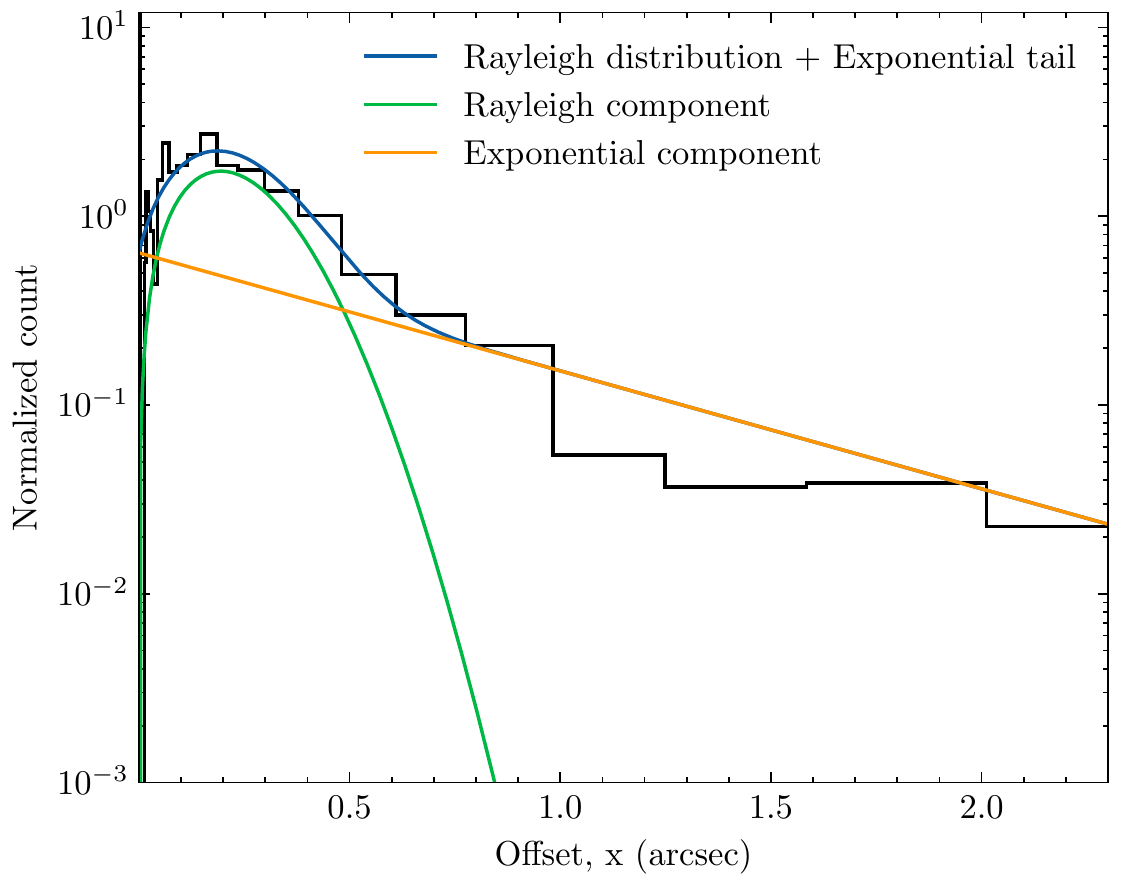}{0.45\textwidth}{(e) AGN with a peak difference magnitude between 19.5 and 23.}
          \fig{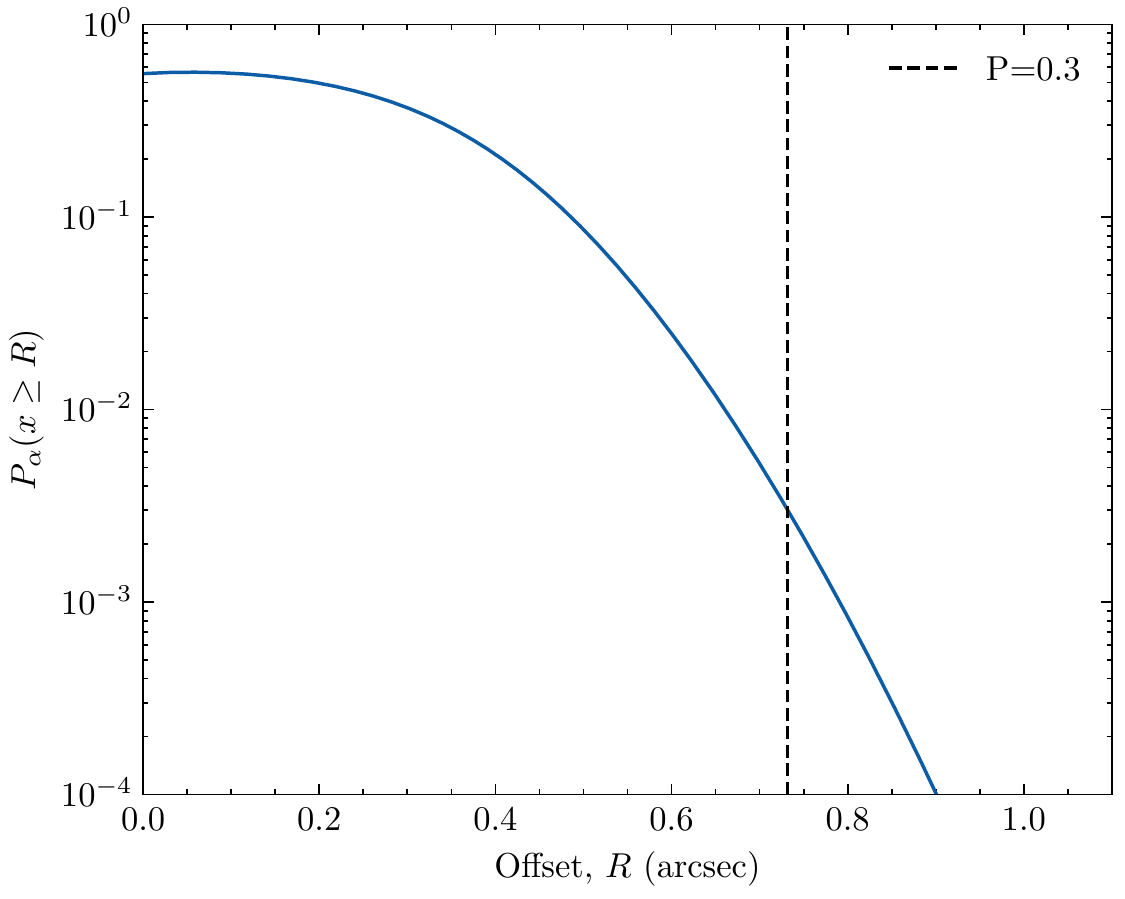}{0.45\textwidth}{(f) 3$\sigma$ offset = 0.732".}}          
\caption{\textbf{Left}: Normalized histogram with logarithmic bins for AGN-host offsets obtained from \texttt{Tractor} modeling. The best-fit model of a mixture distribution with Rayleigh and exponential components are shown. \textbf{Right:} Probability that an offset greater than R is drawn from the Rayleigh component of the mixture distribution shown in (a) instead of the exponential component. The offset where this probability is 0.3\% is shown with a dashed line.}
\label{fig:offset_tractor}
\end{figure*}

We therefore model our AGN-host spatial offset distributions by splitting the AGN sample into 3 different sub-samples based on the peak difference magnitude of the AGN. The ranges in peak-magnitude used to produce the sub-samples are 15-18, 18-19.5, and 19.5-23. These ranges were selected to ensure that each sample was large enough to model the Rayleigh and exponential tail components. The offset distributions of these sub-samples are shown in Figure \ref{fig:offset_tractor}a) c) and e). We fit a mixture distribution consisting of a Rayleigh component $\alpha(x)$ and an exponential component $\epsilon(x)$:

\begin{align}
    P_x&=C \alpha(x) + (1-C) \epsilon(x)\\
    &=C \frac{x}{\sigma^2_{\text{R}}} \text{exp} \big( -\frac{x^2}{2\sigma^2_{\text{R}}}\big) + (1-C) \tau e^{-\tau x}
\end{align}
for offsets $x$, mixture coefficient $C$, Rayleigh width $\sigma_R$ and exponential decay parameter $\tau$. We do this by directly minimizing the log likelihood between the model distribution and the data. 

We first model the exponential decay parameter $\tau$ to fit only offsets $x>1.0$" so that the fit is not heavily biased by low-offset sources which dominate the distribution. We then fix the value of $\tau$ and fit the mixture distribution to the whole sample to find $\sigma_R$ and $C$. The fits are shown in Figure \ref{fig:offset_tractor}a), c) and e), where we can see the Rayleigh component explaining the portion of the distribution which arises from positional uncertainty, and the exponential tail component showing the portion of the distribution which may contain physically real AGN-host offsets.
For a given AGN-host offset $x$, we can calculate the probability that an offset $\geq x$ is drawn from the Rayleigh component as:

\begin{align}
    P_{\alpha}(x>R)&=\frac{C\int_{R}^{\infty} \alpha(x)}{C\int_{R}^{\infty} \alpha(x) + (1-C)\int_{R}^{\infty} \epsilon(x)}
\end{align}
This is shown as a function of offset $x$ in Figure \ref{fig:offset_tractor} b), d) and f). The probability function shows that as the offset increases, it is less likely to be explained by the Rayleigh distribution arising from positional uncertainties and more likely to be part of an exponential tail consisting of possibly real offsets. The spatial offset at which the probability of being drawn from the exponential component of the mixture distribution is 0.3\% is marked by the dashed line in Figure \ref{fig:offset_tractor} b), d) and f).

\begin{deluxetable}{c|ccc|c}
\tabletypesize{\footnotesize}
\tablecolumns{5}
\tablewidth{0pt}
\tablecaption{Selected offset cutoffs \label{table:selectedcutoffs}}
\tablehead{
\colhead{Peak} &\colhead{\texttt{Tractor}}& \colhead{g-band} & \colhead{r-band}& Number\\[-0.25cm]
\colhead{magnitude}&\colhead{modeling}&\colhead{subtraction} &of AGN}
\startdata
19.5-23	& 0.732 & 0.946 & 0.976 & 64	\\
18-19.5	& 0.605 & 1.009 & 0.773 & 164	\\
15-18	& 0.574 & 0.959 & 0.551 & 23	\\
\enddata
\vspace{0.1cm}
\tablecomments{Offset cutoffs (arcseconds) selected for AGN based on their peak magnitude. The first column shows the peak magnitude bin. The second column shows the $>3\sigma$ cutoff for a significant AGN-host offset derived by \texttt{The Tractor}. The third and fourth columns show the uncertainty on the magnitude-weighted transient position derived from ZTF alert packets for r-band and g-band respectively. The fifth column shows the number of AGN which have $>3 \sigma$ \texttt{Tractor} offsets and match the the magnitude-weighted transient position using these cutoffs.}
\end{deluxetable}
Using the probability functions to determine an offset cutoff for the 3 sub-samples, we determine $3\sigma$ offset cutoffs shown in the first column of Table \ref{table:selectedcutoffs}. We select a cutoff of 0.511" for AGN with a peak magnitudes between 15 and 18, 0.773" for AGN with peak magnitudes between 18-19.5, and 0.976" for AGN with peak magnitudes between 19 and 23. 

\subsubsection{Matching of transient positions from \texttt{The Tractor} and ZTF difference images.}
\begin{figure}
    \centering
        \includegraphics[width=0.45\textwidth]{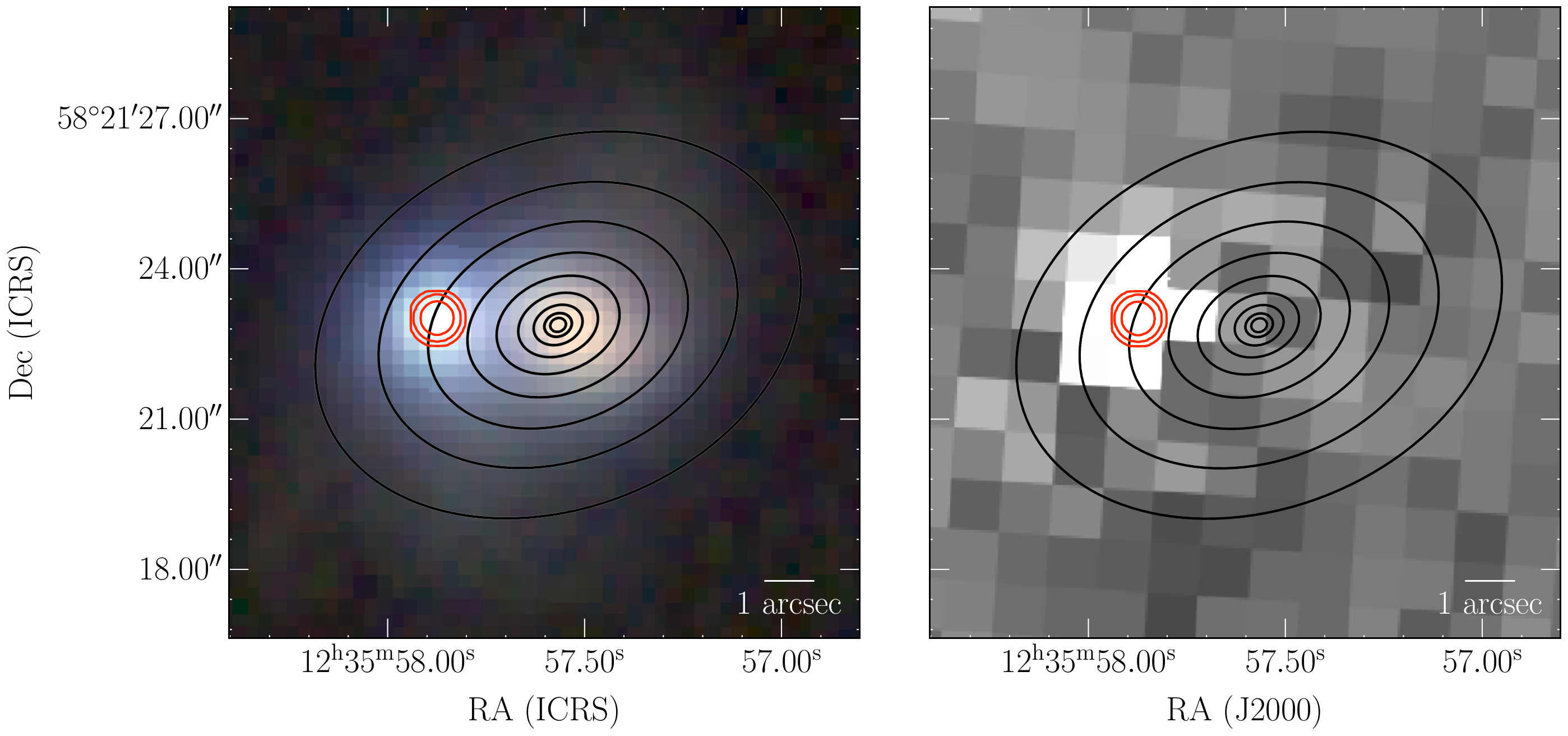}
    \caption{\textit{Left:} Coadded g, r and z-band Legacy Survey images of ZTF18aaxvmpg. \textit{Right:} ZTF image subtraction of ZTF18aaxvmpg when the AGN was close to peak magnitude. Overlaid contours show the best fit \texttt{Tractor} galaxy profile (black) and point source model (red) for a theoretical seeing of 1" derived from ZTF image modeling.}
    \label{fig:subs}
\end{figure}

In order to confirm that the best-fit point source position from \texttt{The Tractor} modeling is consistent with the position of the transient in the ZTF difference images, we calculate the magnitude-weighted position of the transient from the ZTF alert packets containing information about the position and magnitude of each single epoch difference image detection. The weights $1/\sigma_{\text{offset}}^2$ for the magnitude-weighted transient position are calculated using equation 3 from \citet{VanVelzen2018}:
\begin{equation}
\sigma_{\text{offset}} = 0.24 + 0.04(m_{\text{diff}}-20) \end{equation}

In order to determine the uncertainty in the magnitude-weighted transient position from the ZTF alert packets, we undertake the same offset distribution modeling procedure as we do for the \texttt{Tractor} AGN-host offsets. The modeling results for each magnitude binned sub-sample are shown in Figures \ref{fig:offset_r} and \ref{fig:offset_g} in the Appendix. As the distribution of magnitude-weighted offsets is substantially different for g-band images and r-band images, due to the differing contributions of the AGN towards the reference image in different bands, we find cutoffs for the two bands separately. Using the probability functions to determine an offset cutoff for the 3 sub-samples, we determine $3\sigma$ offset cutoffs shown in Table \ref{table:selectedcutoffs}. When checking that the \texttt{Tractor} point source positions are consistent with the magnitude-weighted transient position from the alert packets, we require a match within these selected cutoffs. 

Figure \ref{fig:subs} shows an example ZTF image subtraction for ZTF18aaxvmpg with the \texttt{Tractor} model overlaid. For this object, the \texttt{Tractor} point source position shown in the red contours was consistent with the ZTF transient seen in the image subtraction. This object was therefore considered to be an offset AGN candidate.  

We find that 251 AGN have $>3\sigma$ AGN-host offsets found by \texttt{Tractor} modeling which are consistent with the magnitude-weighted transient position from the alert packets. The breakdown of candidates from each peak magnitude bin is shown in the last column of Table \ref{table:selectedcutoffs}. 

\subsection{Morphological classification of AGN hosts}
In order to confirm the host-AGN offset found in ZTF images and classify the 251 offset AGN based on their morphology we undertook \texttt{Tractor} modeling with deeper, higher resolution images. For this task, we used archival images from the DESI Legacy Imaging Surveys \citep{Dey2019}. The combined DECam Legacy Survey, Mayall z-band Legacy Survey and Beijing-Arizona Sky Survey were taken between 2014 and 2019 and cover declinations from $-18\degr < \delta < +84 \degr$, offering 0.262"/pix resolution and depths of 24.7 , 23.9 and 23.0 for g, r and z bands respectively

We modeled the high resolution coadded g and r and z band Legacy Survey images of each system with a single galaxy profile and an offset point source. We then visually examined the images, \texttt{Tractor} models and residuals to determine if each offset AGN was well modeled as a point source or if there was excess unmodeled emission indicating the presence of a second host galaxy in the system which is centered on the AGN. We separated the sample into 5 categories. The number of objects in each category is shown in Table \ref{table:breakdown}. 

\begin{deluxetable}{lc}
\tabletypesize{\footnotesize}
\tablecolumns{2}
\tablewidth{0pt}
\tablecaption{Classifications of offset AGN \label{table:breakdown}}
\tablehead{
\colhead{Classification} & \colhead{Number}}
\startdata
AGN in galaxy mergers & 52 \\
AGN offset from the stellar bulge of disturbed galaxy & 9 \\
AGN aligned with the stellar bulge of disturbed galaxy  & 21 \\
AGN offset from an undisturbed galaxy & 29 \\
AGN without position confirmation in Legacy Survey modeling & 140 \\
\hline
Total & 251	\\
\enddata
\vspace{0.1cm}
\tablecomments{The breakdown of the complete sample of 251 offset AGN into 5 morphology based classifications. The first row is the number of AGN which have extended galaxy emission around them and appear to be interacting and merging with a second galaxy. The second row shows the number of AGN which are not surrounded by a stellar bulge and appear to be spatially offset from the center of a galaxy with indications of recent merging activity. The third row shows the number of AGN in disturbed, post-merger systems where the stellar bulge is offset from the center of the extended galaxy profile and aligned with the AGN. The fourth row shows the number of AGN which appear to be point sources spatially offset from a undisturbed galaxy and are therefore more likely to be chance coincidences with background galaxies. The fifth row shows the number of AGN where the ZTF position could not be confirmed in archival Legacy Survey images.}
\end{deluxetable}

When the residuals of the galaxy and offset point source model showed a clear stellar bulge surrounding the offset AGN, it was considered likely that there are two galaxies in the system rather than one. If the residuals also showed morphological evidence of merging activity such as tidal structures, we considered the AGN to be part of a galaxy merger. These objects will be discussed further in Section 3.1. 

When the system was well modeled by a single galaxy profile and offset point source and there were no tidal structures indicating recent merging activity it was considered likely to be chance coincidence of an AGN and an unrelated background galaxy. These AGN are discussed in Section 3.2.

For AGN which were well modeled by an offset point source and showed morphological evidence of recent merging activity in the host galaxy residuals, we considered that the AGN may be a candidate for a recoiling SMBH. We consider these objects as recoiling SMBH candidates because the recoiling AGN is expected to be visible for a period of $10^6$ years after recoil while the host galaxy will still show evidence of previous merging activity in its morphology \citep{Blecha2016}. These objects are discussed in Section 3.3. 

When the AGN was in a disturbed, irregularly-shaped host galaxy with a stellar bulge which was offset from the photometric center of the extended galaxy profile and the AGN was aligned with this stellar emission, we did not consider the AGN to be a recoiling SMBH candidate.

The remainder of the 251 objects did not have point source emission from an AGN present in the Legacy Survey images used for \texttt{Tractor} modeling. For these objects, confirmation of the spatial offset discovered in ZTF images requires follow-up with deeper, higher resolution imaging taken when the AGN emission is visible.

\begin{figure}
  \centering
  \includegraphics[width=0.49\textwidth]{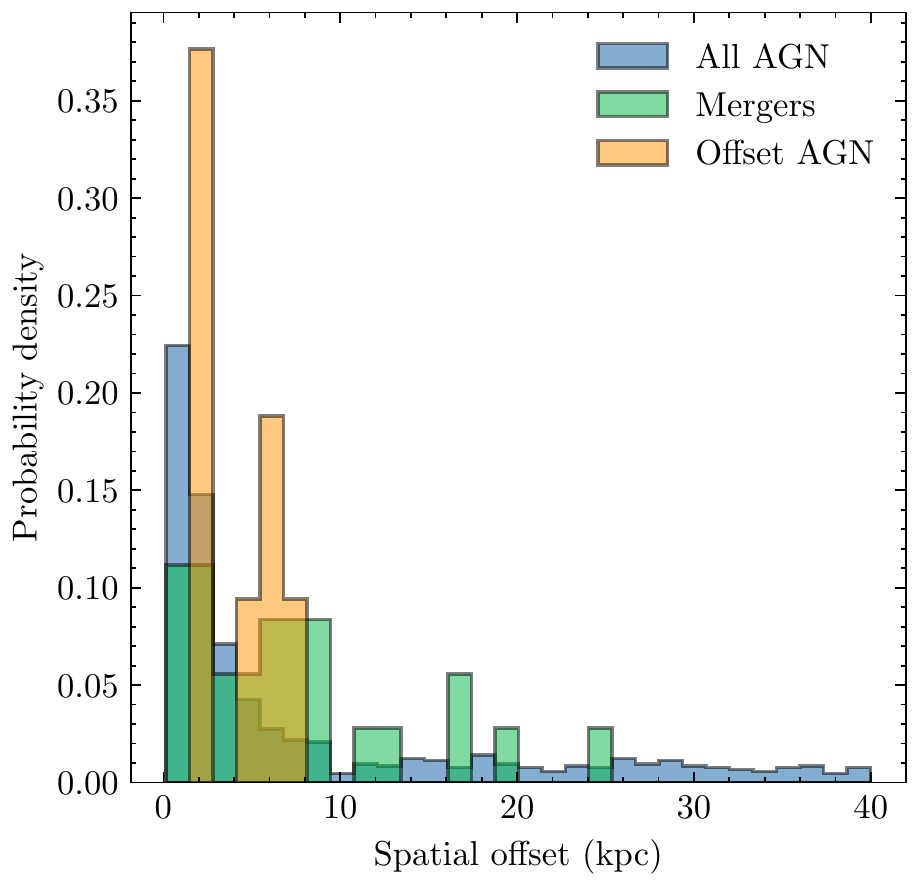}
  \caption{Distribution of physical spatial offsets between AGN and their closest galaxy for 3 samples: 898 AGN from the \texttt{Tractor} modeled sample with available spectroscopic redshifts, 27 of the 52 AGN in mergers with spectroscopic redshifts, and the 9 off-nuclear AGN. A tail of spatial offsets $>10$ kpc can be seen in the blue histogram due to the presence of background galaxies.}
  \label{fig:offkpc}
\end{figure}

The distribution of physical offsets in\,kpc for the whole sample, the AGN in mergers, and the off-nuclear AGN is shown in Figure \ref{fig:offkpc}. The complete AGN sample shows a tail extending beyond 40\,kpc due to chance coincidences with background galaxies. 

\subsection{Spectroscopic analysis of ZTF broad line AGN}
\begin{figure}
  \centering
  \includegraphics[width=0.49\textwidth]{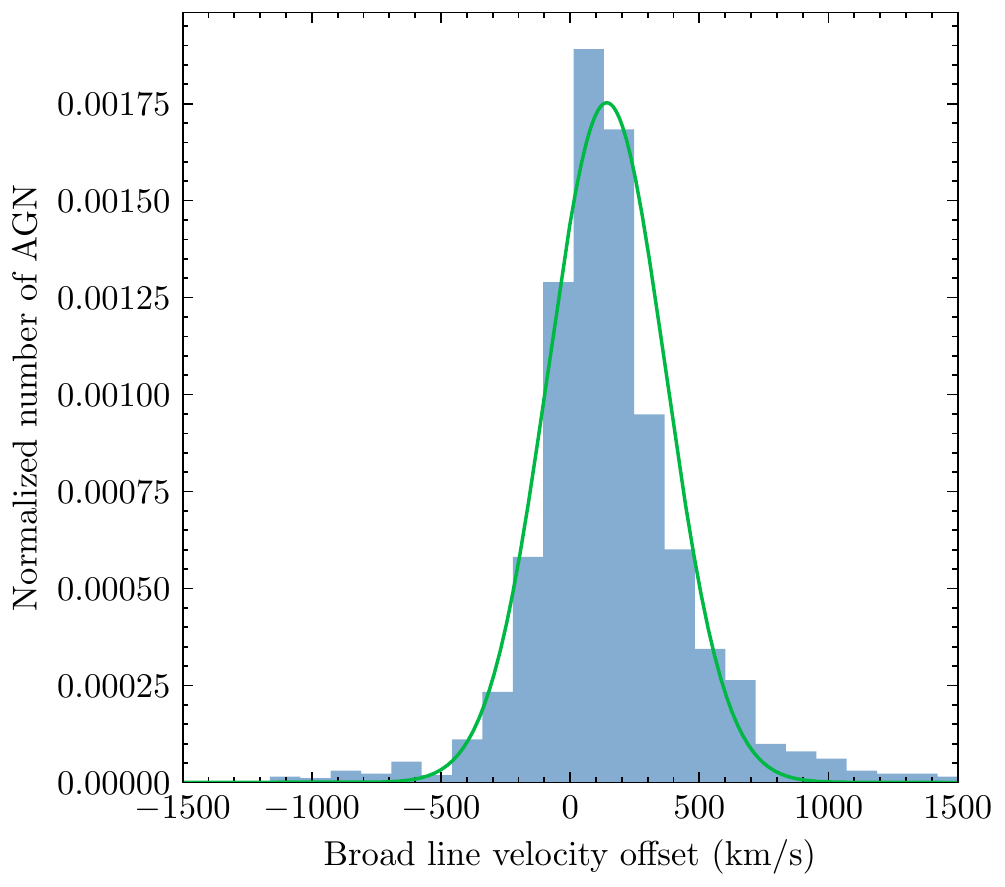}
  \caption{Distribution of Balmer broad line velocities relative to [S\,{\sc ii}] $\lambda$6717, 6731, [N\,{\sc ii}] $\lambda$6550, 6575, [O\,{\sc i}] $\lambda$6302, 6366 and [O\,{\sc iii}] $\lambda$5007, 4959 emission  line velocities found by fitting of archival SDSS spectra with \texttt{pPXF}. Velocities are shown for the 2422 AGN with broad Balmer lines in archival SDSS spectra out of the complete ZTF AGN sample of 5542 objects.}
  \label{fig:veldist}
\end{figure}

In order to model the distribution of broad line velocity offsets in the ZTF AGN sample as a whole, we modeled all 2422 ZTF AGN which had archival SDSS spectra and were classified as broad line AGN by the SDSS DR14 pipeline with Penalized Pixel Fitting (pPXF) \citep{Cappellari2003,Cappellari2017}. This method finds the velocity dispersion of stellar absorption lines using a large sample of high resolution templates of single stellar populations adjusted to match the spectral resolution of the input spectrum. We simultaneously fit the narrow H$\alpha$, H$\beta$, H$\gamma$, H$\delta$,  [S\,{\sc ii}] $\lambda$6717, 6731, [N\,{\sc ii}] $\lambda$6550, 6575, [O\,{\sc i}] $\lambda$6302, 6366 and [O\,{\sc iii}] $\lambda$5007, 4959 emission lines during template fitting. The emission line fluxes are each fit as free parameters but the line widths of the Balmer series are tied, as are the line widths of the forbidden lines. 

In these fits the velocity of whichever broad H$\alpha$, H$\beta$, H$\gamma$, H$\delta$ lines were available within the SDSS spectroscopic wavelength range were allowed to vary up to 3000km s$^{-1}$ from the narrow lines. The velocity of each Balmer broad line was tied to the other Balmer broad lines. 

 We find that the ZTF AGN broad line velocities have a mean displacement of 143 km s$^{-1}$ with a standard deviation of 126 km s$^{-1}$ when the central component is fit with a Gaussian (Figure \ref{fig:veldist}). Our distribution of broad line velocities for variable ZTF AGN is similar to the distribution of H$\beta$ broad line velocities found by \citet{Bonning2007} with a sample of 2598 SDSS AGN where they found a mean displacement of 100km s$^{-1}$ and a standard deviation of 212 km s$^{-1}$. 

Our velocity distribution shows a substantial tail  population with velocities up to $\pm2500$ km s$^{-1}$. The fraction of AGN $f_\nu$ with velocity magnitudes greater than 1000, 1500 and 2000 km s$^{-1}$ are $f_{1000}=0.025$, $f_{1500}=0.009$ and $f_{2000}=0.003$. These values are comparable to those found in the \citet{Bonning2007} sample, where they find fractions of $f_{1000}=0.0035$, $f_{1500}=0.0012$ and $f_{2000}=0.0008$. It therefore appears that a variability selected AGN sample shows a broad line velocity distribution which is typical of a spectroscopically selected AGN sample. 

\section{Results}
\subsection{AGN in galaxy mergers}

\begin{deluxetable*}{lcccclccc}
\tabletypesize{\scriptsize}
\tablecolumns{9}
\tablewidth{0pt}
\tablecaption{Galaxy merger summary \label{table:mergers}}
\tablehead{
\colhead{ZTF name} & \colhead{z} & \colhead{RA} & \colhead{Dec} &\colhead{Host-AGN offset}& \colhead{Host-AGN offset}  &\colhead{AGN flux/ }&\colhead{\# AGN/} &\colhead{Broad line}\\
 &  & (hms) & (dms) &arcseconds)&(kpc) &companion flux& \# spectra&shape}
\startdata
ZTF18aaxvmpg&0.212&12:35:57.810&58:21:21.726&$1.296\pm0.002$&$5.477\pm0.008$&$21.25$&$1/2$&C\\
ZTF18abamzru&-&17:23:27.486&42:21:22.683&$5.231\pm0.001$&$-$&$3.20$&$1/0$&-\\
ZTF18aasxvyo&0.23635&14:17:56.535&25:43:21.506&$5.068\pm0.001$&$19.162\pm0.004$&$24.24$&$1/2$&-\\
ZTF18aaieguy&0.213&13:25:18.144&41:10:09.717&$0.676\pm0.007$&$2.88\pm0.03$&$0.48$&$1/1$&C\\
ZTF19aakjemw&0.147&15:50:07.939&27:28:10.996&$5.435\pm0.002$&$16.184\pm0.006$&$1.16$&$1/2$&2B\\
ZTF18aaifbku&0.067&13:2:58.854&16:24:27.806&$0.985\pm0.002$&$1.371\pm0.003$&$1.75$&$1/1$&C\\
ZTF18aampabj&0.216&16:52:58.864&44:48:45.540&$3.118\pm0.001$&$13.416\pm0.004$&$15.31$&$1/2$&C\\
ZTF19aaagygp&-&01:07:13.787&-11:36:2.998&$3.969\pm0.002$&$-$&$0.34$&$1/0$&-\\
ZTF18abujubn&-&19:08:12.617&45:32:49.643&$8.562\pm0.002$&$-$&$25.97$&$1/1$&MG\\
ZTF18acegbsb&0.037&09:04:36.964&55:36:02.772&$9.027\pm0.009$&$6.992\pm0.007$&$2.14$&$1/2$&RS\\
ZTF19aaozpdm&-&13:37:01.086&20:25:14.343&$1.682\pm0.013$&$-$&$0.88$&$1/0$&-\\
ZTF18aacjltc&0.189&08:12:52.208&40:23:47.253&$3.192\pm0.001$&$12.091\pm0.004$&$6.06$&$2/2$&C\\
ZTF18abtpite&-&23:18:35.915&41:08:0.284&$3.262\pm0.007$&$-$&$0.41$&$1/0$&-\\
ZTF18abvwrxu&-&22:04:07.944&-08:57:24.505&$1.011\pm0.003$&$-$&$5.87$&$1/0$&-\\
ZTF19abaktpb&-&16:42:19.074&03:45:53.037&$8.164\pm0.008$&$-$&$8.5$&$1/0$&-\\
ZTF18aaqjcxl&0.099&07:58:47.235&27:05:16.379&$3.314\pm0.001$&$6.712\pm0.002$&$5.54$&$1/1$&None\\
ZTF18abyoivl&-&00:22:52.018&08:24:0.757&$0.391\pm0.007$&$-$&$0.97$&$1/0$&-\\
ZTF18aabdiug&0.062&12:31:52.060&45:04:43.273&$0.572\pm0.002$&$0.738\pm0.003$&$11.13$&$1/2$&C\\
ZTF19aaviuyv&-&18:56:20.579&37:12:36.076&$2.651\pm0.014$&$-$&$3.48$&$1/0$&-\\
ZTF18aabxczq&0.063&10:38:33.425&46:58:06.741&$0.413\pm0.003$&$0.54\pm0.004$&$1.69$&$1/2$&C\\
ZTF18acvwlrf&0.233&12:50:16.219&04:57:45.074&$1.271\pm0.005$&$5.887\pm0.023$&$0.99$&$1/1$&MG\\
ZTF19aasejqv&0.233&14:13:29.817&26:44:35.232&$1.57\pm0.009$&$7.261\pm0.042$&$2.85$&$1/1$&2B\\
ZTF18aazogyo&0.081&14:56:27.421&30:53:40.225&$5.442\pm0.001$&$9.123\pm0.002$&$13.33$&$1/1$&2B\\
ZTF18aceypvy&0.163&09:51:12.391&31:35:37.084&$2.406\pm0.003$&$7.913\pm0.01$&$9.89$&$1/1$&C\\
ZTF18acbweyd&0.189&10:20:38.565&24:37:12.421&$4.24\pm0.002$&$16.125\pm0.008$&$11.47$&$1/1$&C\\
ZTF18acablce&-&16:30:55.490&72:26:43.352&$1.37\pm0.008$&$-$&$1.77$&$1/0$&-\\
ZTF18abhpvvr&-&00:38:33.041&41:28:53.681&$3.595\pm0.003$&$-$&$9.29$&$1/0$&-\\
ZTF19abfqmjg&-&22:56:41.062&23:02:32.510&$7.402\pm0.003$&$-$&$48.85$&$1/0$&-\\
ZTF18abmqwgr&-&20:27:53.382&14:08:50.604&$1.539\pm0.012$&$-$&$1.33$&$1/0$&-\\
ZTF19aadgbih&0.196&12:46:33.522&45:34:21.773&$0.381\pm0.006$&$1.494\pm0.024$&$1.82$&$1/1$&C\\
ZTF19aalpfan&0.075&13:27:51.414&06:42:49.854&$0.627\pm0.003$&$0.964\pm0.005$&$6.86$&$1/1$&2B\\
ZTF18aawwfep&0.197&08:54:41.735&30:57:54.759&$1.37\pm0.003$&$5.424\pm0.012$&$2.22$&$1/2$&C\\
ZTF19aavxims&-&12:48:55.053&-06:59:54.802&$4.735\pm0.001$&$-$&$31.52$&$1/0$&-\\
ZTF19aaaplct&-&14:57:28.940&08:34:22.879&$2.835\pm0.039$&$-$&$0.97$&$1/0$&-\\
ZTF18aajnqqv&0.081&12:57:41.050&20:23:47.747&$1.747\pm0.004$&$2.901\pm0.007$&$2.83$&$1/2$&C\\
ZTF18abszfur&0.291&22:07:16.099&12:11:03.278&$4.322\pm0.002$&$24.646\pm0.011$&$5.16$&$1/0$&-\\
ZTF19abucbkt&0.16239&01:36:04.252&21:37:25.882&$4.60\pm0.01$&$12.935\pm0.03$&$4.47$&$1/1$&C\\
ZTF18adbhlyb&0.212&11:17:59.188&20:15:19.078&$4.508\pm0.001$&$19.075\pm0.004$&$4.19$&$1/1$&MG\\
ZTF18acxhoij&-&01:12:07.783&-21:04:28.682&$1.099\pm0.045$&$-$&$2.66$&$1/0$&-\\
ZTF18acajwep&-&01:04:05.280&21:22:31.946&$2.906\pm0.001$&$-$&$7.08$&$1/0$&-\\
ZTF19abipoqj&-&22:43:14.796&80:59:27.375&$0.499\pm0.005$&$-$&$0.57$&$1/0$&-\\
ZTF19abpkoou&-&02:34:16.170&05:18:42.732&$3.524\pm0.001$&$-$&$8.38$&$1/0$&-\\
ZTF18abztovy&-&08:29:24.624&34:50:45.655&$1.719\pm0.011$&$-$&$0.86$&$1/0$&-\\
ZTF18acsllgd&-&03:45:45.495&22:23:58.156&$1.041\pm0.025$&$-$&$4.59$&$1/0$&-\\
ZTF19aanxrki&0.114&15:32:27.165&04:19:22.283&$3.557\pm0.064$&$8.304\pm0.149$&$1.29$&$1/2$&C\\
ZTF18aamfuhc&0.086&13:42:34.214&19:13:34.184&$4.845\pm0.001$&$8.608\pm0.002$&$28.39$&$1/1$&MG\\
ZTF18aadwvyr&0.126&08:29:44.346&32:52:21.163&$0.984\pm0.003$&$2.53\pm0.008$&$2.15$&$1/1$&C\\
ZTF19abauzsd&0.285&15:54:32.681&21:43:48.220&$0.483\pm0.007$&$2.699\pm0.039$&$1.03$&$1/1$&RS\\
ZTF18abufbsq&-&23:46:15.513&12:47:07.733&$0.891\pm0.003$&$-$&$1.14$&$1/0$&-\\
ZTF18abzuzrg&0.178&16:55:16.540&32:15:55.145&$1.369\pm0.007$&$4.908\pm0.025$&$1.11$&$1/2$&MG\\
ZTF18abtmcdb&-&01:20:12.473&07:12:58.251&$0.535\pm0.003$&$-$&$3.81$&$1/0$&-\\
ZTF18aauhnby&0.09&12:04:15.954&56:02:58.100&$1.092\pm0.002$&$2.02\pm0.004$&$1.39$&$1/1$&C\\
\enddata
\vspace{0.1cm}
\tablecomments{Summary of the properties of the 52 AGN in galaxy mergers. The host-AGN spatial offset and g-band AGN/host galaxy flux ratio is derived from \texttt{Tractor} modeling of Legacy Survey images. The 8th column lists the number of AGN in the system, confirmed by either spectroscopic narrow emission line ratios or the presence of WISE variability. The second number in this column indicates whether neither the AGN or host galaxy have a spectrum (0), only the AGN has a spectrum (1), or whether there is a spectrum of the host galaxy centroid available as well (2). The last column shows the classification of the shape of the broad line region where we adopt the scheme of \citet{Strateva2003}: prominent red shoulder (RS), prominent blue shoulder (BS), two prominent peaks (2P), two blended peaks (2B), complex multi-Gaussian structure (MS). We denote a Gaussian broad line with a `C' and AGN with no broad line emission with 'None'. If there are no SDSS spectra available for the source or the source does not have broad emission lines, we indicate this with a `$-$'.}
\end{deluxetable*}

52 of our spatially offset AGN were determined to be in a merger with a second galaxy based on the presence of two galaxy nuclei and an interacting morphology visible in Legacy Survey images. The galaxy separations from \texttt{Tractor} modeling of Legacy Survey images ranged from $0.4-9$", which for the 33 AGN with known redshifts, corresponded to physical separations of 0.54 to 24.65\,kpc.

For 14 of these AGN with available SDSS spectra of both galaxies, only ZTF18aacjltc had narrow emission lines consistent with an AGN in both galaxies. The remaining 12 were single AGN, where the companion galaxy did not show narrow AGN emission line ratios. The observed fraction of dual vs offset AGN is consistent with the predictions of \citet{VanWassenhove2012ObservabilityGalaxies}. 

18 galaxy mergers had only one archival SDSS spectrum available, with the fiber centered on the variable AGN. Follow-up spectra of the companion galaxies will be required to determine if these mergers contain one or two AGN. 15 galaxy mergers have no archival SDSS spectra available, as the variable object was classified as an AGN based on optical ZTF and infrared WISE variability. Follow-up spectra of both the AGN and companion galaxy will be required to spectroscopically confirm the presence of one or two AGN.

Aside from appearing in a range of multi-wavelength AGN sample papers, a fraction of our sample of AGN in merging galaxies have been studied in more detail in the literature. ZTF18aasxvyo is a well studied X-ray bright BL Lac object \citep{Halpern19861EGalaxy} and ZTF18aaqjcxl is also a known BL Lac \citep{Plotkin2008AFIRST}. ZTF18acegbsb is a known double peaked emitter \citep{Strateva2003} in a merger, which has been studied in the context of AGN photoionization of companion galaxies \citep{Keel2019AGNDirection}. ZTF18aamfuhc appeared in the same cross-ionization study. ZTF18abhpvvr is a known dual AGN \citep{Huang2014HSTJ0038+4128} and ZTF18aawwfep and ZTF18aajnqqv are also a known AGN pairs from  \citet{Liu2011ACTIVESCALES}.

ZTF18aaifbku, an AGN in the double-lobed galaxy Mark 783, was imaged with the Karl G. Jansky Very Large Array (JVLA) at 5 GHz on September 6 2015 and showed radio emission from a compact core component and an extended component which was 26 kpc long \citep{Congiu2017Kiloparsec-scale783}. The lack of jet emission led the authors to conclude that the radio emission was a relic of previous AGN activity before it entered a quiescent state. This AGN was first observed in ZTF on June 5 2018 and has shown continued variability to r-band magnitudes of 17.07 since then, suggesting that the AGN activity has turned on since on since the radio observations were made in 2015.

\begin{figure}
  \centering
  \includegraphics[width=0.49\textwidth]{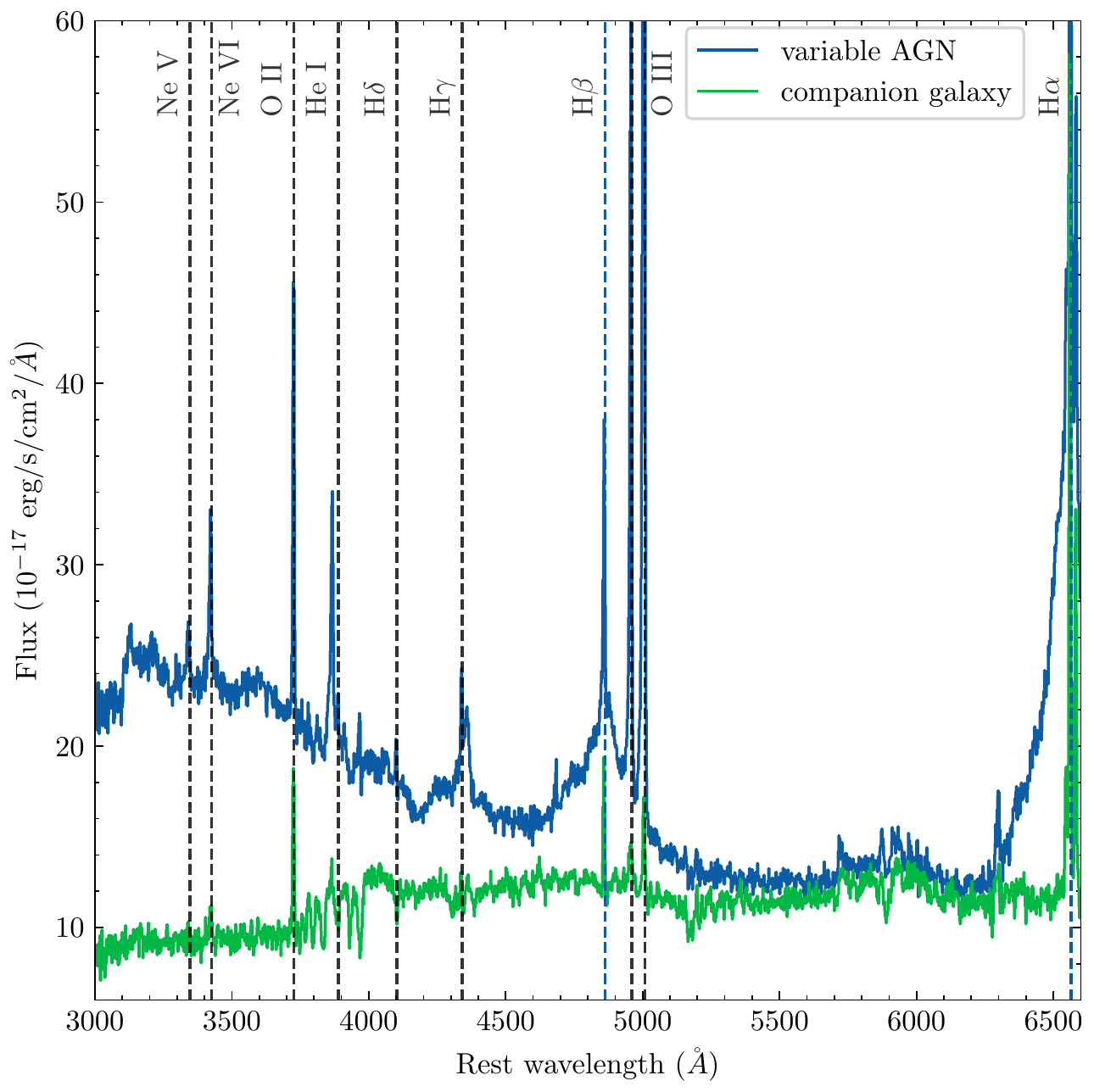}
  \caption{Spectrum of ZTF18aaxvmpg and its offset companion galaxy taken on 2019-10-29 with the DeVeny spectrograph on the Lowell Discovery Telescope. Broad Balmer features can be seen in the AGN spectrum but only narrow emission lines are visible from the companion galaxy.}
  \label{fig:ZTF18aaxvmpg}
\end{figure}

In order to determine if ZTF18aaxvmpg was a dual AGN, we undertook more detailed spectroscopic follow-up. The SDSS spectrum of ZTF18aaxvmpg taken on 2006-05-21 with the fiber centered on the AGN shows the presence of broad Balmer lines but the companion galaxy at 5.5\,kpc from the AGN did not have an archival spectrum. We took a new spectrum of both the host galaxy  and the AGN on 2019-10-29 with the DeVeny spectrograph on the Lowell Discovery Telescope using a 1.5" slit, central wavelength of 5700 Å, a spectroscopic coverage of 3600-8000 Å and a total exposure time of 3200s. The spectrum of the AGN and the companion galaxy is shown in figure \ref{fig:ZTF18aaxvmpg}.

\begin{figure}
  \centering
  \includegraphics[width=0.49\textwidth]{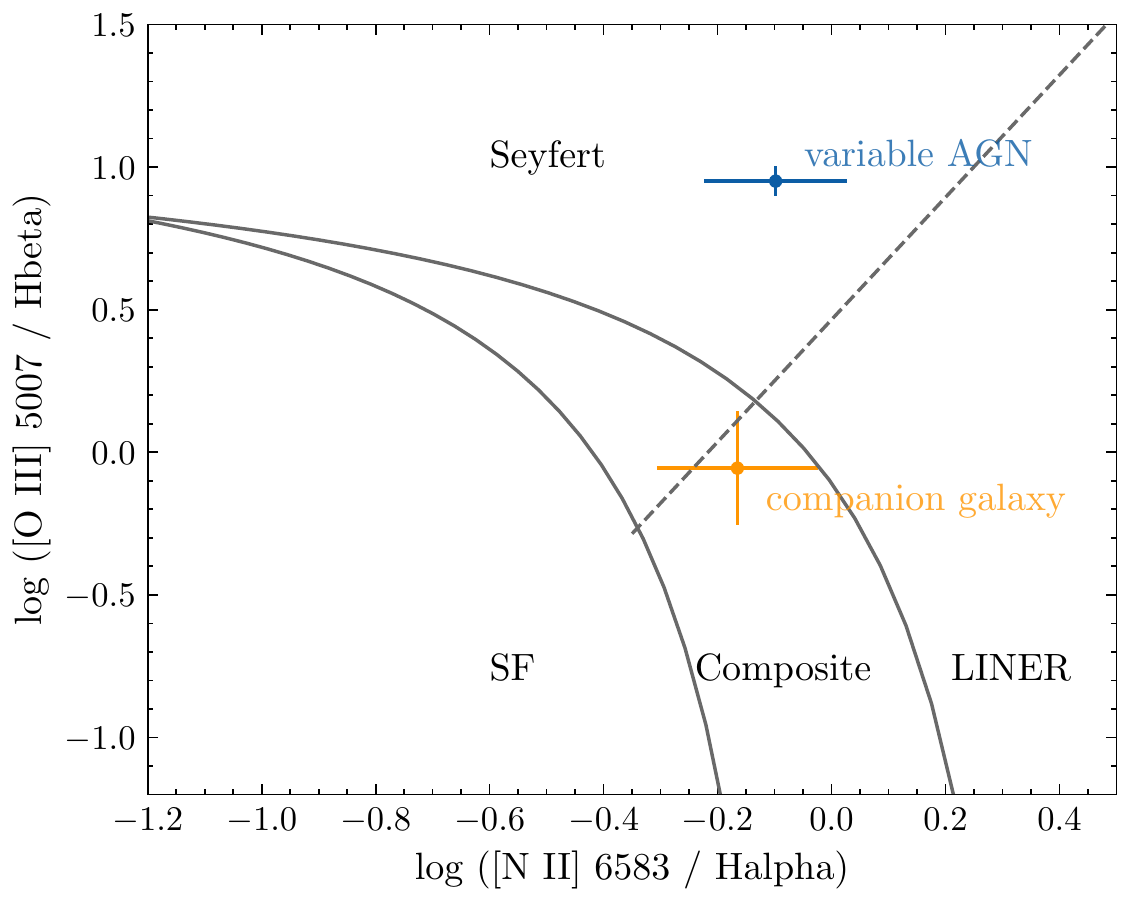}
  \caption{[O\,{\sc iii}] $\lambda$5007, H$\beta$, [N\,{\sc ii}] $\lambda$6573 and H$\alpha$ narrow emission line ratios from pPXF fitting of Deveny spectra taken of  the variable AGN ZTF18aaxvmpg and its companion galaxy overlaid on the AGN classification regions. The ZTF AGN is classified as a Seyfert while the offset galaxy has a combination of star forming and LINER emission.}
  \label{fig:bpt_ZTF18aaxvmpg}
\end{figure}
Using the best-fit line fluxes from pPXF modeling of the spectrum, we determine the AGN and host galaxy types using the [O\,{\sc iii}] $\lambda$5007, H$\beta$, [N\,{\sc ii}] $\lambda$6573 and H$\alpha$ line amplitudes. To separate starburst galaxies, we use the  \citet{Kauffmann2003} condition:
\begin{equation}
    \log(\text{[O\,{\sc iii}]}/\text{H}\beta) < 0.61 / \log(\text{[N\,{\sc ii}]}/\text{H} \alpha) - 0.05) + 1.3
\end{equation}
The composite region is defined to be between the above condition and: 
\begin{equation}
    \log(\text{[O\,{\sc iii}]}/\text{H}\beta) < 0.61 / \log(\text{[N\,{\sc ii}]}/ \text{H} \alpha - 0.47) + 1.19
\end{equation}
The emission line ratios and their classifications for the AGN and companion galaxy are shown in Figure \ref{fig:bpt_ZTF18aaxvmpg}. 
We find that the variable AGN has narrow line emission consistent with a Seyfert, while the host galaxy narrow line emission falls into the Composite/LINER category. Since LINER emission can  be produced by either AGN or hot old stars on galactic scales we must distinguish between the two using a WHAN diagram, which classifies LINERs with a narrow H$\alpha$ equivalent width $>3$ \AA \ and a log$_{10}$([N\,{\sc ii}] $\lambda$6573/ H$\alpha$) flux ratio $>-0.4$ as AGN \citep{CidFernandes2011,Mezcua2020}.
With \texttt{pPXF} we measure the equivalent width of narrow H$\alpha$ to be $3.54\pm0.11$ \AA \ and the log$_{10}$([N\,{\sc ii}] $\lambda$6573/ H$\alpha$) flux ratio to be $-0.165\pm0.097$. This places the spectrum in the `weak AGN' class of the WHAN diagram, suggesting that the system does indeed host a second AGN. We therefore conclude that ZTF18aaxvmpg is part of a dual AGN.

6 of the 27 broad line AGN in mergers for which we have archival SDSS spectra showed double-peaked broad Balmer emission from an unobscured accretion disk, corresponding to 22\% of the sample. The classifications of broad line region shapes of each AGN in a merger is shown in Table \ref{table:mergers}. It will be shown in Section 3.3.1 that 16\% of the broad line ZTF AGN are double-peaked emitters. It is therefore possible that AGN in mergers are more likely to have double-peaked broad lines than normal ZTF AGN.

\begin{figure}
  \centering
  \includegraphics[width=0.49\textwidth]{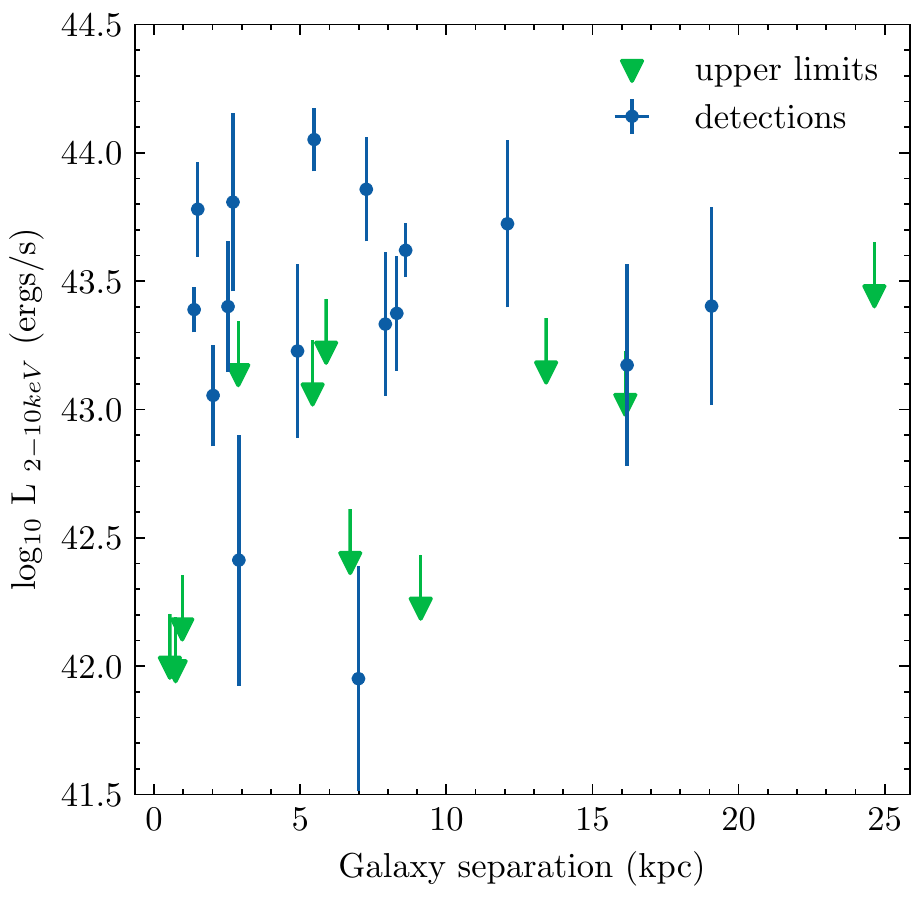}
  \caption{2-10 keV luminosity for 27 AGN with known spectroscopic redshifts as a function of physical galaxy separation. 16 were detected in the second ROSAT All Sky Survey catalog and 11 have non-detections.}
  \label{fig:xraysep}
\end{figure}

\begin{figure}
  \centering
  \includegraphics[width=0.49\textwidth]{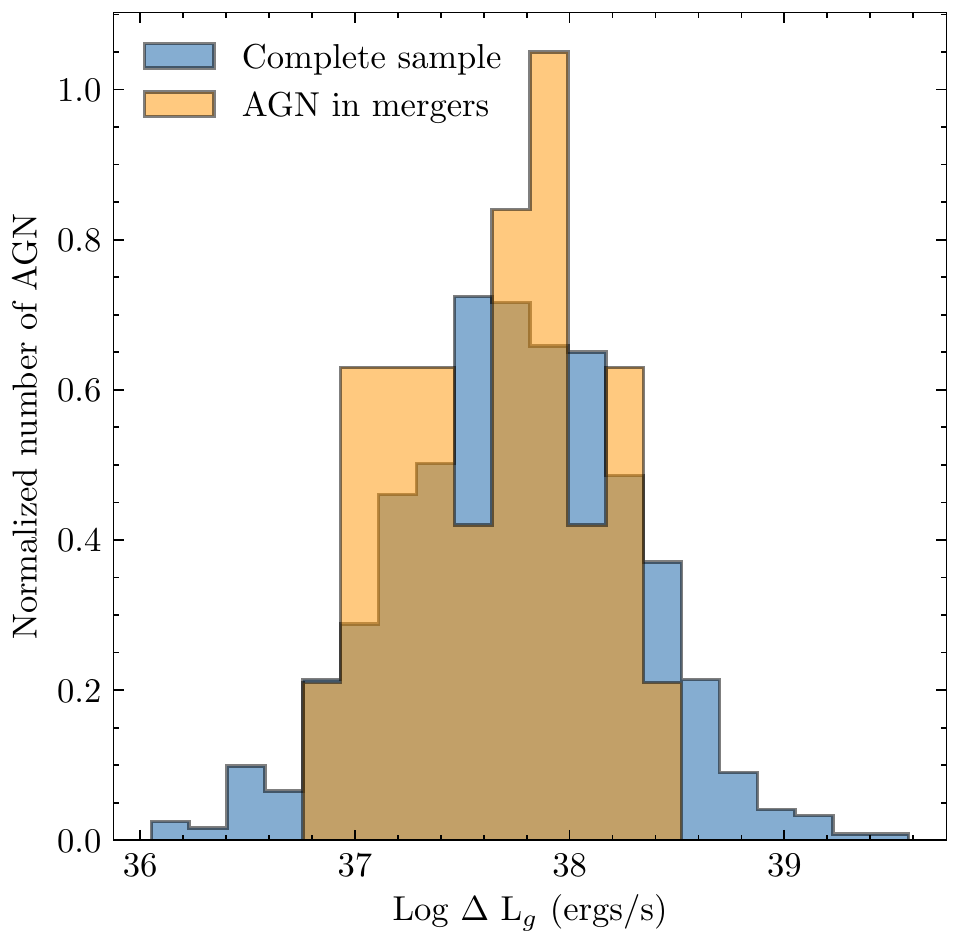}
  \caption{Peak luminosity change relative to the reference image amongst the 27 AGN in mergers and a larger sample of 689 variable AGN which had spectroscopic redshifts available. The larger sample was controlled to undergo the same quality cuts as the merger sample.}
  \label{fig:peakhist}
\end{figure}
In order to determine the X-ray luminosities of the sample, we crossmatched with a 60" radius to the second ROSAT All Sky Survey catalog \citep{Boller2016}. 31 of the 52 AGN had X-ray detections. The X-ray fluxes for the 31 AGN, along with luminosities for those with a known redshift, are shown in the appendix in Table \ref{table:rosat}. 9 of the AGN in mergers had 20cm radio detections in FIRST. 6 of the 10 AGN with detected radio emission were also X-ray bright. The smaller fraction of radio AGN compared to X-ray AGN may be due to delayed triggering of radio emission during the merger \citep{Shabala2017, Skipper2018}.

The relationship between galaxy separation and X-ray luminosity is shown in Figure \ref{fig:xraysep}. We do not find a correlation between galaxy separation and X-ray luminosity in the range of 1-19\,kpc separations represented by our sample. While \citet{Koss2012} found that the X-ray luminosity of AGN pairs decreased with increasing separation up to 90\,kpc, they also did not find a strong correlation at $<20$\,kpc. 

In order to compare the variability of the AGN in mergers with the complete AGN sample, we determined the maximum flux change between the ZTF reference image and single epoch science images for the AGN with spectroscopic redshifts, and determined the corresponding change in luminosity. The distribution of peak luminosity change is shown in Figure \ref{fig:peakhist} for the merger sample and for a larger sample of 689 ZTF AGN, controlled for the quality cuts used to produce the merger sample. The two samples show a similar distribution of peak luminosity. Implementing a K-S test to compare the two samples confirms they are drawn from the same distribution with a p-value of 0.64. 

\subsection{Chance coincidences of AGN and background galaxies}
In our sample of offset AGN we find 29 AGN which are offset from an undisturbed galaxy with a symmetrically shaped profile. While many of these objects are likely to be quasars coinciding with background galaxies, it is possible that a fraction of them are AGN with a real association to the spatially offset host. We therefore calculate the approximate number of chance coincidences with background galaxies that we expect for the sample of 3110 AGN which were modeled by \texttt{The Tractor}. 

To determine the density of background galaxies, we use \texttt{Casjobs} to query the SDSS DR16 catalog to find the number of galaxies with g-band model magnitudes between 15 and 22.8. We choose a g-band limiting magnitude of 22.8 because it is the median effective depth of each set of 30 ZTF images used for \texttt{Tractor} modeling for the AGN sample. We note that the choice of the median limiting magnitude is a significant approximation because there is a wide distribution in effective depths for the \texttt{Tractor} models of the AGN. Future work should take a more systematic approach to reach a consistent limiting magnitude for all objects.

We find that there are 119,364,394 galaxies in the SDSS survey area of 14,555 deg$^2$ within this magnitude range, corresponding to a density of $6.328\times10^{-4}$ per arcsec$^2$. This means that we expect $1.789\times 10^{-2}$ background galaxies in a 3" radius circle around a given AGN. For our sample of 3110 AGN with \texttt{Tractor} modeled offsets, we therefore expect 56 AGN to have unassociated background galaxies within 3 arcseconds. 

As we do not have an excess of AGN which are intrinsically offset from undisturbed host galaxies beyond the estimated number of chance coincidences, we do not have evidence that any of these objects could be recoiling SMBH candidates. The AGN falling under this category require spectra to confirm the host galaxy redshift and provide evidence that they are not chance coincidences.

\subsection{AGN spatially offset from the center of disturbed galaxies}

\begin{deluxetable*}{lccccccccc}
\tabletypesize{\scriptsize}
\tablecolumns{10}
\tablewidth{0pt}
\tablecaption{Properties of the offset AGN sample \label{table:results1}}
\tablehead{
\colhead{ZTF name} & \colhead{z} & \colhead{RA} & \colhead{Dec} &\colhead{Host-AGN offset}& \colhead{Host-AGN offset} & \colhead{BLR velocity} &\colhead{X-ray?}&\colhead{Radio flux}&\colhead{R}\\
 &  & (hms) &(dms) &(arcseconds)&(kpc) &(km s$^{-1}$)& & (mJy/beam)}
\startdata
ZTF19aautrth&0.208&16:30:41.964&30:36:2.448&$1.076\pm0.003$&$4.48\pm0.012$&BS+C&\checkmark&$<0.98$&$<1.56$\\
ZTF19aadgijf&0.262&14:26:14.312&27:29:55.98&$1.197\pm0.005$&$6.195\pm0.026$&$-462.0\pm0.3$&\checkmark&$<1.24$&$<2.05$\\
ZTF18aaxmrom&0.347&16:09:11.257&17:56:16.271&$1.057\pm0.01$&$7.089\pm0.067$&RS+C&\checkmark&$444.48$&1760\\
ZTF19aayrjsx&0.215&23:32:54.463&15:13:5.407&$1.329\pm0.003$&$5.695\pm0.013$&2P+C&\checkmark&$-$&$-$\\
ZTF18aalsidi&0.348&15:53:57.736&47:52:32.015&$0.901\pm0.004$&$6.051\pm0.027$&RS+C&\checkmark&$<0.88$&$<10.32$\\
ZTF18accptjn&0.214&22:12:17.117&3:50:40.531&$0.648\pm0.005$&$2.767\pm0.021$&$344.3\pm0.3$&$\times$&1.12&0.87\\
ZTF18absvcae&0.755&20:49:07.593&5:13:17.362&$1.679\pm0.004$&$22.035\pm0.052$&$-$&$\times$&4.97&51.6\\
ZTF18aaoeobb&0.270&12:48:53.9&34:24:29.448&$0.429\pm0.004$&$2.284\pm0.021$&2B+C&\checkmark&$<0.93$&$<0.95$\\
ZTF19aadggaf&0.266&13:42:06.57&5:05:23.898&$0.488\pm0.004$&$2.554\pm0.021$&$608.2\pm0.3$&\checkmark&3.80&2.89\\
\enddata
\vspace{0.1cm}
\tablecomments{Spectroscopic and multi-wavelength properties of the sample of AGN spatially offset from disturbed host galaxies. The redshifts are based on the position of O\,{\sc iii} lines in the archival SDSS spectrum of the AGN. The spatial offsets are obtained from modeling of point source and galaxy profile positions in the Legacy Survey images with \texttt{Tractor}. For the 3 AGN with Gaussian broad lines the broad line velocities are found by modeling of a Gaussian Balmer broad line series with \texttt{pPXF}. For objects which are poorly modeled by a Gaussian and well-modeled by a double-peaked accretion disk model \citep{Chen1989a} we instead classify the shape of the broad line by adopting the scheme of \citet{Strateva2003}: prominent red shoulder (RS), prominent blue shoulder (BS), two prominent peaks (2P), two blended peaks (2B). If a central Gaussian broad line is required in addition to the double-peaked accretion disk model we indicate this with `+C'. The X-ray detection column shows which objects have detections within 1.0' of the AGN position in the a ROSAT All Sky Survey catalog, and the radio detection column shows the 20cm flux density or upper limit if it was within the coverage of the FIRST radio survey \citep{Helfand2015}. The radio loudness is indicated in the last column.}
\end{deluxetable*}

\begin{figure*}[t]
\begin{minipage}[c]{0.99\textwidth}
 \centerline{\includegraphics[width=7cm]{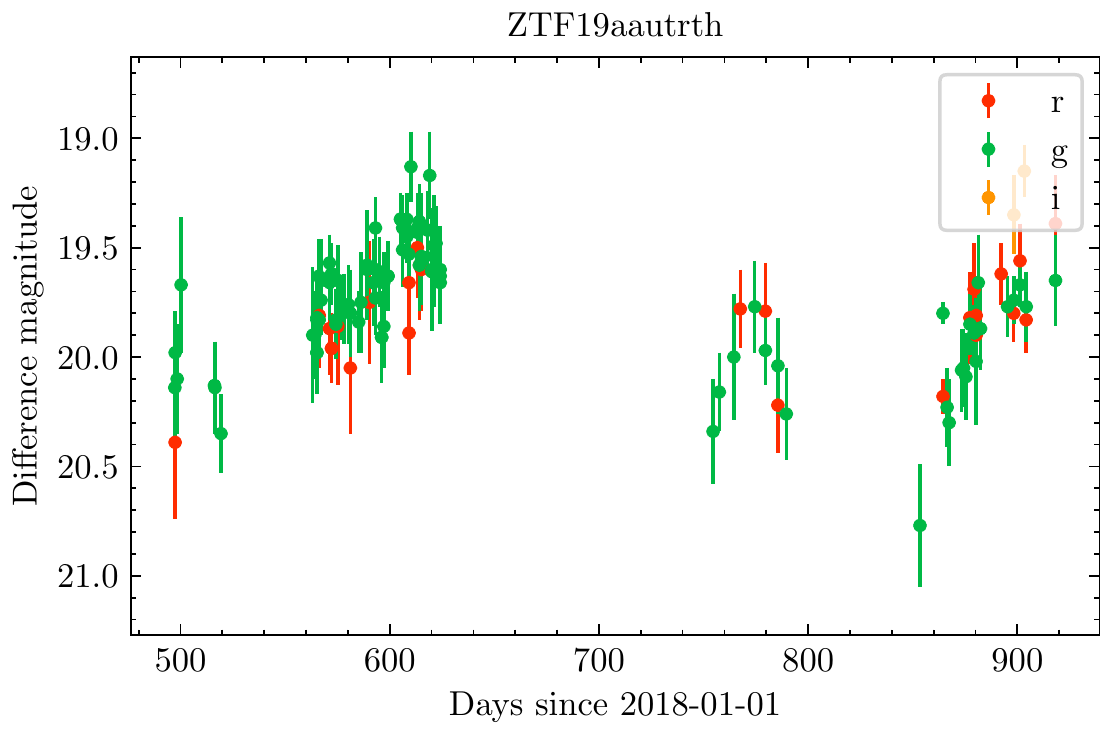}
 \includegraphics[width=7cm]{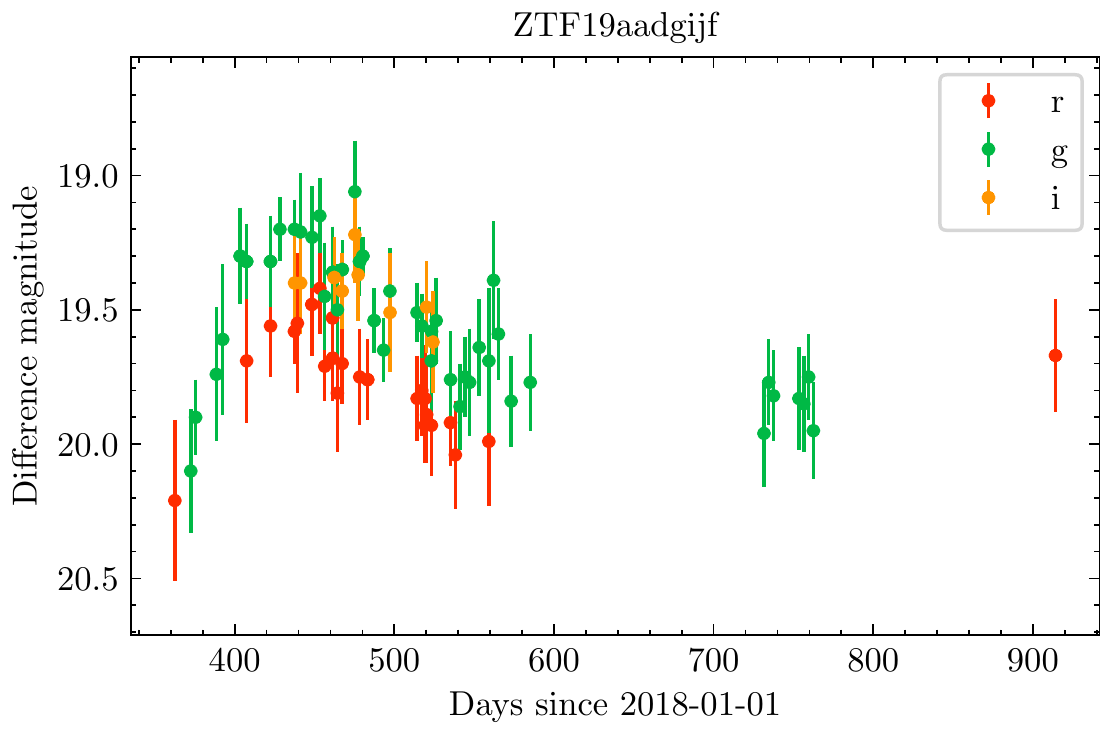}
 }
\end{minipage}\hfill
\begin{minipage}[c]{0.99\textwidth}
 \centerline{\includegraphics[width=7cm]{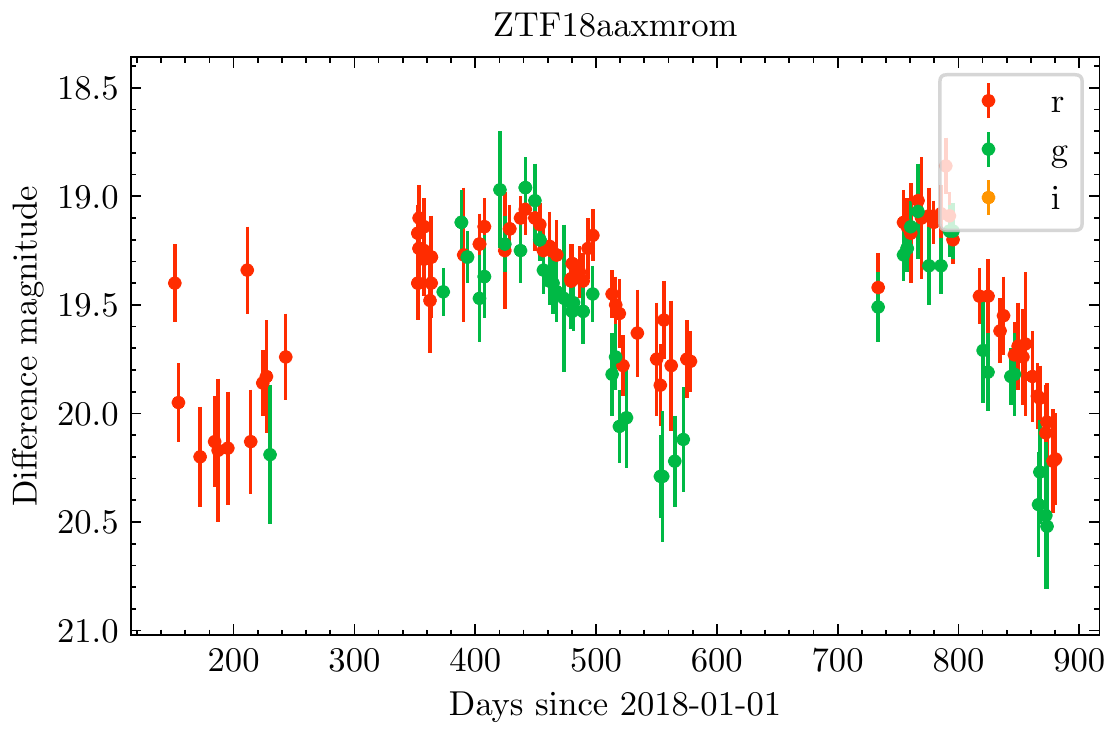}
 \includegraphics[width=7cm]{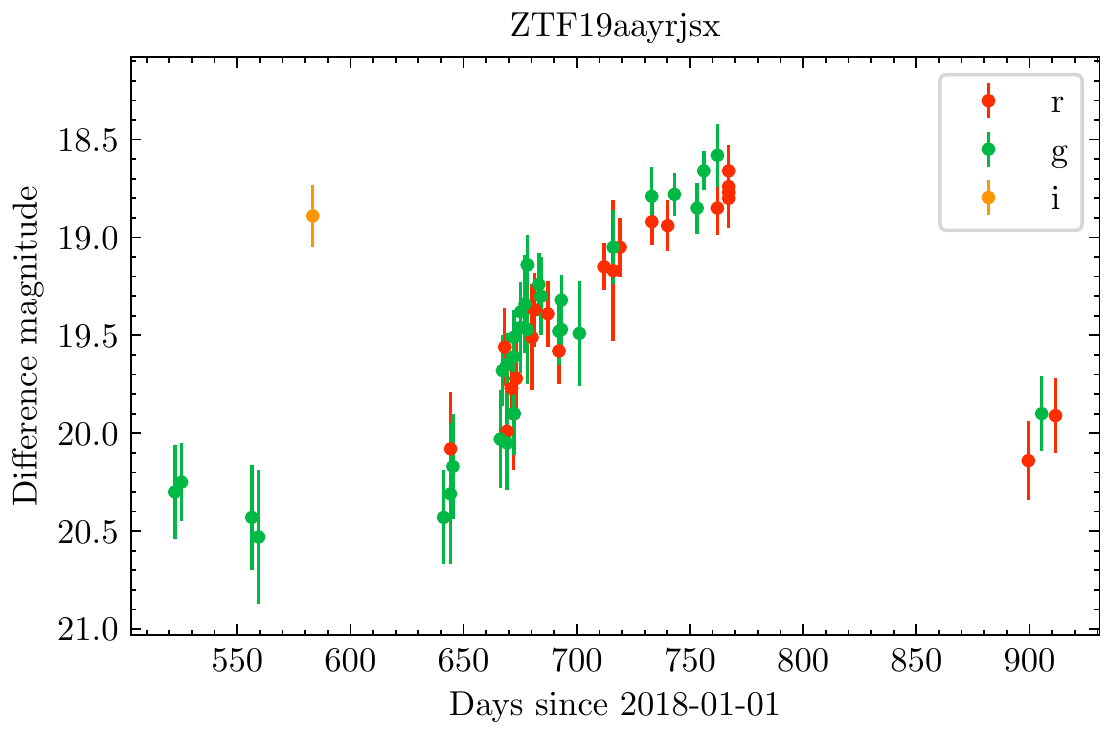}}
\end{minipage}\hfill
\begin{minipage}[c]{0.99\textwidth}
 \centerline{\includegraphics[width=7cm]{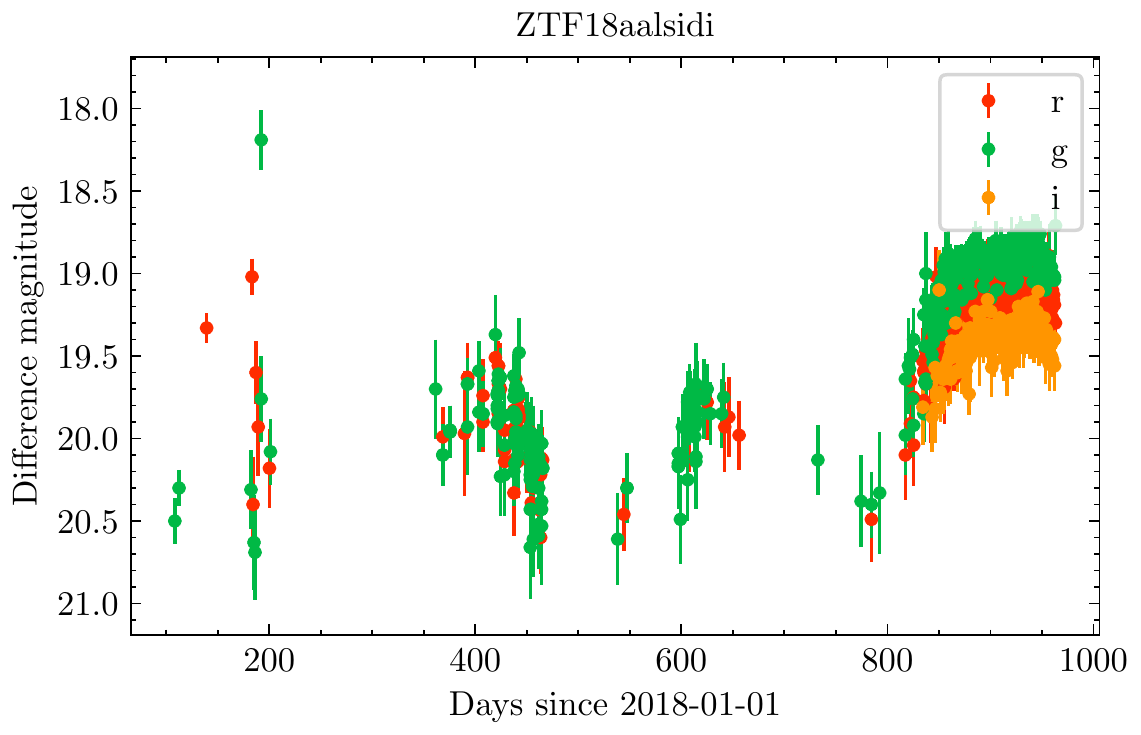}
 \includegraphics[width=7cm]{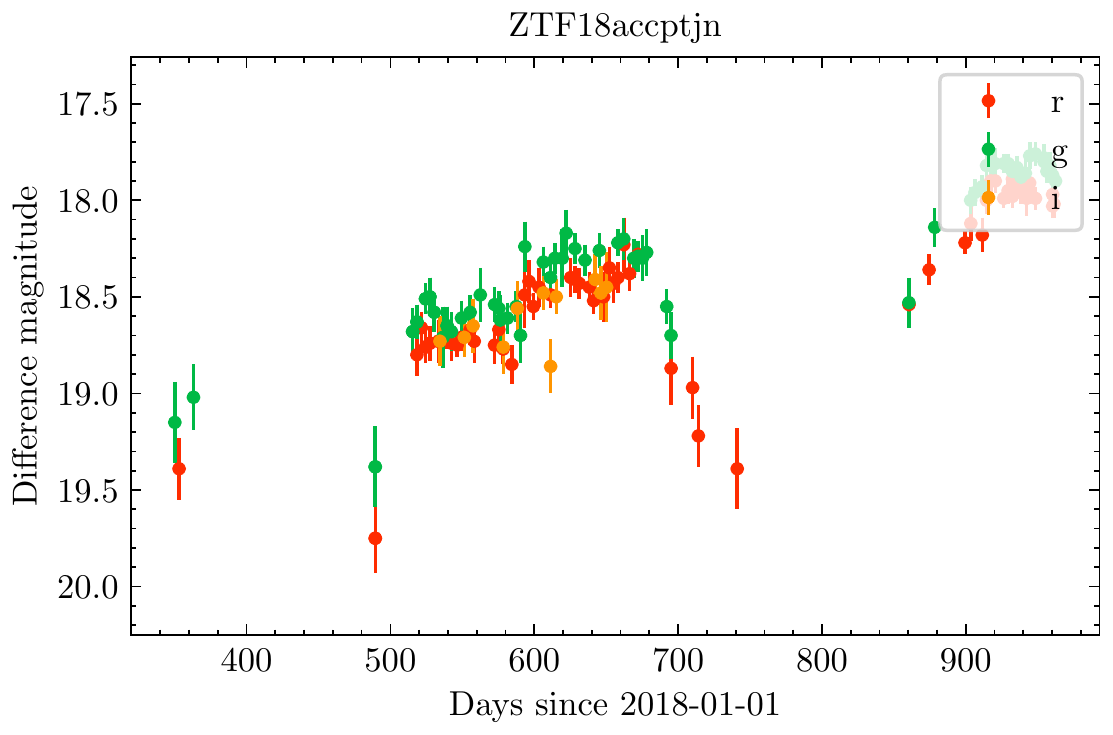}}
\end{minipage}\hfill
\begin{minipage}[c]{0.99\textwidth}
 \centerline{\includegraphics[width=7cm]{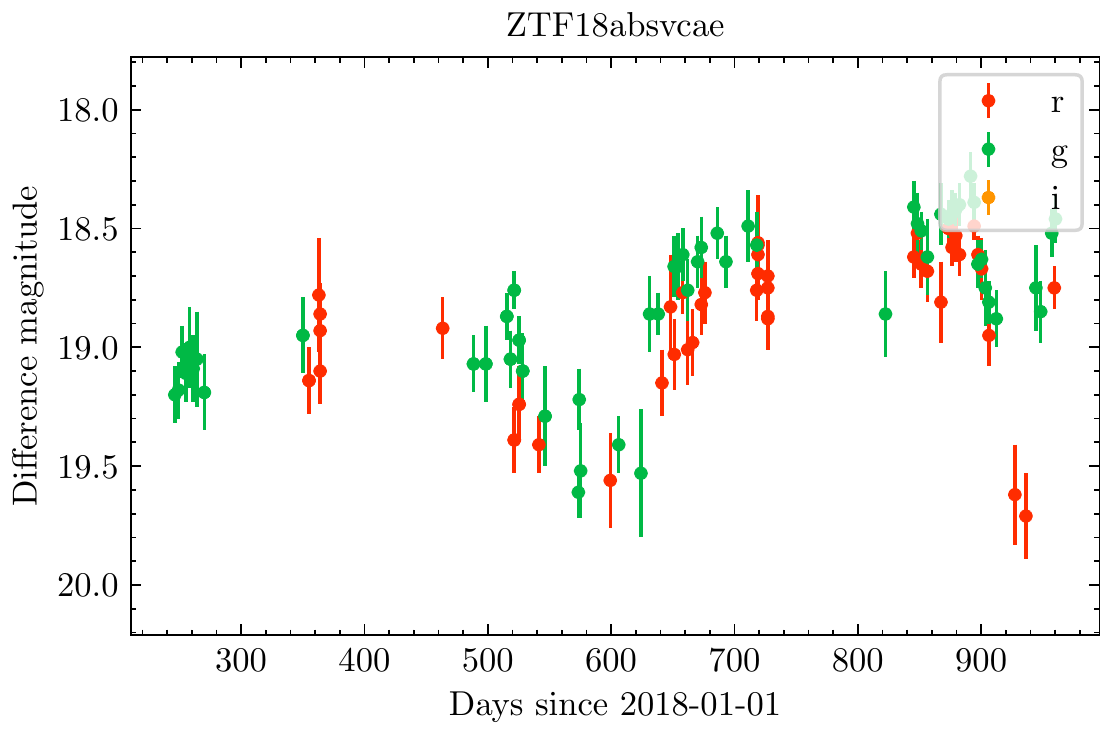}
 \includegraphics[width=7cm]{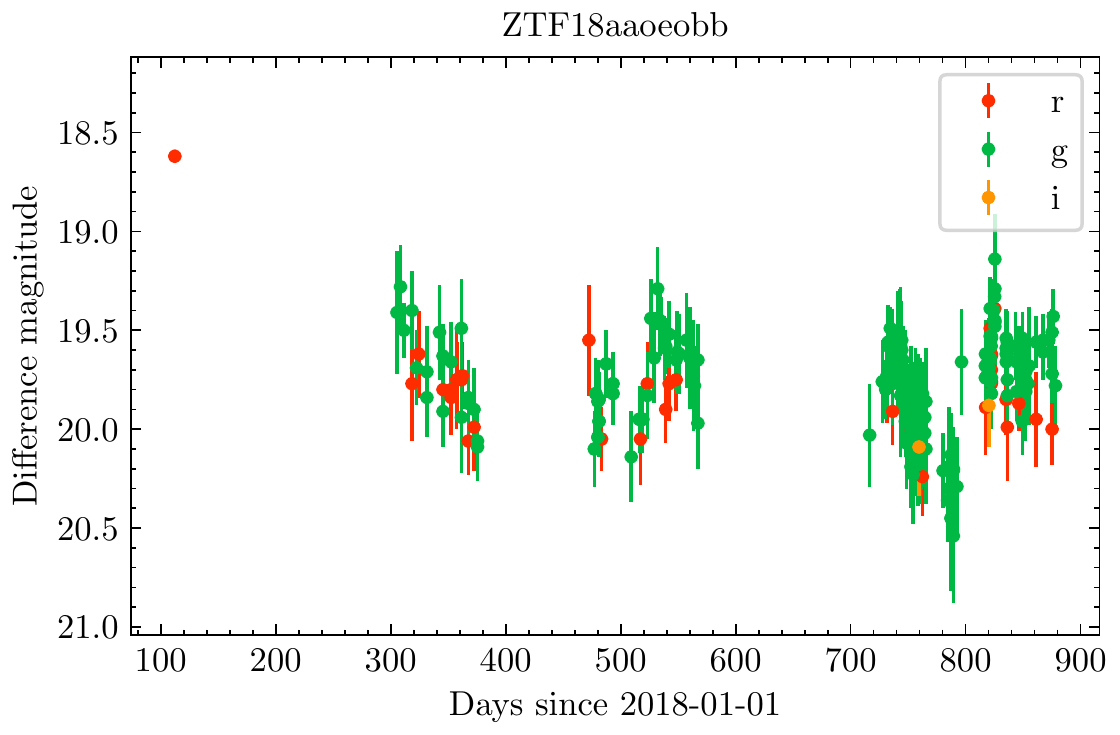}}
\end{minipage}\hfill
\
\begin{minipage}[c]{0.99\textwidth}
 \centerline{\includegraphics[width=7cm]{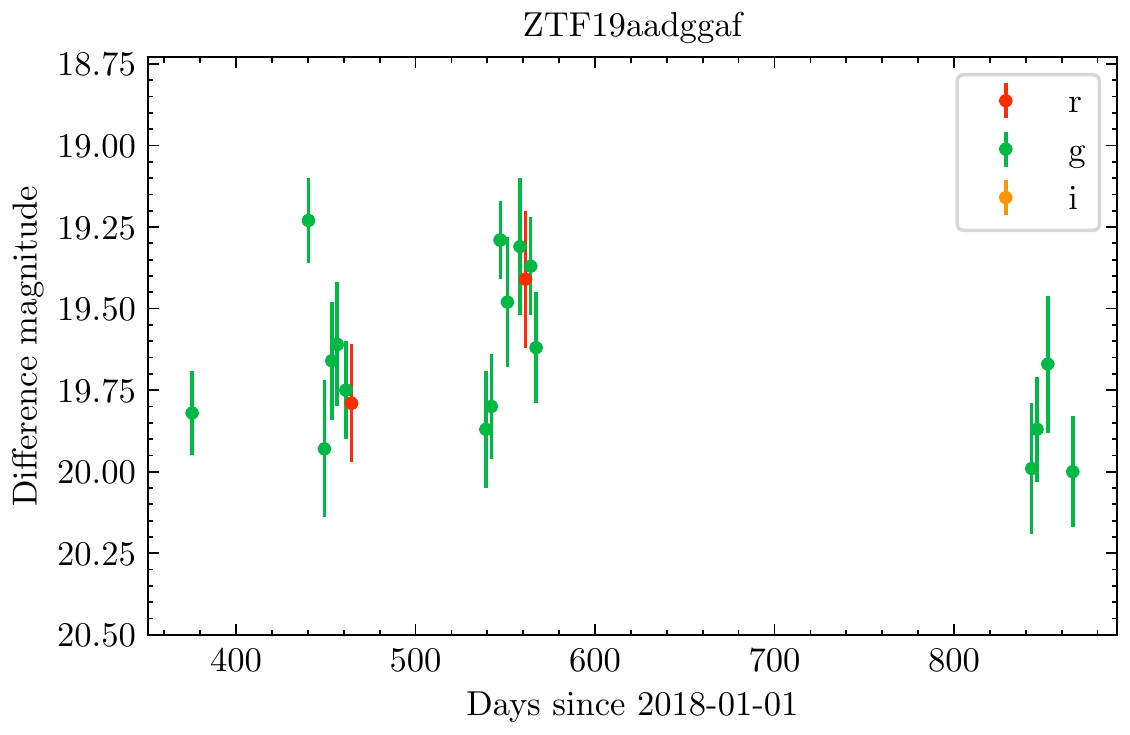}
 \includegraphics[width=7cm]{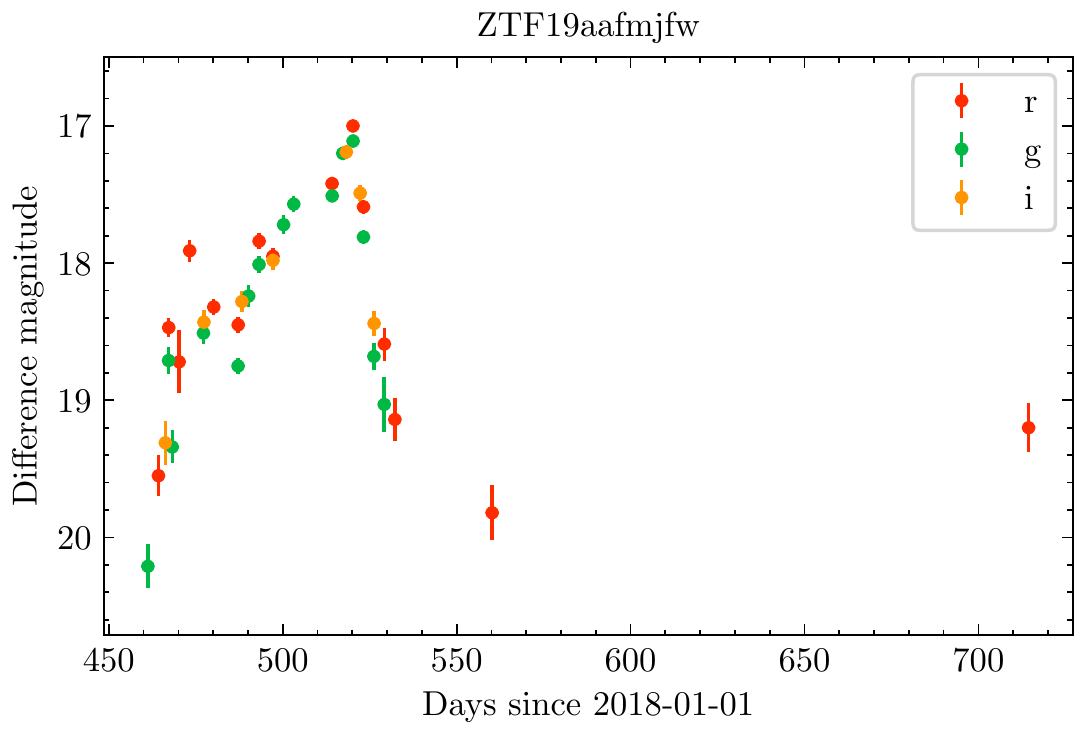}}
\end{minipage}\hfill
 \caption{\small
 ZTF light curves in g, r and i-band for the sample of 9 offset AGN and the known recoiling SMBH candidate in SDSS1133 which rebrightened in ZTF (ZTF19aafmjfw). We show only the $>3\sigma$ detections.}
\label{fig:marshallc}
\end{figure*}

\begin{figure*}
\gridline{\fig{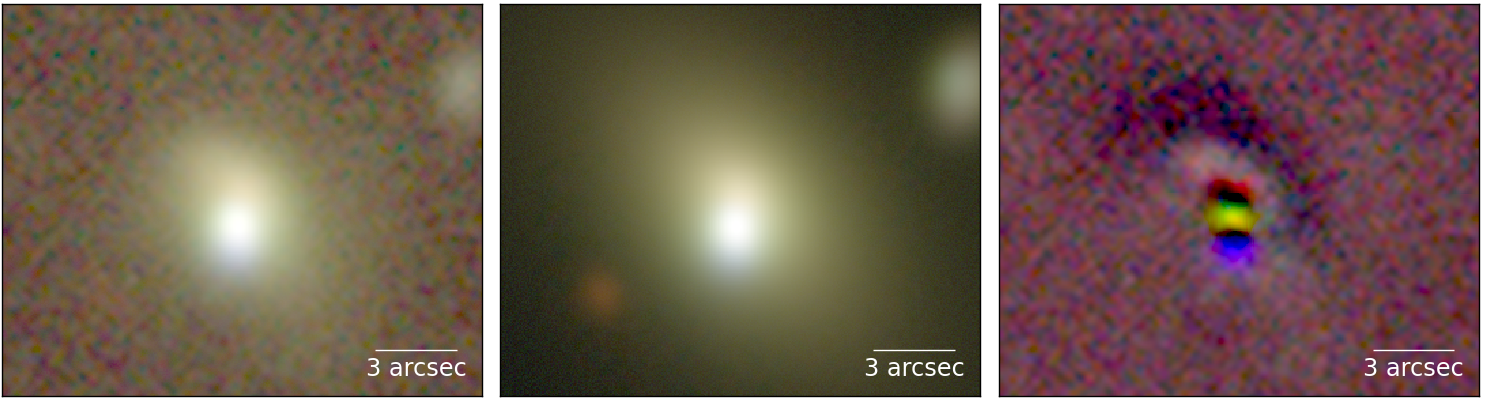}{0.49\textwidth}{(a) ZTF19aautrth}\fig{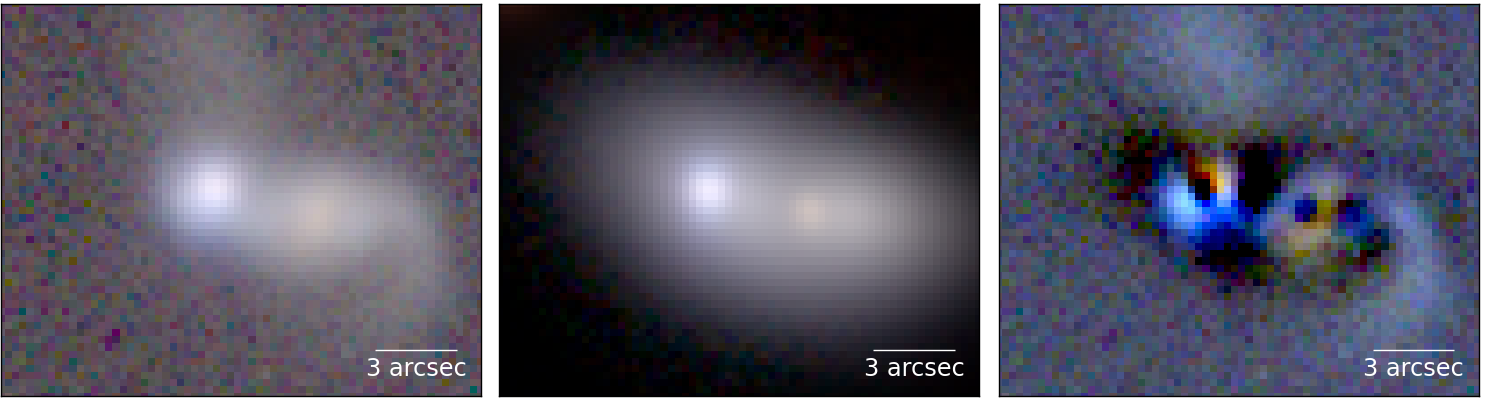}{0.49\textwidth}{(b) ZTF19aadgijf}}
\gridline{\fig{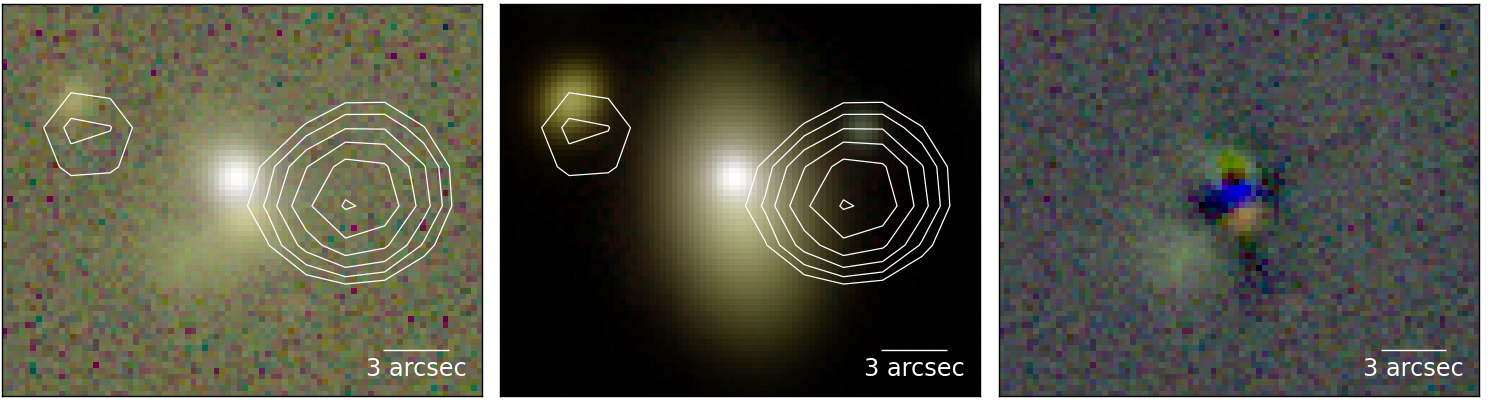}{0.49\textwidth}{(c) ZTF18aaxmrom}\fig{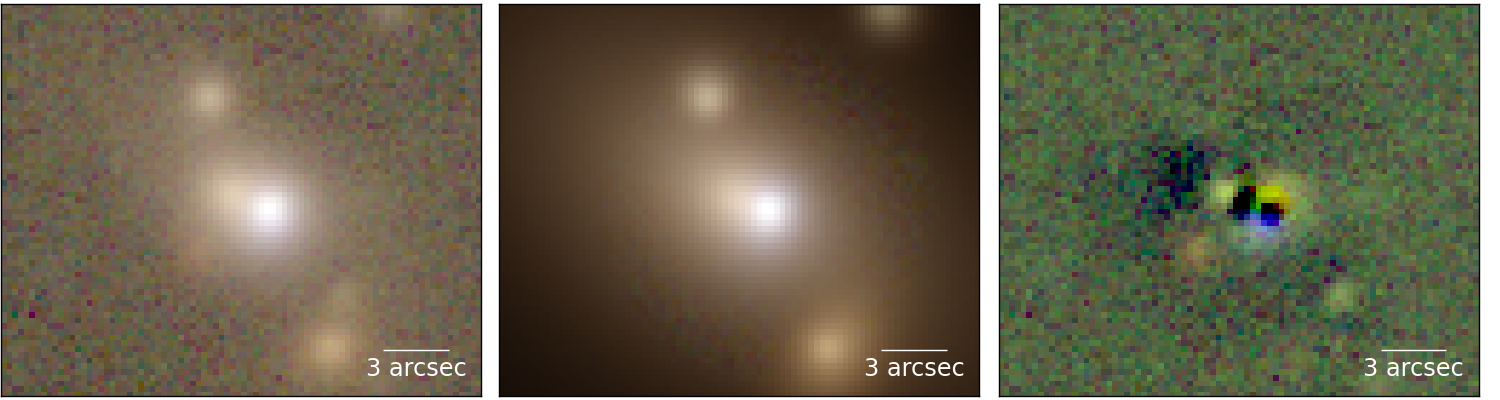}{0.49\textwidth}{(d) ZTF19aayrjsx}}
\gridline{\fig{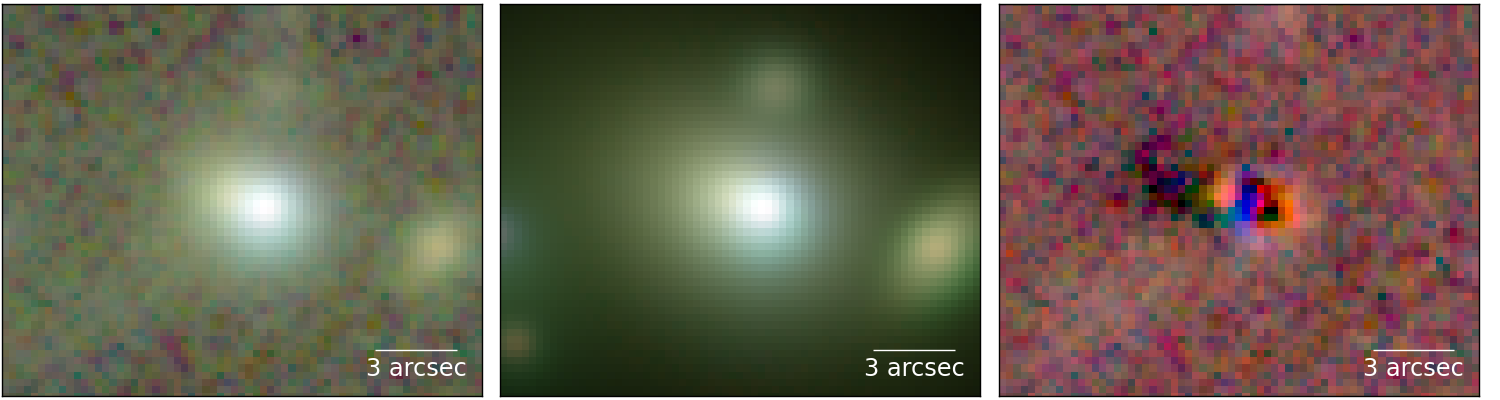}{0.49\textwidth}{(e) ZTF18aalsidi}\fig{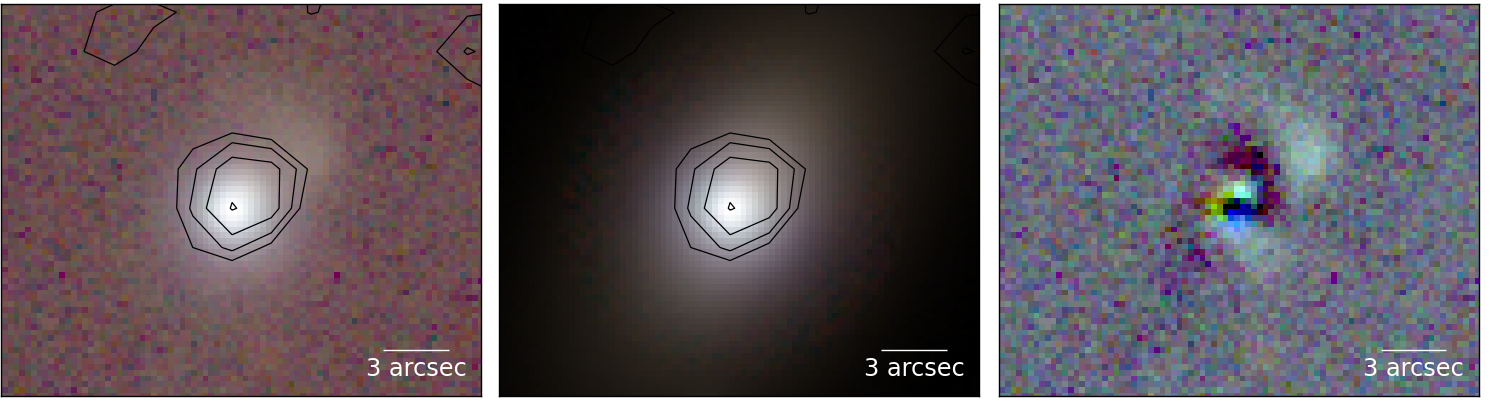}{0.49\textwidth}{(f) ZTF18accptjn}}
\gridline{\fig{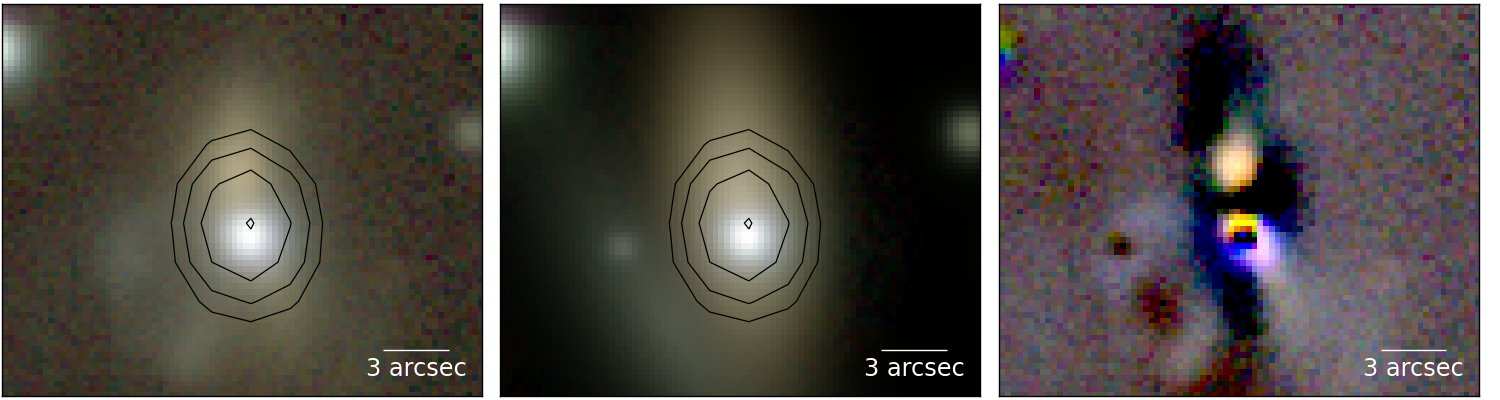}{0.49\textwidth}{(g) ZTF18absvcae}\fig{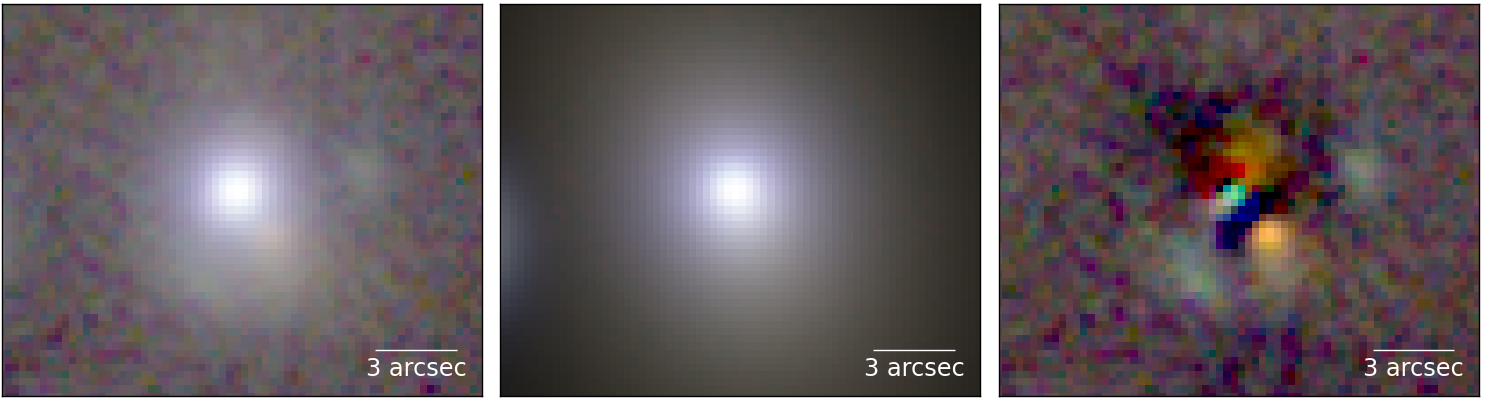}{0.49\textwidth}{(h) ZTF18aaoeobb}}
\gridline{\fig{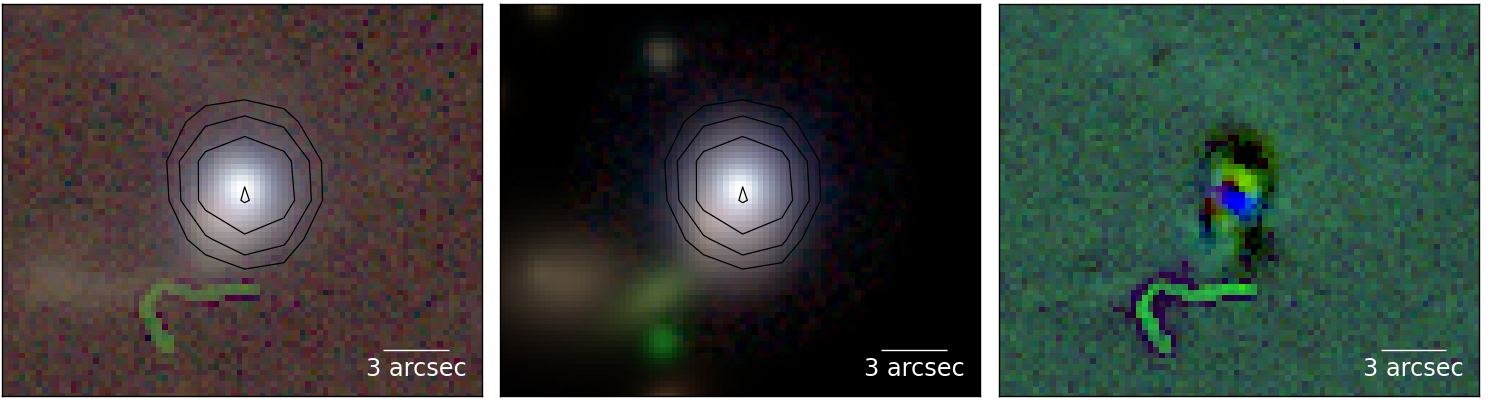}{0.49\textwidth}{(i) ZTF19aadggaf}}
\caption{For each offset AGN: \textit{Left:} Coadded g, r and z band Legacy Survey Images. \textit{Middle:} Corresponding coadded \texttt{Tractor} model. \textit{Right:} Image-model residuals showing galaxy tidal structures. For the 4 AGN with 20cm detections in FIRST (ZTF18aaxmrom, ZTF18accptjn, ZTF18absvcae, ZTF19aadggaf) we overlay contours of the FIRST image.} 
\label{fig:decals}
\end{figure*}

\begin{figure}
  \centering
  \includegraphics[width=0.495\textwidth]{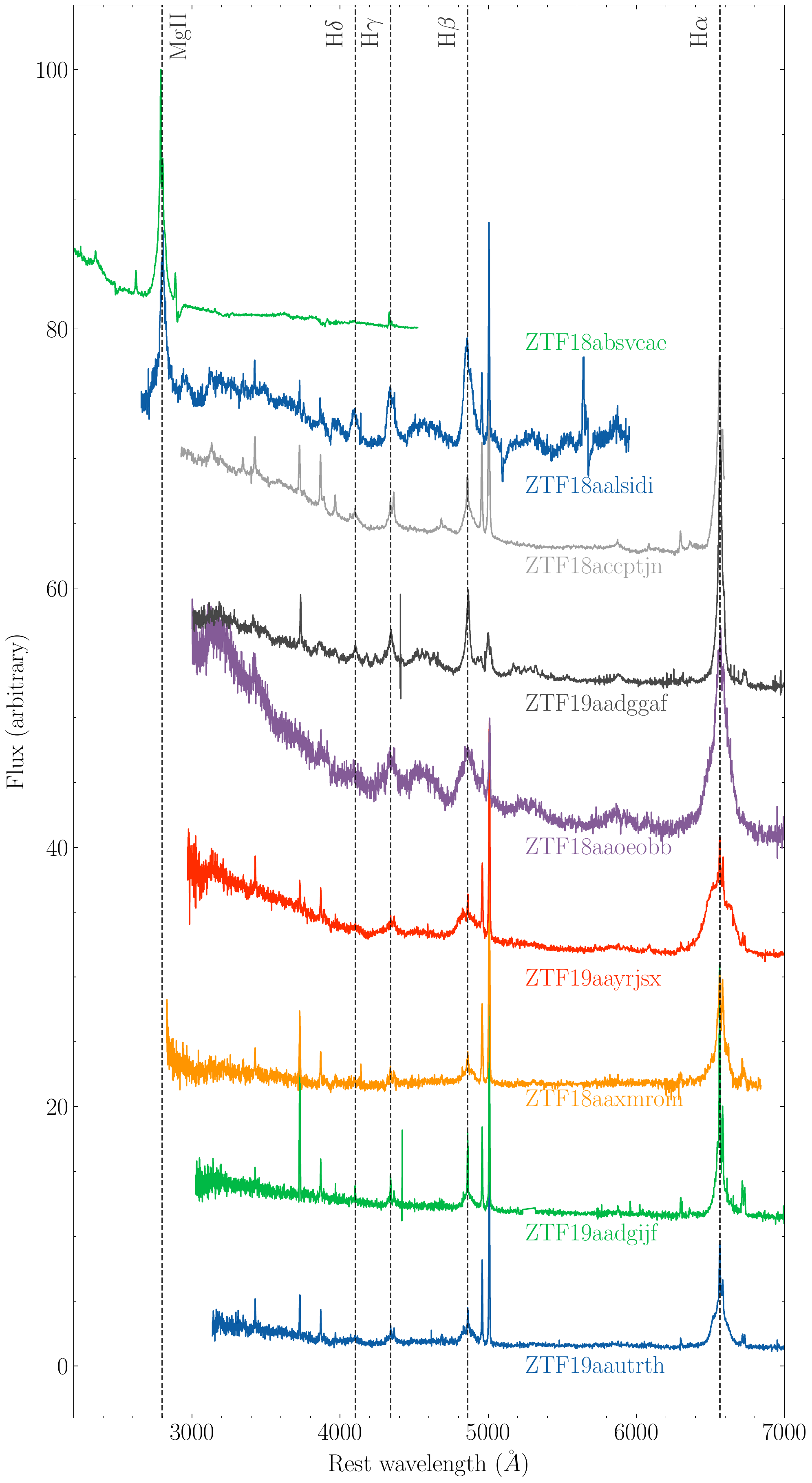}
  \caption{Spectra of the off-nuclear AGN, all from the SDSS archive except for ZTF18absvcae, ZTF18accptjn and ZTF18aalsidi which were observed with the Deveny spectrograph on LDT on 2020-09-13, 2020-09-15 and 2020-10-11 respectively. In all SDSS spectra, the fiber was centered on the AGN rather than the photometric center of the offset host galaxy. The Balmer series is shown with dotted lines.}
  \label{fig:hostspec}
\end{figure}

\begin{figure*}[h]
\begin{minipage}[c]{0.99\linewidth}
\centering
\centerline{
\includegraphics[width=0.49\linewidth, angle=0]{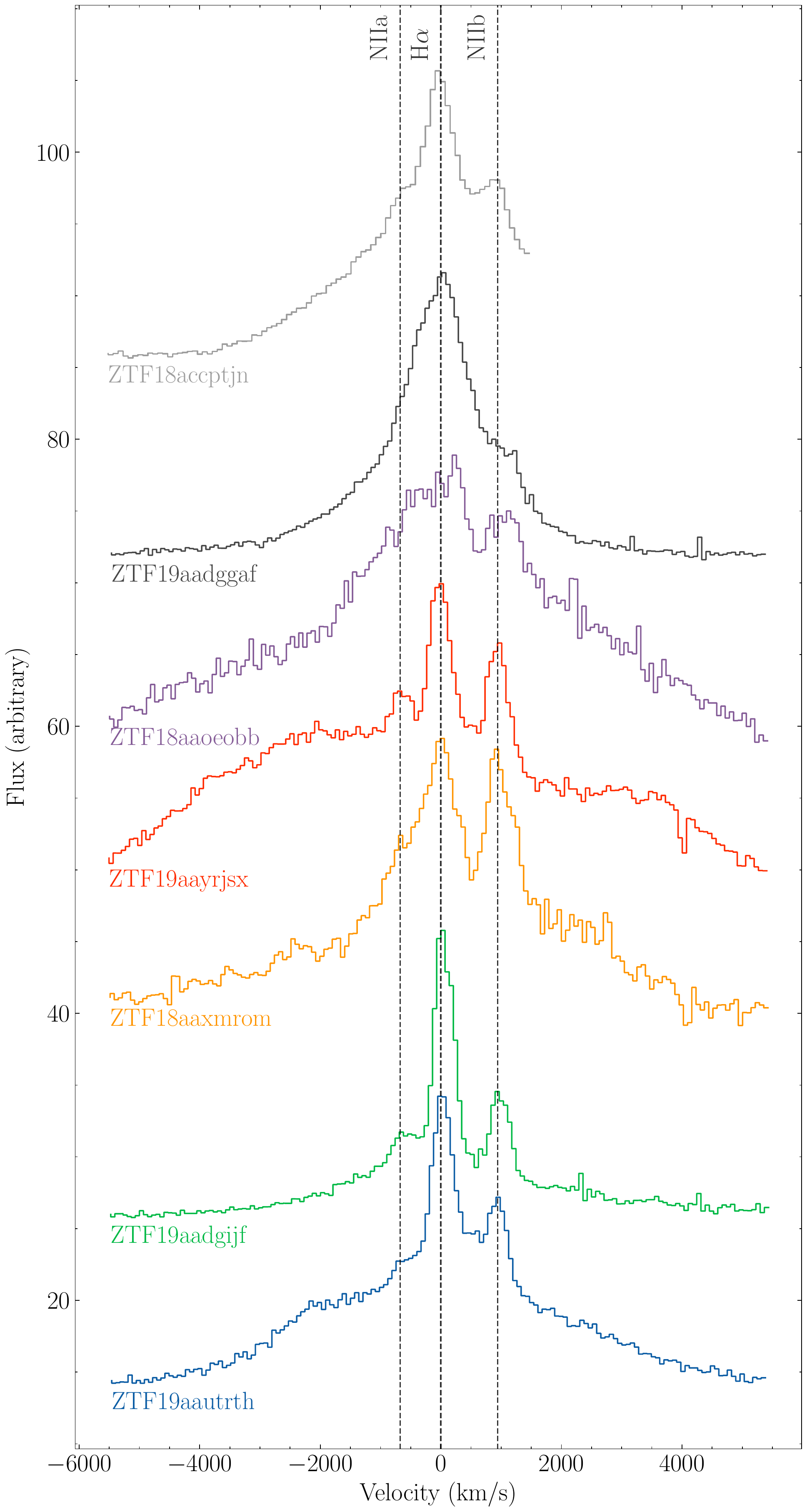}
\includegraphics[width=0.49\linewidth, angle=0]{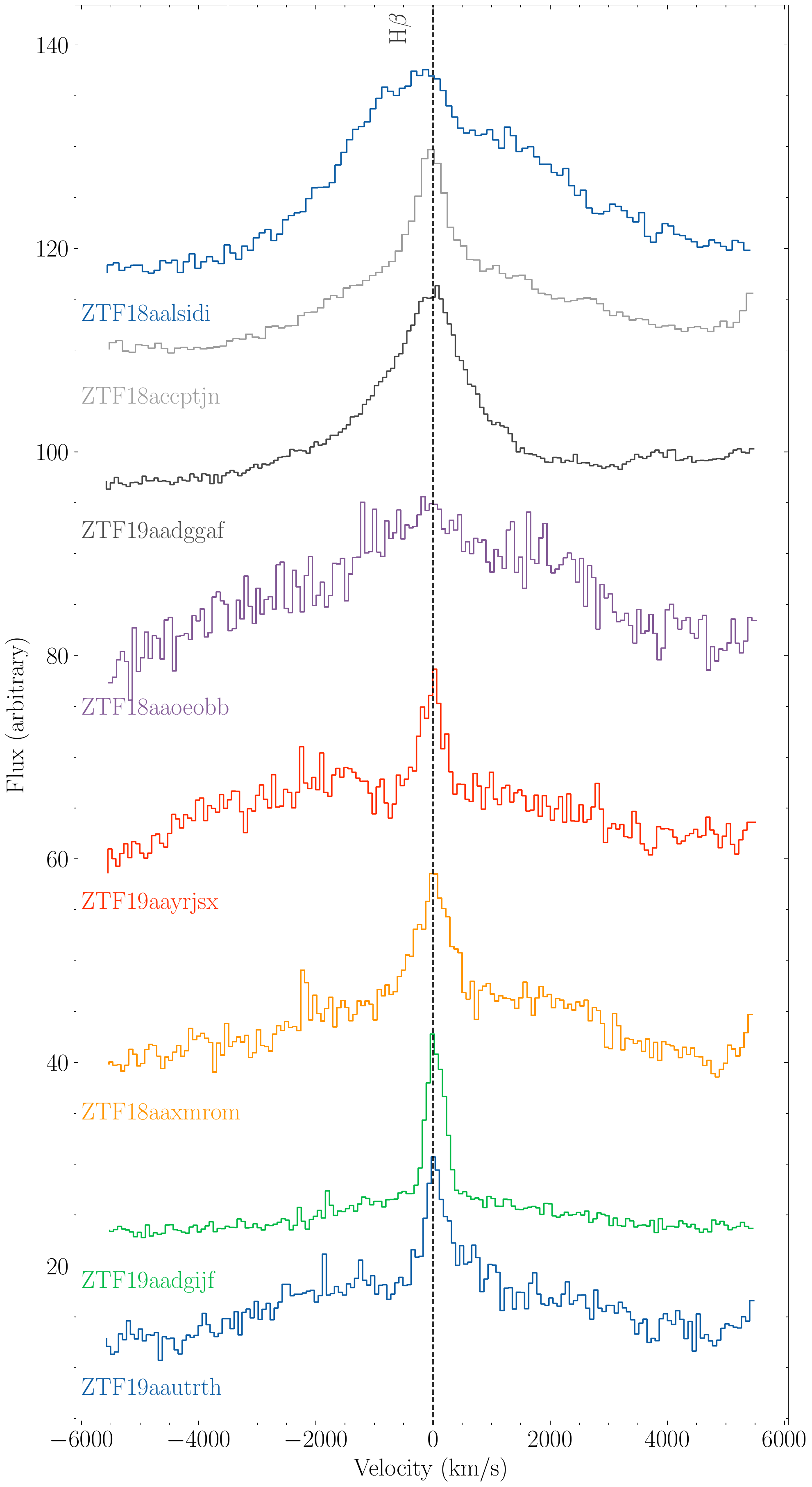}}
\caption{\textit{Left:} H$\alpha$ broad line regions and \textit{Right:} H$\beta$ broad line regions for the off-nuclear AGN candidates. 5 of the 8 off-nuclear AGN for which we have spectra of either the H$\alpha$ or H$\beta$ lines show asymmetric or double-peaked broad Balmer structures (ZTF18aaoeobb, ZTF19aayjrsx, ZTF18aaxmrom, ZTF19aautrth, ZTF18aalsidi). The other three show standard Balmer broad lines (ZTF18accptjn, ZTF19aadggaf, ZTF19aadgijf).}
\label{fig:spec_alpha}
\end{minipage}
\end{figure*}

Our morphological classification scheme finds 9 AGN which are variable point sources spatially offset from a potential host galaxy and do not show evidence of a second stellar bulge around the AGN. The properties of the 9 offset AGN are summarized in Table \ref{table:results1}. Their ZTF light curves are shown in Figure \ref{fig:marshallc}. We show the coadded g-, r- and z-band Legacy Survey images, the best-fit Tractor model and the residuals for each object in Figure \ref{fig:decals}. The residuals all show asymmetric, tidal structures indicative of previous merging activity. Spectra of the offset AGN are shown in Figure \ref{fig:hostspec}.

\begin{figure*}
\gridline{\fig{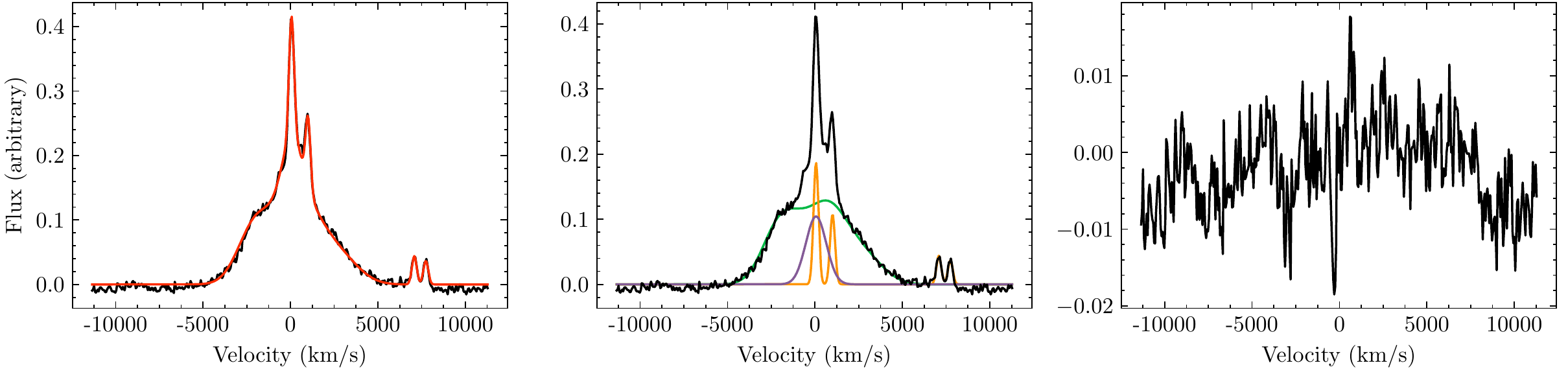}{0.8\textwidth}{ZTF19aautrth}}
\gridline{\fig{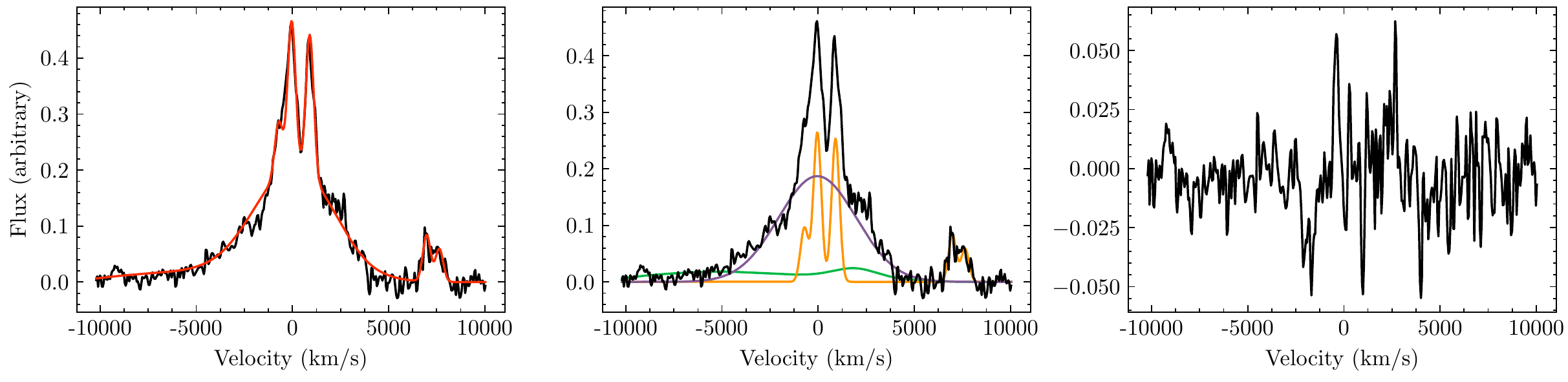}{0.8\textwidth}{ZTF18aaxmrom}}
\gridline{\fig{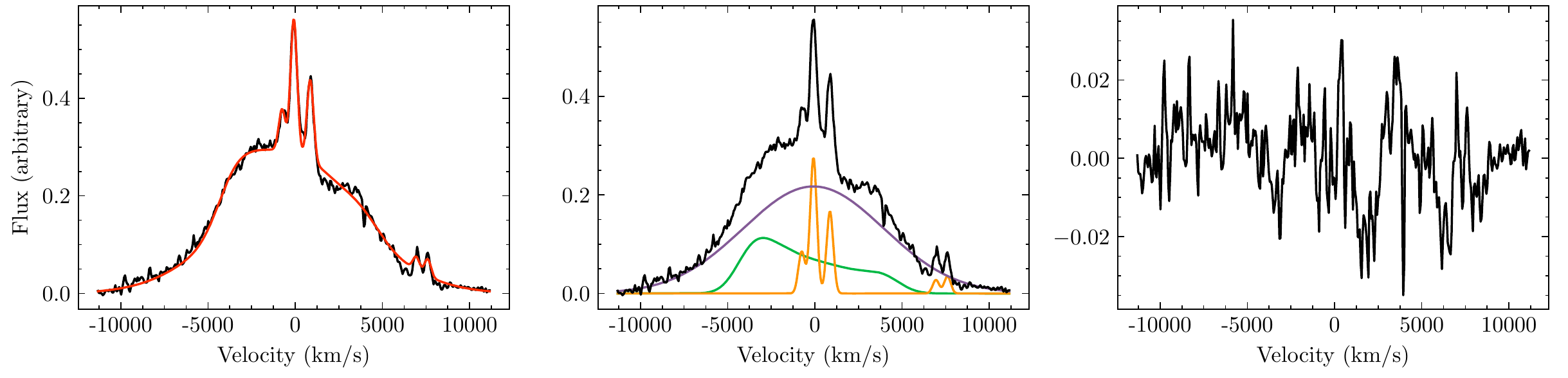}{0.8\textwidth}{ZTF19aayrjsx}}
\gridline{\fig{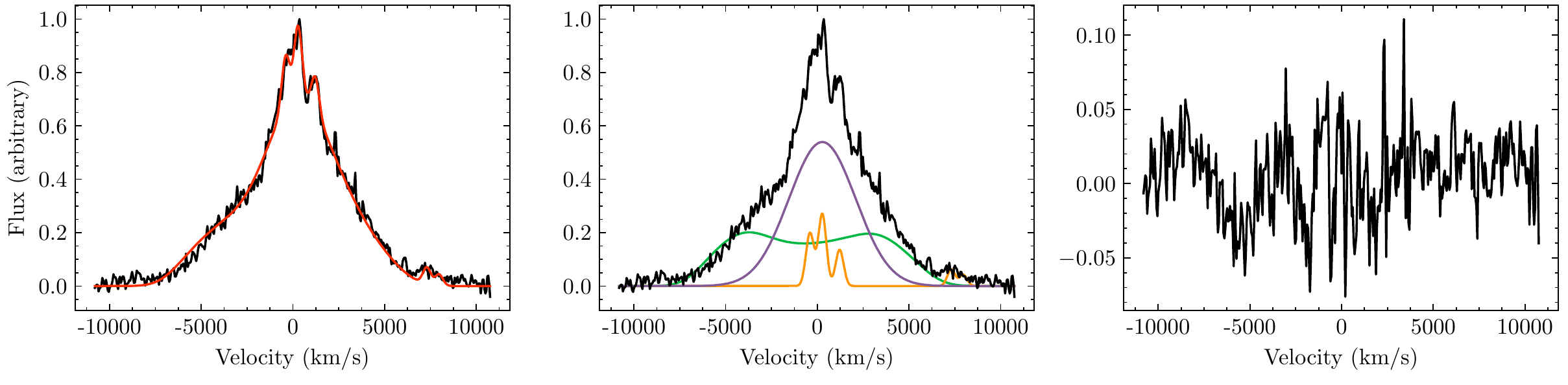}{0.8\textwidth}{ZTF18aaoeobb}}
\gridline{\fig{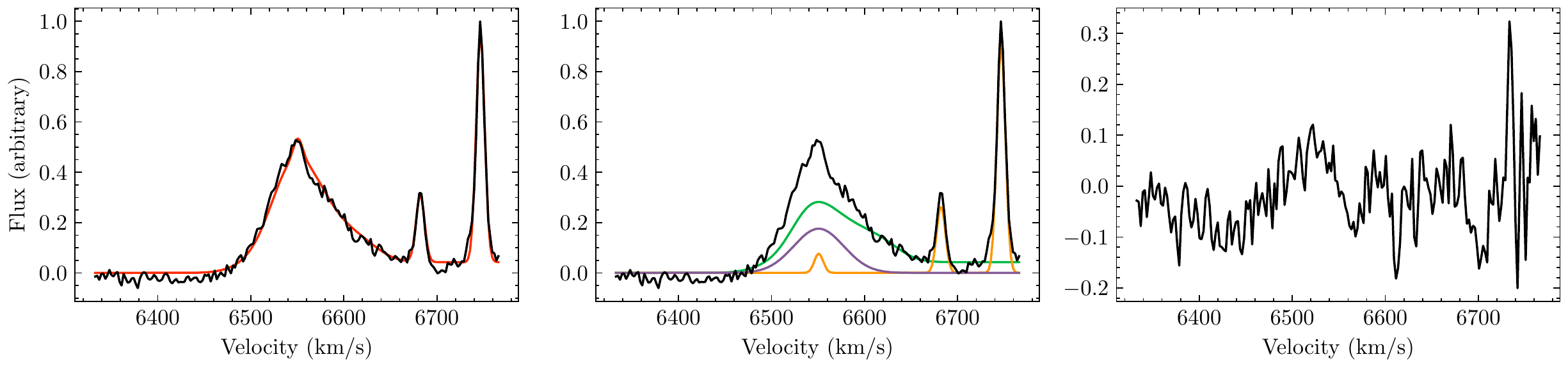}{0.8\textwidth}{ZTF18aalsidi}}
\caption{Best fit models of double-peaked H$\alpha$ accretion disk emission in off-nuclear AGN candidates ZTF19aautrth, ZTF18aaxmrom, ZTF19aayrjsx and ZTF18aaoeobb and H$\beta$ disk emission in ZTF18aalsidi after subtracting a stellar continuum model derived from pPXF. On the left we show the continuum subtracted data in black and the best fit narrow emission line and accretion disk model in red. The middle plots show the separate narrow line (green), accretion disk (orange) and central broad line (purple) components of the fit. The rightmost plots show the error weighted flux residuals after subtracting the best fit model.}
\label{fig:diskfit}
\end{figure*}

\begin{deluxetable*}{c|ccccc}
\tabletypesize{\footnotesize}
\tablecolumns{5}
\tablewidth{0pt}
\tablecaption{Best fit elliptical accretion disk parameters  \label{table:diskfit}}
\tablehead{
\colhead{Disk parameter}&\colhead{ZTF19aautrth}&\colhead{ZTF18aaxmrom}&\colhead{ZTF19aayrjsx}&\colhead{ZTF18aaoeobb}&\colhead{ZTF18aalsidi}}
\startdata
Central component required	& Yes & Yes & Yes & Yes & Yes\\
Inner radius ($\epsilon$)	& 252 & 813&  70	&997 & 21\\
Outer radius ($\epsilon$)	& 608 & 6800& 7995&3841	& 2255\\
Inclination (degrees) & 13.6 &84.9 & 14.0& 47.6 & 12.8\\
Turbulent broadening (c) & 0.0035 & 0.0007& 0.007 & 0.0019 & 0.0042\\
Ellipticity & 0.46 & 0.54 & 0.91 & 0.17&0.93\\
Disk orientation (degrees) & 18.2 & 36.6& 39.0 & 31.6 & 357.1
\enddata
\vspace{0.1cm}
\tablecomments{Best fit accretion disk parameters for the 5 double peaked emitters  from modeling of the H$\alpha$ (ZTF19aadgijf, ZTF19aautrth, ZTF18aaxmrom and ZTF19aayrjsx) and H$\beta$ (ZTF18aalsidi) double-peaked emission after subtracting a stellar continuum model derived from pPXF. Disk parameters were derived from a simple log likelihood minimization fit to illustrate that accretion disk emission models can describe the observed broad line shapes well.}
\end{deluxetable*}

\subsubsection{Broad line region properties}
The details of the H$\alpha$ and H$\beta$ broad line regions of the spatially offset AGN are shown in Figure \ref{fig:spec_alpha}. We do not have H$\alpha$ and H$\beta$ spectra for ZTF18absvcae due to its higher redshift. 3 of the 8 AGN for which we do have spectra of the H$\alpha$ or H$\beta$ lines (ZTF19aadgijf, ZTF18accptjn and ZTF19aadggaf) have Gaussian broad lines. The broad line velocity offsets determined from \texttt{pPXF} fitting are shown in Table \ref{table:results1}. Because of the large distribution of velocity offsets in the overall ZTF AGN sample (Figure \ref{fig:veldist}) and the fact that the ZTF AGN which have extreme $>1500$km\,s$^{-1}$ velocities do not show any evidence of a spatial offset, we do not make any conclusions from the velocities of 200--600 km\,$^{-1}$ magnitude observed for these 3 AGN.

The remaining 5 of the 8 AGN (ZTF18aalsidi, ZTF19aautrth, ZTF18aaxmrom, ZTF19aayrjsx and ZTF18aaoeobb) for which we have spectra of the H$\alpha$ or H$\beta$ lines have asymmetric Balmer broad line regions which are poorly fit by a Gaussian. Adopting the classification scheme of \citet{Strateva2003} we note that ZTF18aalsidi and ZTF18aaxmrom have prominent red shoulders, ZTF18aautrth has a prominent blue shoulder, ZTF19aayrjsx has two prominent peaks which are red and blue-shifted, and ZTF18aaoeobb has two blended peaks which are red and blue-shifted. These classifications are displayed in Table \ref{table:results1}. Such asymmetric structures can arise due to double-peaked emission from an unobscured, relativistic Keplerian accretion disk \citep{Chen1989a,Chen1989b,Eracleous1994Doubled-peakedNuclei}. 

To determine if the velocity-offset peaks observed in the broad Balmer lines of ZTF19aautrth, ZTF18aaxmrom, ZTF19aayrjsx, ZTF18aaoeobb and ZTF18aalsidi can be accounted for by accretion disk emission, we model the broad lines with an elliptical accretion disk model where an inner thick, hot ion torus illuminates a thin outer disk of ionized gas, which has a power law relation between emissivity and radius given by slope $q$ \citep{Chen1989a,Strateva2003}. The model depends on the inner and outer dimensionless gravitational radii of the disk $\epsilon_1$ and $\epsilon_2$, local turbulent broadening parameter $\sigma$, azimuthal angle $\phi$, inclination angle $i$ where 0 degrees is face-on and 90 degrees is edge-on, ellipticity $e$ and disk orientation $\phi_0$. We bound the inner radius to $0<\epsilon_1<1000$, the outer radius to $0<\epsilon_2<8000$, the emissivity slope to $0<q<5$ and the local broadening to $0<\sigma<0.01$, to ensure that we could find good fits to the data with disk shape parameters consistent with the known double-peaked emitter population from \citet{Strateva2003}. All four objects required a central Gaussian broad line in addition to the accretion disk model to produce a good fit to the data, which is not uncommon for double-peaked emitters \citep{Strateva2003}. 

The best fit disk parameters from the H$\alpha$ and H$\beta$ fits of the 5 sources are shown in Table \ref{table:diskfit} and the models are shown in Figure \ref{fig:diskfit}. We note that our simple log likelihood minimization procedure for disk model fitting may not explore the whole parameter space and find the most optimal disk parameters. However, it serves to illustrate that accretion disk emission models can account for the extra flux which is not well described by a Gaussian model. More rigorous fitting with MCMC could be used to better determine disk shape parameters and their uncertainties in the future.

In order to determine if the double-peaked emitter fraction of 63\% amongst spatially offset AGN is substantially different from the entire variable ZTF AGN sample, we applied a multi-Gaussian fitting procedure to the 1923 variable AGN with archival SDSS spectra of the H$\alpha$ region which had broad Balmer lines to search for double-peaked emitters. We used the \texttt{pPXF} fitting procedure described in Section 2.4 to fit two different models. In the first model the spectrum was fit with a single broad Balmer line free to have a velocity offset to the narrow Balmer lines. This was fit at the same time as the narrow emission lines and stellar continuum. In the second model the spectrum was fit with 3 broad lines: one with central velocity tied to the narrow Balmer lines, one with a velocity up to 6000 km s$^{-1}$ relative to the narrow lines, and one with a velocity down to -6000 km s$^{-1}$. The widths of the 3 broad lines were not tied to each other.

To find double-peaked emitters from the fits of these two models, we required the $\chi^2$ improvement from the multiple broad line model compared to the single broad line model to be $>250$. Then to be considered a double-peaked emitter candidate the 2 velocity offset H$\alpha$ broad lines were each required to have peak flux densities of $>33$\% of the narrow H$\alpha$ line and to have a velocity $>500$ km s$^{-1}$ from the narrow line velocity. As the effectiveness of this criteria depended on the relative brightness of narrow and broad lines, and the width of the broad line region, we visually inspected the 275 candidates found via this criteria and rejected 82, leaving 193 double-peaked emitters. We then visually inspected the remaining 1648 spectra to find any objects which may have missed by the criteria. We found 106 double-peaked spectra which were missed by the automatic classification scheme. 

We therefore estimated that 299 of 1923, or ~16\%, of the variable AGN sample are double peaked emitters. This is much larger than the 3.6\% fraction found for SDSS AGN by \citet{Strateva2003} with spectroscopic principal component analysis, suggesting that either variable AGN are more likely to be double-peaked emitters, or that our classification scheme is more likely to classify asymmetric broad line regions as double-peaked emitters than classification schemes used in previous studies. The 16\% fraction of variable ZTF AGN with double-peaked broad Balmer lines is substantially smaller than 63\% fraction seen in the spatially offset AGN sample. 

\subsubsection{Multi-wavelength properties}
7 of the 9 offset AGN were detected in the ROSAT All Sky Survey and 4 AGN (ZTF18aaxmrom, ZTF18accptjn, ZTF18absvcae, ZTF19aadggaf) were detected at 20cm wavelengths in the FIRST radio survey, which uses the VLA to image a footprint largely coincident with SDSS to a detection sensitivity of 1 mJy \citep{Helfand2015}. The radio fluxes for these 4 AGN, as well as upper limits for another 4 which were within FIRST survey coverage but were not detected, are shown in Table \ref{table:results1}. The contours of FIRST radio imaging are overlaid in Figure \ref{fig:decals}. 
ZTF18aaxmrom shows two radio lobes on either side of the galaxy which are likely the result of synchrotron emission from a jet. The other 3 AGN show radio point sources which coincide with the position of the ZTF AGN. The radio loudness\footnote{$R=\frac{L_{5\text{GHz}(\nu)}}{L_{4400\text{\AA}}(\nu)}$} is also shown in Table \ref{table:results1}. Two AGN with radio emission are classified as radio-loud ($R>10$, \citet{Kellermann1989}), 2 are radio-moderate, and 4 others have upper limits indicating that they are not radio loud.  

The recoiling SMBH candidates from \citet{Chiaberge2017} and \citet{Lena2014} also demonstrate radio emission. As noted by \citet{Chiaberge2017}, radio emission from a recoiling SMBH is not surprising given that the rapid spin needed to produce a relativistic jet can be produced by a binary black hole merger \citep{Schnittman2013, Hemberger2013} and that there is a link between radio loud AGNs and major galaxy mergers \citep{Ivison2012, Chiaberge2015}. Assuming that the radio jet axis matches the spin axis of the recoiling SMBH, we would also expect recoil velocity to be preferentially aligned with the radio jet \citep{Lena2014}.

\subsubsection{Confirming the nature of the offset AGN}
There are two main alternative explanations for the nature of these offset AGN. Firstly, they may be chance coincidences with disturbed background galaxies. Secondly, they may be AGN in mergers with compact or undermassive host galaxies such that extended emission around the AGN is very faint. Given the predicted frequency of SMBHs in tidally stripped nuclei which may appear as offset AGN \citep{Voggel2019TheUniverse}, it would not be surprising if a large fraction of our objects are in fact AGN in compact galaxies merging with a larger galaxy. 

If the distribution of expected velocity and spatial offsets of the recoiling SMBH population follows the simulated distribution of \citet{Blecha2016}, we would expect the velocity offsets of our sample to range from 300 to 3000 km s$^{-1}$ given their large spatial offsets. Spectroscopic fitting (Section 3.3.1) shows that the 3 AGN with Gaussian broad lines have velocity offsets ranging from 344-608 km/s (see Table \ref{table:results1}) and the remainder of the sample with double-peaked emission have spectra consistent with a broad line region and accretion disk close to rest velocity (Figure \ref{fig:diskfit}). It is possible for large separation recoiling SMBHs to show small line of sight velocities, depending on their orbital trajectory at the time. However, if our objects are indeed all recoiling SMBHs, it is surprising that a fraction of the 9 do not have larger $>600$ km/s line of sight velocities. The observed velocities therefore argue against the recoil hypothesis for these objects.

Much further work is therefore required to investigate the recoiling SMBH hypothesis for any of these offset AGN. In order to rule out that they are not AGN in mergers with undermassive host galaxies, Hubble Space Telescope (HST) IR imaging could be used to search for a second extended region of old stellar emission around the offset AGN. HST imaging would be well complemented by IFU observations to map the positions of the narrow line emission relative to the AGN broad line emission \citep{Chiaberge2018}.

Chandra X-ray imaging to search for a second obscured AGN would also be essential for these candidates \citep{Comerford2017}. If we observe just one X-ray point source, we can produce upper limits on the X-ray luminosity of a second AGN and show that they rule out the presence of an obscured AGN using the [O\,{\sc iii}] flux from the host center. This would strongly favor the recoiling black hole hypothesis.

As spatial offsets greater than 1.5\,kpc are only possible in simulations where misaligned SMBH binary spins can occur \citep{Blecha2016}, if even a subset of our candidates could be confirmed as recoiling SMBHs, their extreme spatial offsets could demonstrate the existence of SMBH binaries with misaligned spins. 

\subsection{SDSS 1133 (ZTF19aafmjfw)}

SDSS1133, a variable object and recoiling SMBH candidate discovered by \citet{Koss2014} was identified by our search pipeline during a dramatic flare in ZTF, labeled ZTF19aafmjfw. This object has a long history of variability and the alternative scenarios proposed for its origin include: an LBV which exploded as a type IIn supernova in 2001, an LBV which continues to exhibit giant eruptions, or an offset AGN with flaring and stochastic variability \citep{Koss2014}. It was first observed in 1950 with the 103aO DSS plate at a magnitude of 18.6 and was again observed at comparable magnitudes in 1994 and 1999.  It was then discovered to be flaring to $g=16.4$ in SDSS on 2001-12-18 and 2002-04-01 but had faded to $g=18.7$ by 2003-03-09. It was observed at a minimum of $g=20.18$ with PS1 on 2012-02-22 after which it brightened to $g=19.3$ on 2014-01-20 \citep{Koss2014}.

\begin{figure}
\gridline{\fig{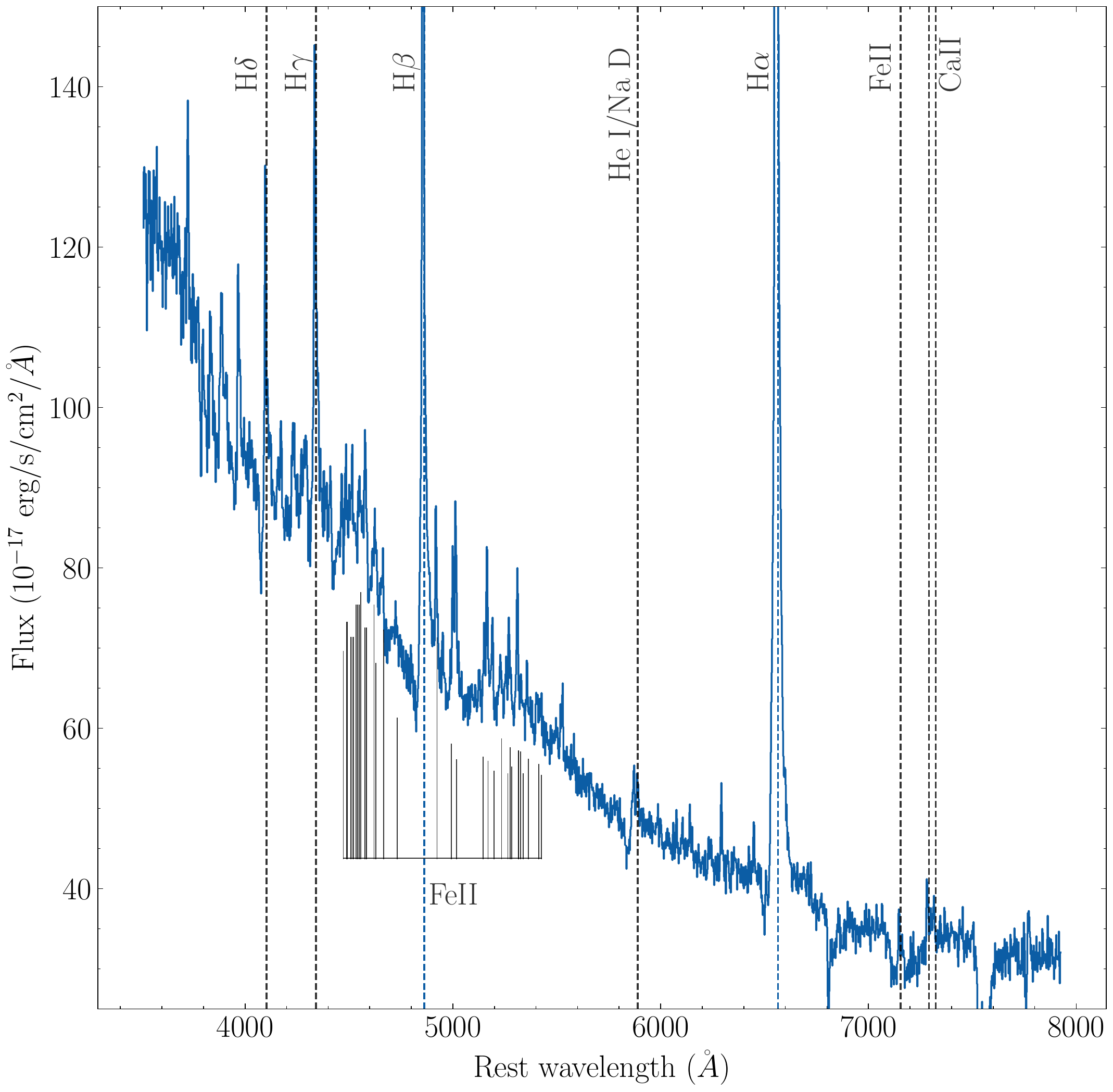}{0.47\textwidth}{Complete spectrum}}
\gridline{\fig{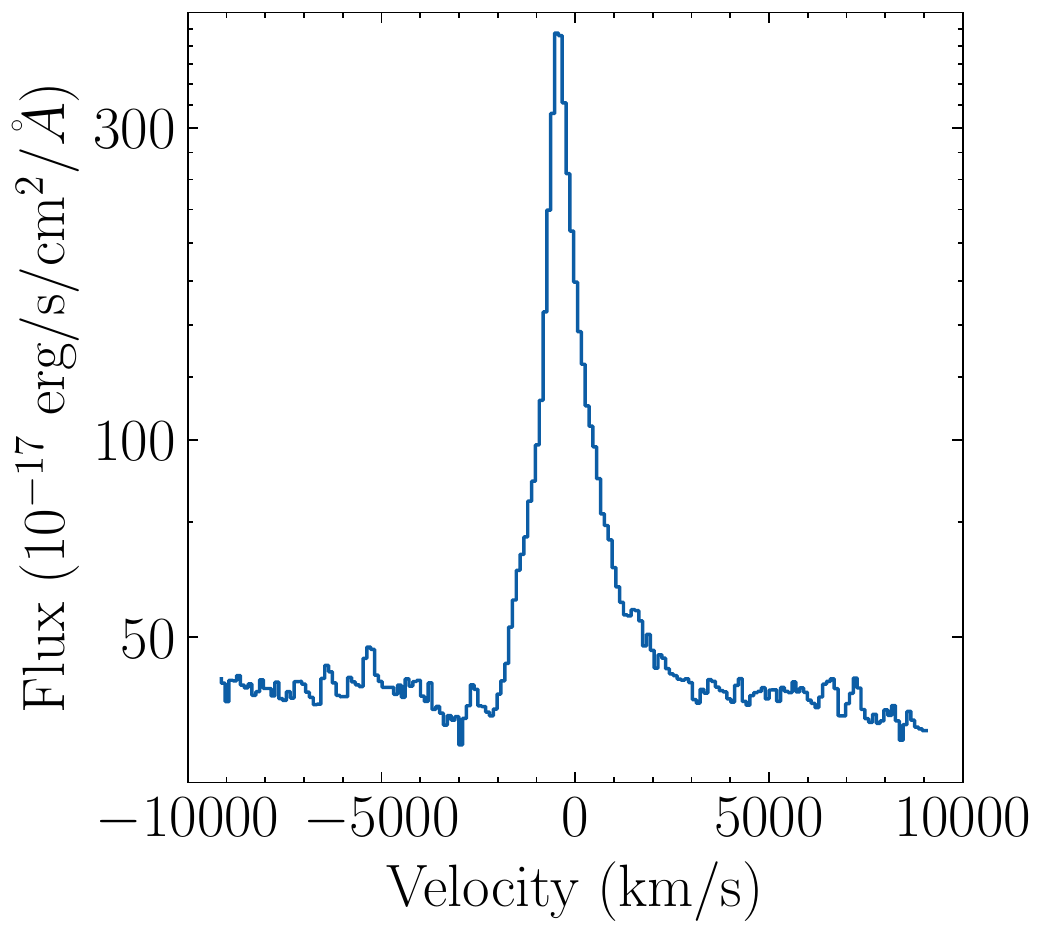}{0.24\textwidth}{H$\alpha$ broad line region.}
\fig{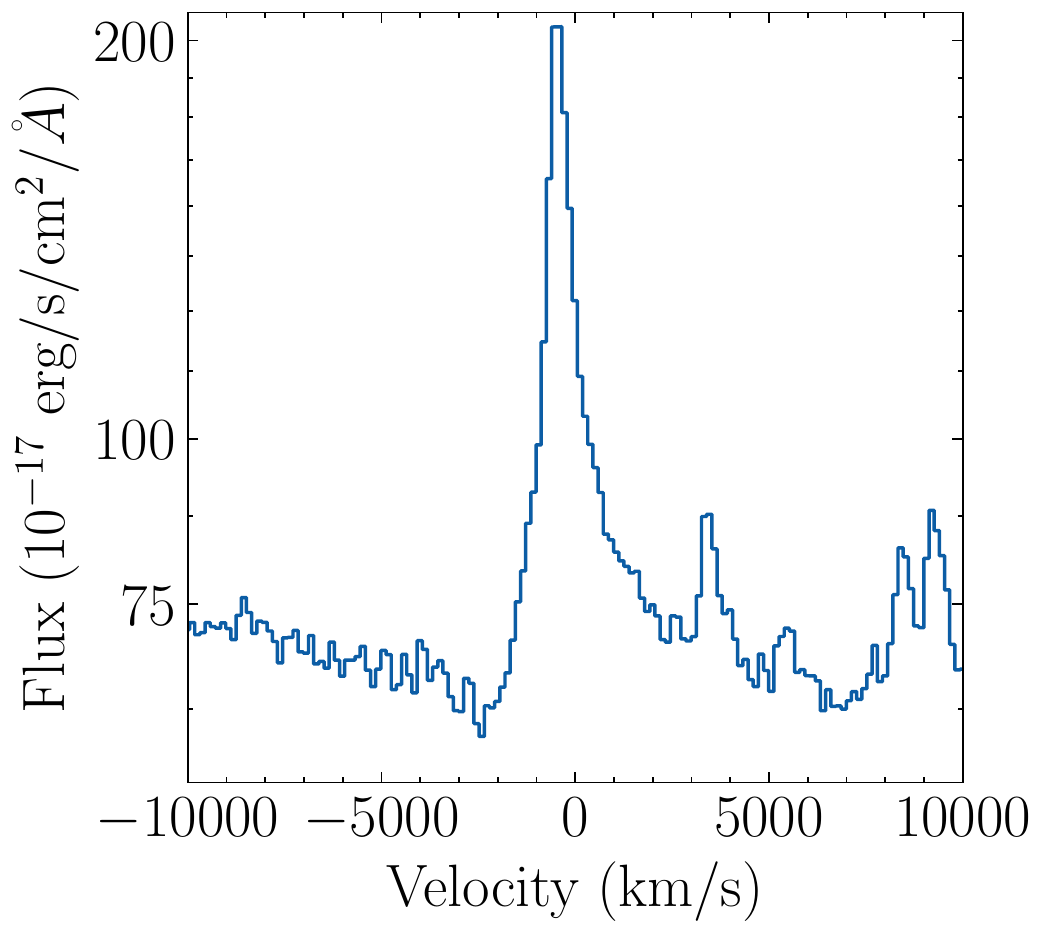}{0.24\textwidth}{H$\beta$ broad line region.}}
    \caption{Spectrum of the flare ZTF19aafmjfw from SDSS1133 taken on 2019-05-29 with the Deveny spectrograph on the Lowell Discovery Telescope. Subplot (a) shows the complete spectrum, while (b) and (c) show close-ups of H$\alpha$ and H$\beta$ in log scale to emphasize the P-cygni profile structure.}
    \label{fig:Koss_DCT}
\end{figure}

We searched for this object in archival data from the Catalina Real-time Transient Survey \citep{Drake2009FirstSurvey} and found that this object demonstrated another small scale flaring event in 2014. After showing no detectable V-band activity between 2006-02-03 and 2014-05-12, it brightened by V=0.22 between 2014-04-28 and 2014-06-05. After a gap in observations it had faded again by 2014-12-27. 

This object showed no evidence of variability in ZTF prior to 2019-04-07, when it became detectable at $m_g=20.21$ on 2019-04-07, flared to $m_g=17.00$ on 2019-06-05, and faded again by 2019-12-16. The ZTF flare ZTF19aafmjfw is shown in Figure \ref{fig:marshallc}. The 3 magnitude change in brightness seen in ZTF was of comparable scale to the SDSS flare in 2001. The ZTF rebrightening of this object suggests that the transient emission from 2001 to 2013 was not from a Type IIn supernova with a variable star progenitor. 

We obtained a spectrum of ZTF19aafmjfw with the Deveny spectrograph on LDT on 2019-05-29 when the object was bright. The spectrum showed the return of broad H$\alpha$ and H$\beta$ absorption lines blue-shifted over a 2000 to 8000 km s$^{-1}$ range (Figure \ref{fig:Koss_DCT}), as had been observed in December 2013 and January 2014 \citep{Koss2014}. It also showed the return of [Fe\,{\sc II}] $\lambda$7155 and [Ca\,{\sc II}] $\lambda \lambda$ 7291, 7324 lines -- features which have been seen only occasionally in AGN (e.g. \citet{Phillips1976The1}) spectra, late-time supernovae spectra \citep[e.g.][]{Filippenko1997OPTICALSUPERNOVAE,Pastorello2019A2018cnf} and outbursting LBVs \citep{Solovyeva2019New4736}. 

The 100 day timescale, 3-4 magnitude flux change and high velocity absorption lines make this outburst comparable to the transient SN2009ip, a supersonic stellar explosion from a $>60$\(\textup{M}_\odot\) star \citep{Smith2010,Foley2011TheProgenitors}. The presence of broad Balmer, He I and Na D lines and a very blue continuum also links ZTF19aafmjfw to SN2009ip. However, SN2009ip did not show  the forest of Fe group lines and [Ca\,{\sc II}] $\lambda \lambda$ 7291, 7324 emission lines which we observe in ZTF19aafmjfw. Such Fe and Ca features were observed in the stellar outburst UGC 2773 OT2009-1, which is likely to be either a giant LBV eruption or extreme S Dor variability with circumstellar dust eruption \citep{Smith2010,Foley2011TheProgenitors}. However, UGC 2773 OT2009-1 had a cooler temperature and weaker Balmer emission than SN2009ip, suggesting that it was a subsonic outburst rather than an explosion.

As ZTF19aafmjfw has spectroscopic features in common with SN2009ip which could indicate a high velocity explosion (blueshifted Balmer absorption lines, bright and broad Balmer emission and a blue continuum) and other spectroscopic features in common with UGC 2773 OT2009-1 which could indicate a stellar outburst within a circumstellar dust envelope (FeII and CaII emission lines), the observed spectroscopic features may be due to a non-terminal, supersonic LBV outburst in a dusty circumstellar environment rather.

We note, however, that the presence of such a rare and massive star in a dwarf galaxy which does not show recent star formation in Keck adaptive optics and HST imaging is very unlikely \citep{Koss2014}, and while the observed combination of spectroscopic features would be unusual for a standard AGN, they may be induced by a recoil event. We therefore cannot yet completely rule out the recoil hypothesis in favor of an LBV or supernova imposter explanation.

\section{Conclusions}

We have described a novel search strategy for the discovery of spatially offset AGN from recoiling SMBHs and ongoing galaxy mergers. This strategy uses ZTF difference imaging to find variable AGN and \texttt{Tractor} forward modeling to determine the AGN position across multiple ZTF epochs and in deep, high resolution Legacy Survey images. 

We have found a sample of 52 AGN in galaxy mergers and a sample of 9 AGN which may be spatially offset from their host galaxies. 5 of the 8 offset AGN for which we have spectra of the H$\alpha$ or H$\beta$ broad line regions show irregularly shaped broad Balmer lines with velocity offset broad peaks. These structures may arise due to emission from an unobscured, relativistic  Keplerian accretion disk around the AGN. The 63\% fraction of double-peaked emitters in the offset AGN sample is much larger than the 16\% observed for the whole ZTF AGN sample. The remaining 3 offset AGN with spectra have Gaussian broad lines with velocities ranging in magnitude from 200 to 600 km/s relative to the narrow emission lines.

Our search strategy detected the variable object and recoiling SMBH candidate SDSS1133 \citep{Koss2014} when it re-brightened by 3 magnitudes in ZTF. Follow-up spectra showed the return of a blue continuum, blue-shifted absorption lines and [Fe\,{\sc II}] $\lambda$7155 and [Ca\,{\sc II}] $\lambda \lambda$ 7291, 7324 forbidden lines, suggesting that the source may be either a luminous blue variable star which continues to show non-terminal outbursts or a recoiling AGN which continues to show variability.

Further multi-wavelength follow-up is required to confirm that our recoiling SMBH candidates are not AGN with undermassive hosts in mergers or chance coincidences with disturbed background galaxies. If even a subset of our offset AGN candidates are confirmed to be recoiling black holes, their large spatial offsets could show that SMBH binaries with misaligned spins are able to form. Such binaries may be detectable at a later stage of evolution by the Laser Interferometer Space Antenna (LISA) and would provide strong constraints on models of SMBH binary formation.  The success of our variability-based search strategy with ZTF suggests that future searches for offset AGN with the Vera Rubin Observatory \citep{Ivezic2019LSST:Products} may yield large populations of recoiling SMBH candidates and AGN in mergers.

\section{Acknowledgements}
We thank the anonymous reviewer for their helpful feedback. S. Gezari is supported in part by NSF grant 1616566. This work relied on observations obtained with the Samuel Oschin Telescope 48-inch and the 60-inch Telescope at the Palomar Observatory as part of the Zwicky Transient Facility (ZTF) survey. ZTF is supported by the National Science Foundation under Grant No. AST-1440341 and a collaboration including Caltech, IPAC, the Weizmann Institute for Science, the Oskar Klein Center at Stockholm University, the University of Maryland, the University of Washington, Deutsches Elektronen-Synchrotron and Humboldt University, Los Alamos National Laboratories, the TANGO Consortium of Taiwan, the University of Wisconsin at Milwaukee, and Lawrence Berkeley National Laboratories. Operations are conducted by COO, IPAC, and UW. SED Machine is based upon work supported by the National Science Foundation under Grant No. 1106171. This work was supported by the GROWTH project funded by the National Science Foundation under Grant No 1545949. MMK acknowledges generous support from the David and Lucille Packard Foundation.

This research used resources of the National Energy Research Scientific Computing Center, a DOE Office of Science User Facility supported by the Office of Science of the U.S. Department of Energy under Contract No. DE-AC02-05CH11231. P.E.N. acknowledges support from the DOE under grant DE-AC02-05CH11231, Analytical Modeling for Extreme-Scale Computing Environments.

These results made use of the Lowell Discovery Telescope (LDT) at Lowell Observatory. Lowell is a private, non-profit institution dedicated to astrophysical research and public appreciation of astronomy and operates the LDT in partnership with Boston University, the University of Maryland, the University of Toledo, Northern Arizona University and Yale University. The upgrade of the DeVeny optical spectrograph has been funded by a generous grant from John and Ginger Giovale.

This project used data obtained with the Dark Energy Camera (DECam), which was constructed by the Dark Energy Survey (DES) collaboration. Funding for the DES Projects has been provided by the U.S. Department of Energy, the U.S. National Science Foundation, the Ministry of Science and Education of Spain, the Science and Technology Facilities Council of the United Kingdom, the Higher Education Funding Council for England, the National Center for Supercomputing Applications at the University of Illinois at Urbana-Champaign, the Kavli Institute of Cosmological Physics at the University of Chicago, Center for Cosmology and Astro-Particle Physics at the Ohio State University, the Mitchell Institute for Fundamental Physics and Astronomy at Texas A\&M University, Financiadora de Estudos e Projetos, Fundacao Carlos Chagas Filho de Amparo, Financiadora de Estudos e Projetos, Fundacao Carlos Chagas Filho de Amparo a Pesquisa do Estado do Rio de Janeiro, Conselho Nacional de Desenvolvimento Cientifico e Tecnologico and the Ministerio da Ciencia, Tecnologia e Inovacao, the Deutsche Forschungsgemeinschaft and the Collaborating Institutions in the Dark Energy Survey. The Collaborating Institutions are Argonne National Laboratory, the University of California at Santa Cruz, the University of Cambridge, Centro de Investigaciones Energeticas, Medioambientales y Tecnologicas-Madrid, the University of Chicago, University College London, the DES-Brazil Consortium, the University of Edinburgh, the Eidgenossische Technische Hochschule (ETH) Zurich, Fermi National Accelerator Laboratory, the University of Illinois at Urbana-Champaign, the Institut de Ciencies de l'Espai (IEEC/CSIC), the Institut de Fisica d'Altes Energies, Lawrence Berkeley National Laboratory, the Ludwig-Maximilians Universitat Munchen and the associated Excellence Cluster Universe, the University of Michigan, the National Optical Astronomy Observatory, the University of Nottingham, the Ohio State University, the University of Pennsylvania, the University of Portsmouth, SLAC National Accelerator Laboratory, Stanford University, the University of Sussex, and Texas A\&M University.

\software{
Ampel \citep{Nordin2019},
Astropy \citep{Robitaille2013Astropy:Astronomy,Price-Whelan2018ThePackage}, 
catsHTM \citep{Soumagnac2018},
extcats (\url{github.com/MatteoGiomi/extcats}),
GROWTH Marshal \citep{Kasliwal_2019},
The Tractor \citep{2016ascl.soft04008L}.
}

\bibstyle{aasjournal}
\bibliography{references_linked}

\appendix
\begin{figure*}
\gridline{\fig{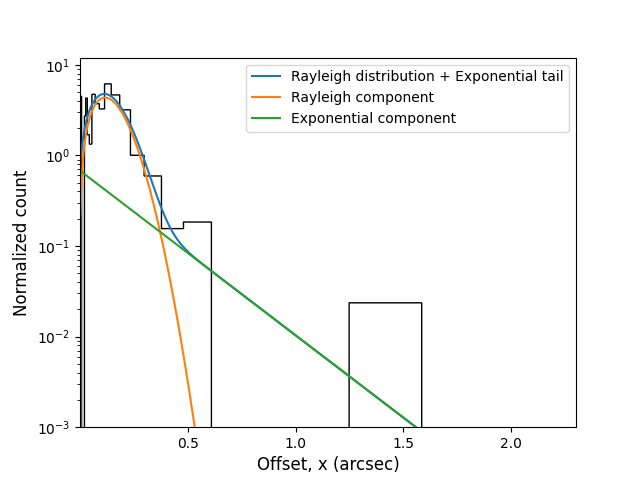}{0.45\textwidth}{(a) AGN with a peak difference magnitude between 15 and 18.}
          \fig{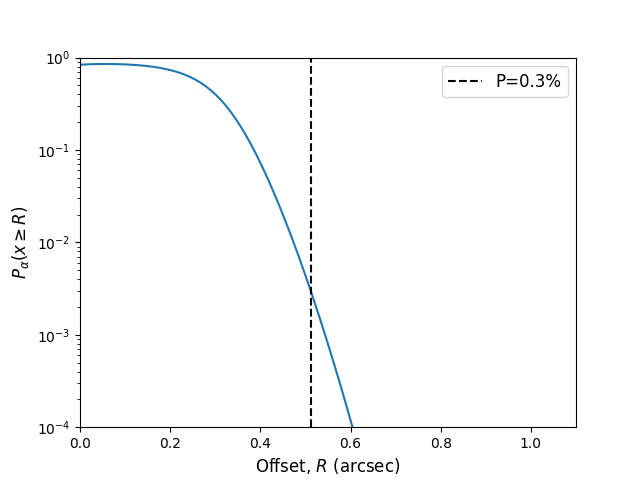}{0.45\textwidth}{(b) 3$\sigma$ offset = 0.551".}}
\gridline{\fig{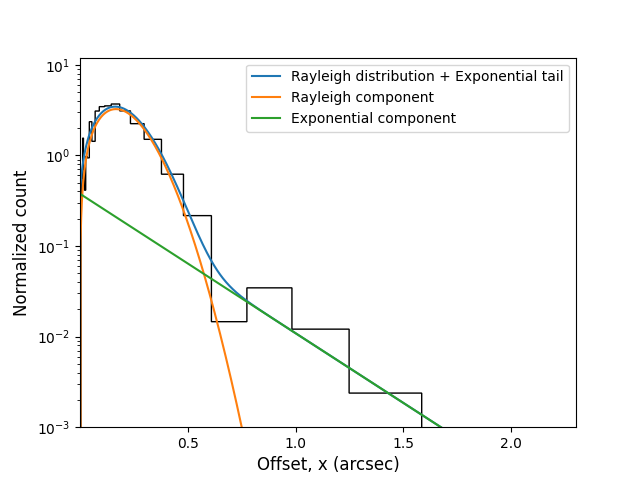}{0.45\textwidth}{(c) AGN with a peak difference magnitude between 18 and 19.5.}
          \fig{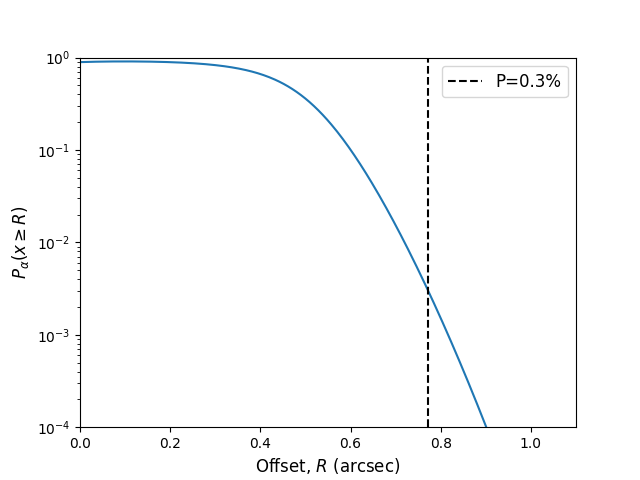}{0.45\textwidth}{(d) 3$\sigma$ offset = 0.773".}}
\gridline{\fig{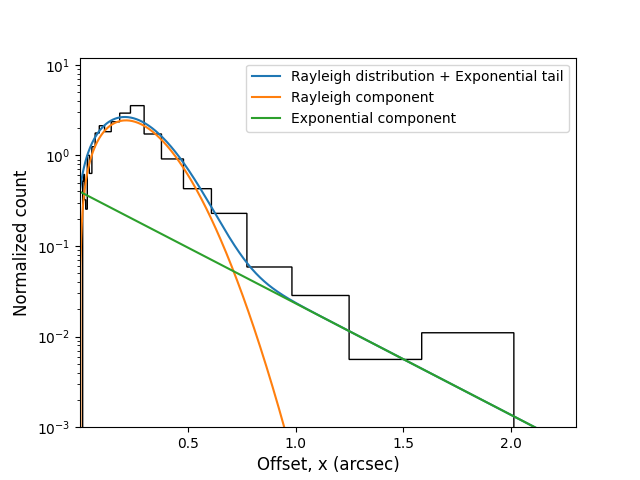}{0.45\textwidth}{(e) AGN with a peak difference magnitude between 19.5 and 23.}
          \fig{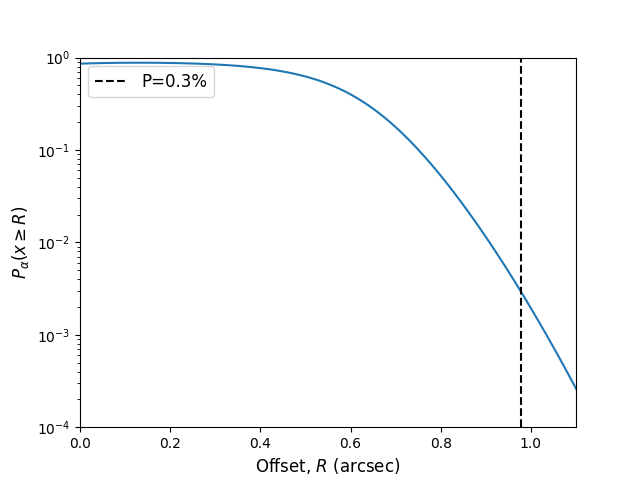}{0.45\textwidth}{(f) 3$\sigma$ offset = 0.976".}}          
\caption{\textbf{Left}: Normalized histogram with logarithmic bins for magnitude weighted AGN-host offsets obtained from r-band ZTF difference images. The best-fit model of a mixture distribution with Rayleigh and exponential components are shown. \textbf{Right:} Probability that an offset greater than R is drawn from the Rayleigh component of the mixture distribution shown in (a) instead of the exponential component. The offset where this probability is 0.3\% is shown with a dashed line.}
\label{fig:offset_r}
\end{figure*}

\begin{figure*}
\gridline{\fig{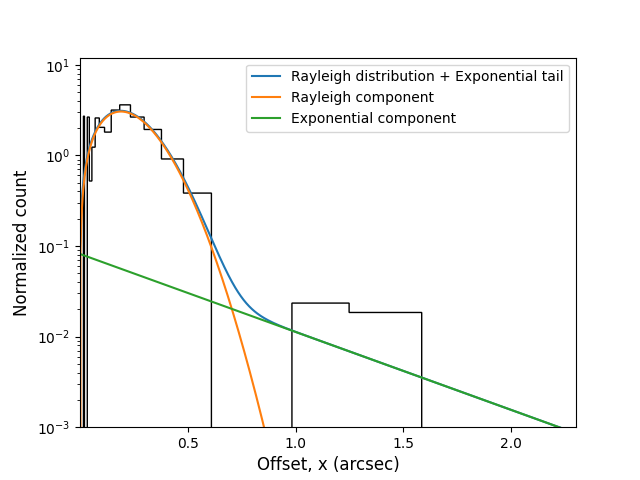}{0.45\textwidth}{(a) AGN with a peak difference magnitude between 15 and 18.}
          \fig{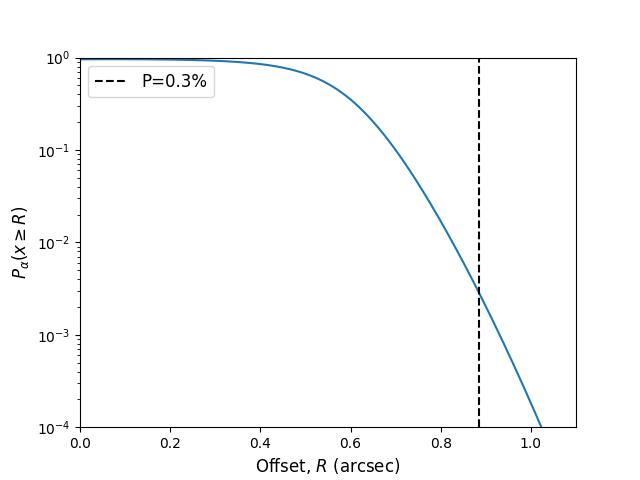}{0.4\textwidth}{(b) 3$\sigma$ offset = 0.959".}}
\gridline{\fig{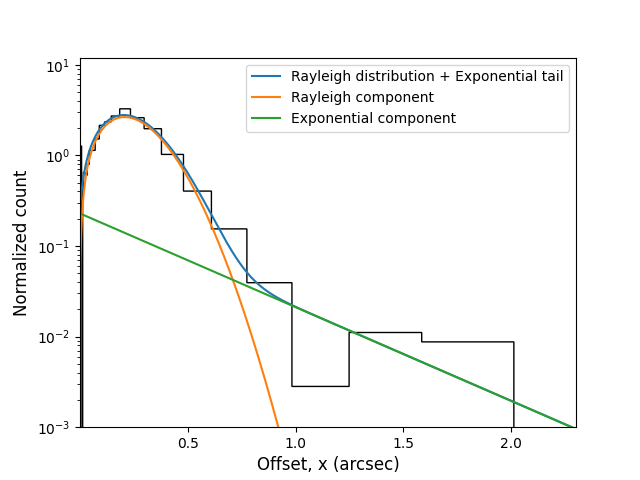}{0.4\textwidth}{(c) AGN with a peak difference magnitude between 18 and 19.5}
          \fig{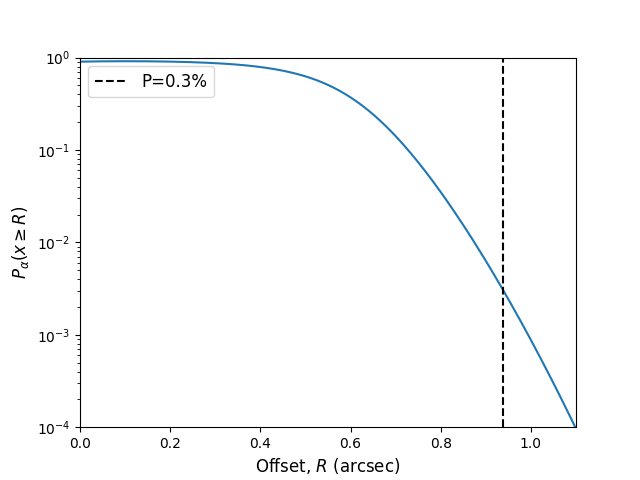}{0.4\textwidth}{(d) 3$\sigma$ offset = 1.009".}}
\gridline{\fig{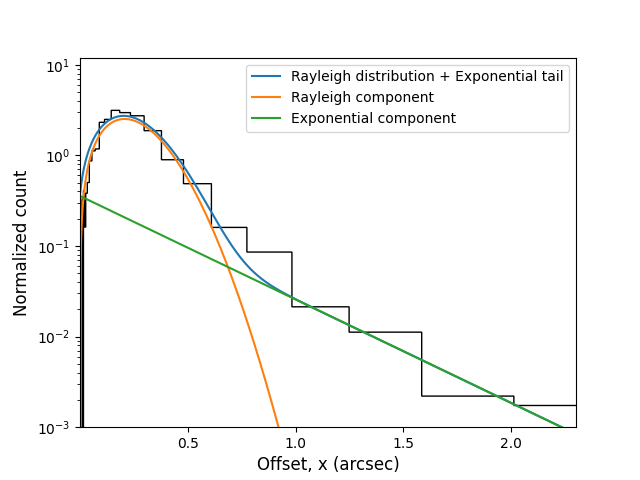}{0.4\textwidth}{(e) AGN with a peak difference magnitude between 19.5 and 23.}
          \fig{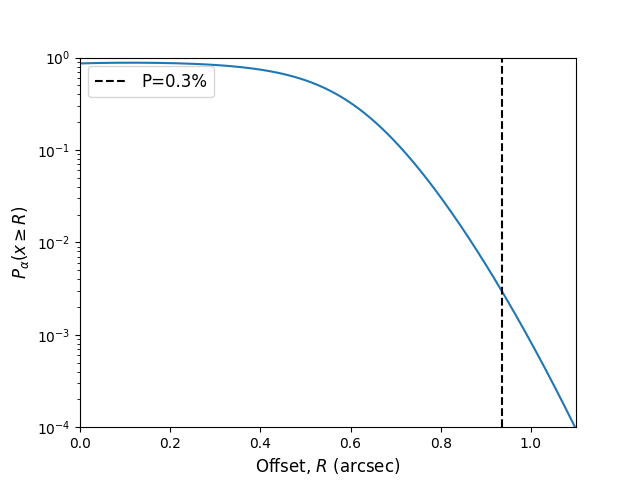}{0.4\textwidth}{(f) 3$\sigma$ offset = 0.946".}}          
\caption{\textbf{Left}: Normalized histogram with logarithmic bins for magnitude weighted AGN-host offsets obtained from g-band ZTF difference images. The best-fit model of a mixture distribution with Rayleigh and exponential components are shown. \textbf{Right:} Probability that an offset greater than R is drawn from the Rayleigh component of the mixture distribution shown in (a) instead of the exponential component. The offset where this probability is 0.3\% is shown with a dashed line.}
\label{fig:offset_g}
\end{figure*}

\begin{deluxetable*}{ccccc}
\tabletypesize{\scriptsize}
\tablecolumns{5}
\tablewidth{0pt}
\tablecaption{Galaxy merger X-ray and radio detections \label{table:rosat}}
\tablehead{
\colhead{ZTF name} & \colhead{ROSAT flux} & \colhead{L$_{\mathrm{2-10keV}}$} & \colhead{FIRST flux} & \colhead{L$_{\mathrm{5GHz}}$}\\
  & \colhead{(cts/s)} &\colhead{($\times 10^{44}$ ergs/s)}& \colhead{($\times 10^{24}$ ergs/cm$^2$/s)} &\colhead{($\times 10^{55}$ ergs/s)}}
\startdata
ZTF18aaxvmpg&$0.196\pm0.022$&$1.126\pm0.129$&$<1.39$&$<18.552$\\
ZTF18abamzru&$<0.05$&$-$&$<1.33$&$-$\\
ZTF18aasxvyo&$1.728\pm0.06$&$-$&$24.3\pm1.36$&$-$\\
ZTF18aaieguy&$<0.05$&$-$&$<1.54$&$<20.944$\\
ZTF19aakjemw&$0.031\pm0.011$&$0.149\pm0.054$&$<1.46$&$<8.666$\\
ZTF18aaifbku&$0.302\pm0.025$&$0.245\pm0.02$&$185.3\pm1.5$&$208.297\pm1.686$\\
ZTF18aampabj&$<0.05$&$-$&$24.1\pm1.34$&$335.637\pm18.662$\\
ZTF19aaagygp&$0.04\pm0.012$&$-$&$-$&$-$\\
ZTF18abujubn&$<0.05$&$-$&$-$&$-$\\
ZTF18acegbsb&$0.034\pm0.014$&$0.009\pm0.004$&$12.8\pm0.97$&$4.206\pm0.319$\\
ZTF19aaozpdm&$<0.05$&$-$&$<1.39$&$-$\\
ZTF18aacjltc&$0.051\pm0.015$&$0.529\pm0.159$&$<1.76$&$<18.156$\\
ZTF18abtpite&$0.024\pm0.01$&$-$&$-$&$-$\\
ZTF18abvwrxu&$0.163\pm0.03$&$-$&$<1.42$&$-$\\
ZTF19abaktpb&$<0.05$&$-$&$-$&$-$\\
ZTF18aaqjcxl&$<0.05$&$-$&$<1.47$&$<3.706$\\
ZTF18abyoivl&$<0.05$&$-$&$<1.1$&$-$\\
ZTF18aabdiug&$<0.05$&$-$&$37.4\pm1.43$&$35.732\pm1.366$\\
ZTF19aaviuyv&$0.051\pm0.01$&$-$&$-$&$-$\\
ZTF18aabxczq&$<0.05$&$-$&$<1.45$&$<1.43$\\
ZTF18acvwlrf&$<0.05$&$-$&$11.6\pm1.51$&$192.822\pm25.1$\\
ZTF19aasejqv&$0.086\pm0.016$&$0.721\pm0.135$&$<1.34$&$<22.161$\\
ZTF18aazogyo&$<0.05$&$-$&$<1.44$&$<2.41$\\
ZTF18aceypvy&$0.047\pm0.012$&$0.215\pm0.055$&$<1.39$&$<10.349$\\
ZTF18acbweyd&$<0.05$&$-$&$<1.36$&$<14.168$\\
ZTF18acablce&$0.022\pm0.005$&$-$&$-$&$-$\\
ZTF18abhpvvr&$0.072\pm0.014$&$-$&$-$&$-$\\
ZTF19abfqmjg&$<0.05$&$-$&$-$&$-$\\
ZTF18abmqwgr&$0.077\pm0.015$&$-$&$-$&$-$\\
ZTF19aadgbih&$0.092\pm0.016$&$0.603\pm0.102$&$24.9\pm1.5$&$279.113\pm16.814$\\
ZTF19aalpfan&$<0.05$&$-$&$<1.48$&$<2.056$\\
ZTF18aawwfep&$<0.05$&$-$&$<1.39$&$<15.897$\\
ZTF19aavxims&$0.041\pm0.015$&$-$&$<1.52$&$-$\\
ZTF19aaaplct&$0.165\pm0.025$&$-$&$<1.48$&$-$\\
ZTF18aajnqqv&$0.02\pm0.009$&$0.026\pm0.012$&$36.5\pm1.47$&$59.825\pm2.409$\\
ZTF18abszfur&$<0.05$&$-$&$<1.5$&$<41.386$\\
ZTF19abucbkt&$<0.05$&$-$&$-$&$-$\\
ZTF18adbhlyb&$0.036\pm0.013$&$0.253\pm0.09$&$<1.34$&$<17.931$\\
ZTF18acxhoij&$0.086\pm0.016$&$-$&$-$&$-$\\
ZTF18acajwep&$<0.05$&$-$&$-$&$-$\\
ZTF19abipoqj&$<0.05$&$-$&$<1.05$&$-$\\
ZTF19abpkoou&$0.043\pm0.016$&$-$&$<1.08$&$-$\\
ZTF18abztovy&$<0.05$&$-$&$<1.54$&$-$\\
ZTF18acsllgd&$0.031\pm0.012$&$-$&$-$&$-$\\
ZTF19aanxrki&$0.068\pm0.014$&$0.237\pm0.049$&$<1.44$&$<4.96$\\
ZTF18aamfuhc&$0.341\pm0.033$&$0.417\pm0.04$&$<1.42$&$<2.693$\\
ZTF18aadwvyr&$0.06\pm0.014$&$0.252\pm0.059$&$<1.02$&$<4.353$\\
ZTF19abauzsd&$0.027\pm0.009$&$0.643\pm0.206$&$<1.5$&$<39.366$\\
ZTF18abufbsq&$0.048\pm0.013$&$-$&$21.5\pm1.39$&$-$\\
ZTF18abzuzrg&$0.028\pm0.009$&$0.169\pm0.053$&$<1.45$&$<13.173$\\
ZTF18abtmcdb&$0.032\pm0.01$&$-$&$<1.11$&$-$\\
ZTF18aauhnby&$0.1\pm0.018$&$0.113\pm0.021$&$<1.53$&$<3.168$\\
\enddata
\vspace{0.1cm}
\tablecomments{X-ray and radio properties of the 52 ZTF AGN in galaxy mergers. We show the 2-10keV flux in cts/s for the AGN with detections in the second ROSAT All-Sky Survey and conversions to luminosity for those with a spectroscopically  confirmed redshift. We also the 20cm flux from the FIRST survey, including upper limits where available, and the corresponding luminosities.}
\end{deluxetable*}
\end{document}